\documentclass[journal]{IEEEtran}

\makeatletter
\def\ps@headings{%
\def\@oddhead{\mbox{}\scriptsize\rightmark \hfil \thepage}%
\def\@evenhead{\scriptsize\thepage \hfil \leftmark\mbox{}}%
\def\@oddfoot{}%
\def\@evenfoot{}}
\makeatother 
\usepackage{mathrsfs}
\usepackage{amsfonts}
\usepackage{array}       % ????¡Àšª??tabular?Dm{1cm}?š¹šº1š®??1?¡À?š®?D
\usepackage{amssymb}
\usepackage{amsmath}
\usepackage{graphicx}
\usepackage{multirow}   % ??DD¡Àšª??
\usepackage{amsthm}

\usepackage[lined,ruled,commentsnumbered]{algorithm2e}
\usepackage{epstopdf}
\usepackage{amssymb,amsmath}
\usepackage{multirow,url}
\usepackage{color,cite}
\usepackage{amsthm}
\theoremstyle{plain}
\newtheorem{theorem}{Theorem}
\newtheorem{lemma}{Lemma}

\newtheorem{definition}{Definition}
\newtheorem{proposition}{Proposition}

\ifodd 1212
\newcommand{\rev}[1]{{\color{blue}#1}} %revise of the text
\newcommand{\com}[1]{\textbf{\color{red} (COMMENT: #1) }} %comment of the text
\newcommand{\comg}[1]{\textbf{\color{green} (COMMENT: #1)}}
\newcommand{\response}[1]{\textbf{\color{green} (RESPONSE: #1)}} %response to comment
\else
\newcommand{\rev}[1]{#1}

\newcommand{\com}[1]{}
\newcommand{\comg}[1]{}
\newcommand{\response}[1]{}
\fi
\def\I{\mathcal{I}}
\def\J{\mathcal{J}}
\def\K{k}
\def\P{p}
\def\D{d}
\def\Da{\xi}
\def\Db{\varepsilon}
\def\hDa{\hat{\Da}}
\def\VDa{\sigma_{\Da}}
\def\VDb{\sigma_{\Db}}
\def\EDa{\mu_{\Da}}
\def\EDb{\mu_{\Db}}

\def\EX{\mathbb{E}}
\def\dt{\mathrm{d}}

\def\db{\textsc{db}}
\def\ms{\textsc{ms}}
\def\sym{\textsc{sym}}
\def\asy{\textsc{asy}}

\def\Utot{\Pi}				% \pi^{\mathrm{c}}_{\mathrm{sym}}
\def\Ums{\pi_{\ms}}				% \pi^{\mathrm{m}}
\def\Udb{\pi_{\db}}				% \pi^{\mathrm{m}}
\def\Udba{\bar{\pi}_{\db}}	
\def\Umsmin{\pi_{\ms}^{\mathrm{min}}} %\Ums^{\mathrm{min}}

\def\Kso{\K^{\circ}}	 	% K^{\mathrm{c}}_{\mathrm{sym}}
\def\Kdbsym{\K_{\textsc{(i)}}^{\sym}}	 % K^{\mathrm{d}}_{\mathrm{sym}}
\def\Kdbasy{\K_{\textsc{(i)}}^{\asy}} %K^{\mathrm{d}}_{\mathrm{asy}}
\def\Kmssym{\K_{\textsc{(ii)}}^{\sym}}	 % K^{\mathrm{m}}_{\mathrm{sym}}
\def\Kmsasy{\K_{\textsc{(ii)}}^{\asy}} %K^{\mathrm{m}}_{\mathrm{asy}}

\def\CTRdb{\Psi_{\textsc{(i)}}}
\def\Kdbctr{\K_{\textsc{(i)}}^{*}}
\def\Pdbctr{\P_{\textsc{(i)}}^{*}}
\def\phidb{\phi_{\textsc{(i)}}}

\def\CTRms{\Psi_{\textsc{(ii)}}}
\def\Kmsctr{\K_{\textsc{(ii)}}^{*}}
\def\Pmsctr{\P_{\textsc{(ii)}}^{*}}
\def\phims{\phi_{\textsc{(ii)}}}

\def\Nset{\mathcal{N}}  %% WSDs set
\def\n{n}     %% the index of WSD

\def\TDa{\Da_{0}}   %% total contract demand
\def\TDb{\Db_{0}}    %% total burst demand

\def\TKdb{K_{0}}  %% total reserved spectrum
\def\OTKdb{K_{\textsc{(i)}}}  %% actual reserved spectrum
\def\OTKms{K_{\textsc{(ii)}}}

\def\AUdb{\Udb^{\textsc{a}}}   %% additional revenue of db

\def\Pr{\text{Pr}}
\def\TD{D_{0}}   %% total demand

\def\ce{c^{\textsc{ex}}}

\def\c{c}

\def\bbd{spectrum}
\begin{document}

%\title{Contract-Based {\bbd} Reservation in Cognitive Radio Networks: A Database Model under Demand Uncertainty}
%\title{Contract-based {\bbd} Reservation in TV White Space Networks}
\title{Spectrum Reservation Contract Design in TV White Space Networks}
%\title{A Business Model for Geo-location Database Driven Cognitive Radio Networks}
%\title{White Space Broker by Geo-location Database}

\author{Yuan~Luo,~\IEEEmembership{Student~Member,~IEEE,}
		Lin~Gao,~\IEEEmembership{Member,~IEEE,}
        and~Jianwei~Huang,~\IEEEmembership{Senior Member,~IEEE}\\
%        (\textbf{\emph{Invited Paper}})% <-this % stops a space
\IEEEcompsocitemizethanks{
%        \IEEEcompsocthanksitem
%		This work is supported by the General Research Funds (Project Number CUHK 412713 and 14202814) established under the University Grant Committee of the Hong Kong Special Administrative Region, China.
		\IEEEcompsocthanksitem
		Yuan~Luo and Jianwei~Huang {(corresponding author)} are with Network Communications and Economics Lab (NCEL),
		Department of Information Engineering, The Chinese University of Hong Kong, HK,
		E-mail: \{ly011, jwhuang\}@ie.cuhk.edu.hk.
		\IEEEcompsocthanksitem
		Lin~Gao is with the Harbin Institute of Technology (HIT) Shenzhen Graduate School, E-mail: gaolin@hitsz.edu.cn.
\IEEEcompsocthanksitem
Part of the results have appeared in IEEE GLOBECOM 2012 \cite{yuan2012}.
 }
\vspace{-5mm}
 }

\maketitle
%\thispagestyle{empty}
%\vspace{-18mm}
\begin{abstract}
	In this paper, we study a broker-based TV white space market, where unlicensed white space devices (WSDs) purchase   white space spectrum from TV licensees via a third-party geo-location database (DB), which serves as a spectrum broker, reserving spectrum from TV licensees and then reselling the reserved spectrum to WSDs.
We propose a contract-theoretic framework for the database's spectrum reservation under demand stochasticity and information asymmetry. 
In such a framework, the database offers a set of contract items in the form of reservation amount and the corresponding payment, and each WSD chooses the best contract item based on its  private information. 
We systematically study the optimal reservation contract design (that maximizes the database's expected profit) under two different risk-bearing schemes: DB-bearing-risk and WSD-bearing-risk, depending on who (the database or the WSDs) will bear the risk of over reservation.
Counter-intuitively, we show that the optimal contract under DB-bearing-risk leads to a higher profit for the database and a higher total network profit. 
%	Given the optimal contract design and WSDs' private information realizations and reservation choices, we show that the database can further improve its total expected profit (from all WSDs) by taking advantage of the reservation reuse among WSDs.
%	Our numerical results show that with such an optimization, the database can improve its total profit up to $12\%$ (with a $30$ WSDs), without decreasing  WSDs' performance.
		
	\begin{IEEEkeywords}
		TV White Space Networks, {\bbd} Reservation, Contract Theory, Game Theory
%		\vspace{-2mm}
	\end{IEEEkeywords}
	
\end{abstract}
%\IEEEpeerreviewmaketitle

%\footnotetext[1]{This work is supported by the General Research Funds
%(Project Number CUHK412710 and CUHK412511) established under the
%University Grant Committee of the Hong Kong Special Administrative
%Region, China. Author emails: \{ly011,\ lgao,\ jwhuang\}@ie.cuhk.edu.hk.}

% \IEEEpeerreviewmaketitle

%\input{Section_introduction}
%!TEX root = main_CR_Database.tex
%SourceDoc main_CR_Database.tex

\section{Introduction}\label{sec:intro}

\subsection{Background and Motivations}

Nowadays, radio spectrum is becoming more congested and scarce with the explosive development of wireless services and networks.
%Secondary utilization of under-utilized licensed spectrums can effectively improve spectrum efficiency and alleviate spectrum scarcity.
%Dynamic spectrum access is a promising approach for improving the spectrum utilization efficiency and alleviating the spectrum scarcity.
%\footnote{For example, the UHF band between around 500MHz and 800MHz licensed for broadcast TV.}
%As an example, FCC has granted the secondary utilization of certain UHF band licensed for broadcast TV (called \emph{TV white spaces}) by the
%unlicensed devices (called \emph{white space devices}, WSDs) \cite{federal2008second,
%Ofcom2007ddr,ghosh2011cross}.
Dynamic spectrum sharing can effectively improve the spectrum efficiency and alleviate the spectrum scarcity,
by allowing unlicensed secondary devices access to the licensed spectrum in an opportunistic manner.
%by opening the licensed spectrum for secondary utilization.
\emph{TV white space network} is one of the promising paradigms of dynamic spectrum sharing \cite{federal2010second, ofcom2012}, where unlicensed devices (called white space devices, WSDs) exploit the un-used or under-utilized broadcast television spectrum (called TV white spaces, TVWS\footnote{For convenience, we simply use \emph{spectrum} to represent TVWS in this paper.}) opportunistically.

In order to fully utilize TVWS while not harming licensed devices, regulatory bodies (e.g., FCC in the US and Ofcom 
in the UK)
%and standardization organizations (e.g., IEEE  and  ECC)
have advocated a database-assisted spectrum access solution,
which relies on a third-party white space database called \emph{geo-location}
\cite{ofcom2012, federal2010second}.\footnote{{Based on the database-assisted solution proposed by the regulators, The IEEE 802.22 \cite{IEEE80222}, CEPT ECC \cite{ECC2010dractreport}, and ETSI \cite{ETSIreport} proposed corresponding standards for WSDs operating in a database-assisted TV white space network.}}
In this solution, WSDs obtain the available spectrum information through querying the geo-location database, instead of performing spectrum sensing.
More specifically, WSDs periodically report their location information and other optional information (e.g., spectrum demand) to a geo-location database, and then the database returns the available spectrum in the respective locations and time periods to WSDs.

In general, there are two types of different TV white space spectrum resources. 
The first type is the TV spectrum \emph{not} registered to any TV licensee or Programme Making and Special Events (PMSE) at a particular location.
%	, hence is the public resource at that location.
This type of spectrum is usually for the open and shared usage among unlicensed WSDs, according to the regulators' policies \cite{federal2010second}.
%	To provide the information about these un-registered spectrum, the geo-location database needs to house a global repository of TV licensees and maintains
%	the up-to-date spectrum occupation information of TV licensees, with which it can identify the un-registered spectrum at any given time and location.
The second type is the TV spectrum already registered to some TV licensees and PMSE, but \emph{not} fully utilized by those licensees. 
Hence, the licensees can temporarily lease these idle spectrum to unlicensed WSDs for the exclusive usage. 
In such a secondary spectrum market, the geo-location database can act as an intermediary (e.g., a broker) between the licensees (sellers) and the WSDs (buyers), due to its proximity to both sides of the market.\footnote{\rev{This model is currently employed by real-world geo-location database operators such as SpectrumBridge (\url{https://spectrumbridge.com/}) in the US and COGEU (\url{http://www.ict-cogeu.eu/}) in Europe.}}

In this work, we focus on the secondary sharing and trading of the second type spectrum resource, i.e., those registered but under-utilized spectrum. 
Such spectrum can be exclusively used by a WSD (with the permission of the licensees), hence are particularly suitable for supporting applications that require a high QoS.
%Similar to that employed by SpectrumBridge \cite{spectrumbridge},

\vspace{-2mm}
\subsection{Market Model and Problem}

\begin{figure}[t]
\vspace{-2mm}
  \centering
  \includegraphics[width=3.5in]{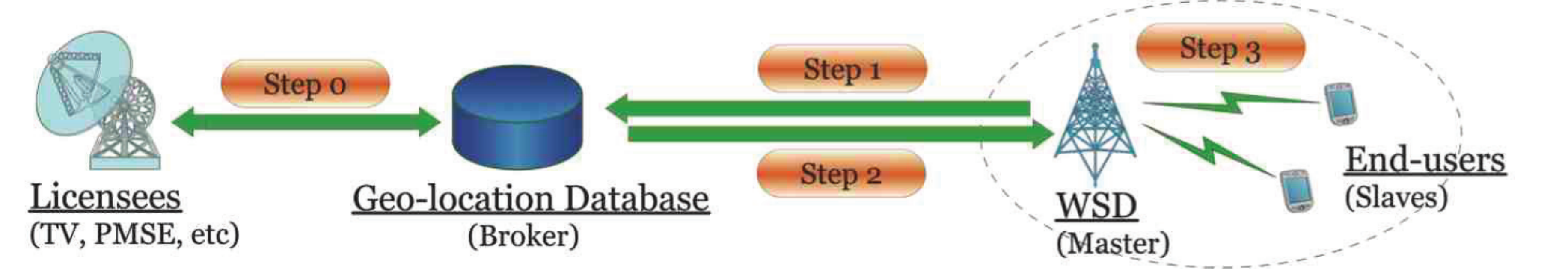}
\vspace{-4mm}
  \caption{Broker-based spectrum reservation market. In Step 0,
the geo-location database (broker) reserves spectrum from TV licensees and PMSE for every reservation period (e.g., one day).
In Step 1, each WSD (master) reports its location and demand in every access period (e.g., one hour).
In Step 2, the database sells the corresponding spectrum to the WSD in every access period.
In Step 3, the WSD serves end-users (slaves) in every access period using the obtained spectrum.
Notice that Steps 1-3 will occur repeatedly within every reservation period, as one reservation period consists of many access periods.}
% At Step 0, the database updates TV licensees' spectrum occupation information periodically.
% At Step 1, the WSD reports its location to the database.
% At Step 2, the database returns the available spectrum to the WSD.
%  At Step 3, the WSD serves end-users by using the returned spectrum.
%  At step 0, the database reserves {\bbd} from TV licensees for every 2 hours. At steps 1 and 2, the WSD reports the necessary information (e.g., location, {\bbd} requirement, etc.), and then the database computes and returns the available {\bbd} to the WSD, in every minute. At step 3, the WSD serves the end-users with the allocated {\bbd} in every minute.
%the access period is much smaller than the reservation period.
%\com{Yuan: In such case, we may just replace figure 1 with figure 2?}
\vspace{-3mm}
\label{fig:system-mdoel}
\end{figure}

Specifically, we study a broker-based secondary spectrum market, where TV licensees lease their idle spectrum to  unlicensed WSDs via a \emph{spectrum broker} acted by a geo-location database.
%\footnote{{Some existing studies have considered the broker-based secondary market in the cognitive radio literature (e.g., \cite{niyato2009}-\cite{Qian2011}). These prior studies mainly focused on competition issue in secondary market. None of them considered the problem of spectrum reservation under demand uncertainty and information asymmetry. Without a good understanding of this issue, it is difficult to utilize the TV spectrum efficiently and to have large-scale commercial deployment of the network.
%%\cite{niyato2009,niyato2008,Hossain2009,niyato2008,niyato2012,Han2011,Qian2011}.
%}}.
As a broker, the database purchases spectrum from TV licensees in advance, and then resells the leased spectrum to WSDs.
Figure \ref{fig:system-mdoel} illustrates such a broker-based spectrum reservation market model.
%Notice that TV licensees may not want to be involved into the  spectrum trading process frequently.
{As the TV towers have fixed locations and TV programs have well planned schedules, the reservation period of TV spectrum can be relative long \cite{bogucka}.
	Thus, we model and analyze a \emph{spectrum reservation market}, where the database reserves spectrum from TV licensees in advance for a relatively large time period (e.g., more than one day), called the \emph{reservation period}. Then, within each reservation period, the database sells the reserved spectrum to WSDs periodically with a relatively small time period ({e.g., one hour}), called the \emph{access period}. Namely, the spectrum reservation decision is made at the beginning of the reservation period, which consists of many access periods.} \footnote{{Please refer to Section \ref{sec:reservation_market_detailed} for the detailed model.}}

In such a spectrum reservation market, the database needs to reserve spectrum in advance, without knowing the actual future demands from WSDs.
Therefore, an important problem for the database in this market is:
%In this paper, we want to study the following spectrum reservation problem:
\begin{itemize}
\item
\emph{How much bandwidth should the database reserve for each WSD, aiming at maximizing the database's profit}?
\end{itemize}
The problem is challenging due to the demand stochasticity and the information asymmetry.
% (i) the spectrum demand (from WSDs as well as their served end-users) is uncertain, and moreover (ii) this demand information is asymmetric between the database and WSDs.

\emph{(i) Demand Stochasticity}.
Due to the stochastic nature of end-users' activities and requirements, each WSD's spectrum demand (for serving its end-users) is a random variable, and cannot be precisely predicted by the WSD or the database in advance.
Therefore, there is inevitably a risk of {reservation mismatch}, e.g., spectrum over-reservation or under-reservation.  Accordingly, the database's {\bbd} reservation decision depends on the \emph{risk-bearing} scheme, namely, who will bear the risk of {{\bbd} over-reservation}: the database (called  {DB-bearing-risk}) or the WSD (called  {WSD-bearing-risk})?
In the former case, the WSD only pays for the spectrum it actually purchases in every access period; while in the latter case, the WSD has to pay for the reserved spectrum(even if it is more than actually needed) in every access period.\footnote{\rev{
	The DB-bearing-risk scheme is widely used in manufacturing outsourcing systems such as {\cite{ cachon2001contract,ozer2006}}, while the WSD-bearing-risk scheme is widely used in many Newsvendor models and   practical retailing markets such as {\cite{niyato2008}}.}
}
%\footnote{Both risk-bearing schemes are widely used in practice.
%The DB-bearing-risk scheme is widely used in manufacturing outsourcing systems such as {\cite{ cachon2001contract,ozer2006}}, where the outsourced manufactures (similar as the database in this work) need to reserve resource (such as building factories and investing production lines) to complete the future orders from the outsourcers (similar as WSDs in this work).
%The WSD-bearing-risk scheme is widely used in many Newsvendor models and   practical retailing markets such as {\cite{niyato2008}}.
%}
%We will study the {{\bbd} reservation} solutions under both risk-bearing schemes systematically.

  \emph{(ii) Information Asymmetry}.
The above mentioned demand information is asymmetric between the database and WSDs.
Due to the proximity to end-users, the WSD usually has more information (i.e., with less uncertainty) about the spectrum demand than the database.
%Thus, there is a willingness for the database to obtain the WSD's private information (about the demand) in order to make better reservation decisions.
This implies that the database can potentially make a better reservation decision, if it is able to know the WSD's private information regarding the demand.
However, without proper \emph{incentives}, the WSD may not be willing to share its private information with the database. {As will be shown in Section 5, the WSD may even report a false information to the database intentionally, as long as such a misreport can increase the WSD profit.}

\subsection{Results and Contributions}

% More specifically, we focus on the database's spectrum reservation problem in a TV white space network with one database and multiple WSDs (possibly co-located), where each WSD serves a distinct set of end-users. 
%, and solve the spectrum reservation problem under demand stochasticity and information asymmetry.
%\rev{first focus on the interaction between the database and a WSD} and solve the above spectrum reservation problem under  demand stochasticity and information asymmetry systematically. 
We propose a \emph{contract-theoretic} {\bbd} reservation framework, in which the database offers a list of contract items in the form of reservation amount and the corresponding payment, and each WSD chooses the best contract item based on its private demand information (from its served end-users).
We first study the \emph{incentive compatible} contract design, under which each WSD will disclose its private demand information credibly, by choosing the contract item intended for its private information.
With the incentive compatibility, we further derive the optimal {\bbd} reservation contracts that maximize the database expected profit under both DB-bearing-risk and WSD-bearing-risk schemes.
%%(Sections \ref{sec:contract-DB} and \ref{sec:contract-master}).
%Moreover, to evaluate the performance of the proposed optimal contracts, we also provide the {\bbd} reservation solutions under information symmetry (in Sections \ref{sec:asymm_database-sym} and \ref{sec:asymm_WSD-sym})
%and information asymmetry without information sharing
%(in Sections \ref{sec:asymm_database-asym} and \ref{sec:asymm_WSD-asym})
%as benchmarks.
%%\footnote{Under the WSD-bearing-risk scheme, each WSD will make the reservation decision in the benchmark solution, as it needs to bear the risk of over-reservation.}
For clarity, we summarize the key  results {regarding the optimal contract design} in Table~\ref{table:key-result}.

As far as we know, this is the first paper that systematically studies the contract-based {\bbd} reservations under different risk-bearing schemes for TV white space markets. 
The proposed market model, together with the derived {\bbd} reservation solutions, can offer the proper economic incentives for the database operator, and support the practical and commercial deployment of TV white space networks.
The main contributions of this paper are summarized as follows.

\begin{itemize}
%  \item \emph{Practical significance}:
%We model a broker-based spectrum reservation market, and study the database's optimal {\bbd} reservation systematically.

        %under demand uncertainty and information asymmetry.

\item \emph{Novel modeling and solution techniques}:
        We study a generic spectrum reservation market under demand stochasticity and information asymmetry, and propose a contract-theoretic reservation framework, which ensures that WSDs disclose their private   information truthfully, and meanwhile maximizes the database profit.

% \item \emph{Multiple risk-bearing schemes}:
%        We consider two different risk-bearing schemes, depending on who (the database or WSDs) will bear the risk of {\bbd} over-reservation.  This renders our work ...

% \item \emph{Multiple information scenarios}: We consider two information scenarios, depending on whether the database and WSDs experience the same degree of uncertainty about the market demand. This renders our work ...
% \item \emph{Demand uncertainty and information asymmetry}: We derive the optimal {\bbd} reservations under demand uncertainty and (demand) information asymmetry, for both risk-bearing schemes.
%  \item  \emph{Under information symmetry}, the database and the WSD hold the same degree of uncertainty about the market demand. We derive the benchmark optimal centralized  {\bbd} reservations.
%  \item  \emph{Under information asymmetry}, the WSD has less uncertainty about demand (due to his proximity to end-users) than the database. We propose contract-based {\bbd} reservation mechanisms, which ensures the WSD shares his local information with the database credibly.

\item \emph{Optimal contract design}:
    %To ensure WSDs credibly share their exclusive demand information with the database (under information asymmetry), we propose contract-based {\bbd} reservation mechanisms, and further derive the optimal reservation contracts for both risk-bearing schemes.
    We analytically derive the optimal {\bbd} reservation contract design  under   DB-bearing-risk and WSD-bearing-risk schemes, and numerically compare their performances.
%    We also compare these optimal contracts with benchmark solutions via numerical studies.
Through these numerical comparisons, we characterize the impacts of risk-bearing scheme,  demand stochasticity, and information asymmetry on the {\bbd} reservation solutions.

%\item \emph{General reservation framework}: Our proposed {\bbd} reservation framework focuses only on the reservation decision and the necessary economic incentives, while does not need to alter the real spectrum trading process.
%    This implies that our results are compatible with (and thus applicable to) many existing spectrum market scenarios.

%  \item \emph{Performance comparison}: Simulations show that under information asymmetry, the optimal {\bbd} reservation contract improves both the database profit and the social welfare (without sacrificing the WSDs' benefits), comparing to those without information sharing.
 \item \emph{Numerical results and insights}:
Our numerical results show that the optimal contract under the DB-bearing-risk scheme can achieve a higher  database profit and a higher total network profit, compared to the optimal contract under the WSD-bearing-risk scheme.
%the database can achieve a higher maximum profit under the scheme that it bears all the risk.
%This provides the following important insight: it is not only individually better but also socially better to leave the over-reservation risk to the less risk-averse individual (i.e., the database in our model).
The intuition is that the WSD is more risk-averse than the database.
%Therefore, our results provide the following important insight for a general reservation problem: \emph{it is not only individually better, but also socially better to leave the over-reservation risk to the less risk-averse decision maker}.
\end{itemize}

\begin{table*}[t]
\centering
\caption{Key Results in This Paper}
\label{table:key-result}
\footnotesize
\begin{tabular}{|p{2.7cm}||p{1.7cm}|p{2.2cm}|p{6cm}|c|c|}
\hline
\centering \textbf{Risk-Bearing}
&
\multicolumn{2}{|c|}{\textbf{Information \& Sharing}}
& \centering \textbf{Spectrum Reservation Decision}
& \textbf{Solution}
& \textbf{Section}
\\
\hline
\hline
& \centering \textbf{Symmetry} {\scriptsize(\emph{Benchmark})}
& \centering ---
&
The database makes the reservation decision based on the WSD's knowledge about demand.
%\footnote{In fact, the database makes decision based its own knowledge about demand, which is assumed to be same to the WSD's knowledge under information symmetry.}
 & $\Kdbsym
 %= \Da + {G^{-1} \Big( \frac{w-c}{w}  \Big)}
 $ in Eq.~(\ref{sym_dencentral_bandwidth})
  &    \ref{sec:asymm_database-sym}
   \\
 \cline{2-6}
\centering \textbf{DB-Bearing-Risk}
{\scriptsize(\emph{Scheme I})}
&
\centering \textbf{Asymmetry} {\scriptsize(\emph{Benchmark})}
&
\centering No Sharing
& The database makes the reservation decision based on its own knowledge about demand.
 & $
\Kdbasy
%=  (F \times G )^{-1} \Big( \frac{w-c}{w}  \Big),
$ in Eq.~(\ref{asym_decentral_bandwidth})
 &     \ref{sec:asymm_database-asym}
   \\
 \cline{2-6}
 & \centering \textbf{Asymmetry}
 {\scriptsize(\emph{Our Focus})}
&
\centering Credibly Sharing via Contract
& The database offers a \textbf{spectrum reservation contract}, and each WSD chooses a proper contract item (reservation-payment pair).
& $
 %\CTRdb^* = \{ \langle
 \Kdbctr
 %(\Da),{\Pdbctr}(\Da) \rangle \}_{\forall \Da}
 $ in Theorem \ref{thrm:optimal-contract}
 &    \ref{sec:contract-DB}
   \\
\hline
%\multirow{3}*{\textbf{WSD-Bearing-Risk}}
&
\centering \textbf{Symmetry}
{\scriptsize(\emph{Benchmark})}
&
\centering ---
& Each WSD makes the reservation decision based on its knowledge about demand.
 & $\Kmssym
$ in  Eq.~(\ref{sym_dencentral_bandwidth2})
 &    \ref{sec:asymm_WSD-sym}
   \\
 \cline{2-6}
\centering \textbf{WSD-Bearing-Risk}
{\scriptsize(\emph{Scheme II})}
 &
\centering \textbf{Asymmetry}
{\scriptsize(\emph{Benchmark})}
&
\centering No Sharing
& Each WSD makes the reservation decision based on its knowledge about demand.
 & $\Kmsasy $ in Eq.~(\ref{asym_decentral_bandwidth2})
 &    \ref{sec:asymm_WSD-asym}
   \\
 \cline{2-6}
 & \centering \textbf{Asymmetry}
 {\scriptsize(\emph{Our Focus})}
&
\centering Credibly Sharing via Contract
& The database offers a \textbf{spectrum reservation contract}, and each WSD chooses a proper contract item (reservation-payment pair).
 &
 $\Kmsctr$ in Theorem \ref{thrm:optimal-contract-AD}
 &    \ref{sec:contract-WSD}
   \\
   \hline
\end{tabular}
\vspace{-4mm}
 \end{table*}

The rest of this paper is organized as follows. In Section
\ref{sec:related}, we review the related literature.
In Section \ref{sec:model}, we present the system model.
In Sections \ref{sec:integrated}, we provide the integrated optimal reservation solution as a benchmark.
In Sections \ref{sec:asymm-no} and \ref{sec:asymm-contract}, we study the decentralized {\bbd} reservations without information sharing and with information sharing (via contract), respectively.
%In Section \ref{sec:contract_multiple_short}, we study the optimal aggregate reservation strategy of the database after knowing WSDs' contract item choices.
We provide numerical results in Section \ref{sec:simu}, and finally conclude in Section \ref{sec:conclusion}.

%\input{Section_literature}
%!TEX root = main_CR_Database.tex
%SourceDoc main_CR_Database.tex

\section{Related Work}\label{sec:related}

In the recent regulator's policy \cite{ECC2010dractreport}, the databases are allowed to determine their own pricing schemes for operating the TVWS. 
This motivates researchers to study the economic issues in TVWS\cite{feng2013database, luo2013,luo2014wiopt,luo2014SDP,luo2015information,luo2015magazine,luo2015INFOCOM}. 
In \cite{feng2013database}, Feng \emph{et al.} studied the hybrid pricing scheme for the database manager.
In \cite{luo2013}, Luo \emph{et al.} studied the pricing strategy of oligopoly competitive WSDs.
However, none of the existing work considered the bandwidth reservation problem under information asymmetry.
Some recent studies \cite{luo2014wiopt,luo2014SDP,luo2015information,luo2015magazine,luo2015INFOCOM} proposed the pure and hybrid information models for TV white spaces, which focus on unlicensed TV white space.
%Thus, we focus on the {business modeling} for a geo-location database in this paper. In particular, we formulate a white space market, and study the bandwidth reservation problem in this market.~~~~~~~~~~~

Our work is related to the supply chain contract design in  the operations management and marketing science literature.
Supply chain contract is  widely used as a mechanism to coordinate production quantity and pricing, so that the performance of decentralized supply chain is close or the same as that of an integrated one.
%(see \cite{larivier1999contract} for recent surveys).
%In \cite{corbett2004contract}, Corbet \emph{et al.} used the bilateral monopoly setting to analyze three contracts under full and incomplete information in the setting of deterministic demand.
In \cite{cachon2001contract}, Cachon \emph{et al.} considered the stochastic nature of demand and prescribed analytical remedies for credible information sharing between a supplier and a manufacturer.
$\ddot{O}$zer \emph{et al.} in \cite{ozer2006} extended Cachon's work and further examined how a supplier can screen buyers's private information by offering a menu of contracts.
However, the above work considered the case where the contract designer bears all of the risk of over-reservation. We consider both cases where the contract designer (the database) and the buyer (WSD) bears the risk of over-reservation, respectively.

Recently, the concept of contract was also introduced into the spectrum trading model (e.g. \cite{gao2011contract,duan2013contract,sheng2013contract}).
%In \cite{kalathil2011contract},
%%authors
%%focus on single \rev{licensed user} and single \rev{unlicensed user}, and
%Kalathil \emph{et al.} proposed a contract-based spectrum sharing mechanism between a single licensed user and single unlicensed user to avoid possible manipulating in an auction.
%%However, implementations of TVWS services are likely to start with point-to-multipoint deployments \cite{fitch2011cr}.
%%The contract scheme proposed by Kalathil \emph{et al.} is not applicable to the database-assistance network.
In \cite{gao2011contract}, Gao \emph{et al.} proposed a quality-price contract for the spectrum trading in a monopoly spectrum market.
In \cite{duan2013contract}, Duan \emph{et al.} proposed a contract-based cooperative spectrum sharing mechanism to promote the cooperation of a primary user and a secondary user.
In \cite{sheng2013contract}, Sheng \emph{et al.} proposed a contract for a primary license holder to sell its excess spectrum capacity to potential secondary users.
In this paper, we propose a contract-based mechanism for the spectrum reservation problem.
In our model, the demand of a WSD consists of two parts: one is unknown by both the database and the WSD, and the other is only known by the WSD (hence is the WSD's private information).
Thus, the optimal contract design needs to consider not only  the  truthful information disclosure of the WSD, but also the  uncertainty of demand for both the database and the WSD.
This makes our contract design much more challenging than existing contract designs.
%In \cite{yuan2012}, Yuan \emph{et al.} proposed a similar spectrum reservation contract for the database in TV white space networks. However, they only considered the case where the database bears the risk of over-reservation, while we consider both cases where the database and the WSD bears the risk, respectively.

%Compared with the above works, our model considers a more general and more practical demand modeling.

%, often in the forms of pricing (e.g.,\cite{wang2008price}, \cite{niyato2008price}), contract(e.g., \cite{kalathil2011contract}, \cite{gXao2011contract} ), and auction (e.g., \cite{huang2006aunction}-\cite{zhou2008aunction}). \emph{Pricing} is often used when the seller knows precisely the value of the resource being sold. \emph{Contract} becomes a suitable approach when the seller only knows limited information (e.g., distribution) of the buyers' information (e.g., demand, valuations of the resource). \emph{Auction} is more effective in the case where the seller has no knowledge about the value of the resource. Concerning the database driven white space network structure (i.e., TVBD needs to report partial information to geo-location database), contract-based spectrum trading is the most favorable.

%\input{Section_model}
%!TEX root = main_CR_Database.tex
%SourceDoc main_CR_Database.tex

\section{System Model}\label{sec:model}

\subsection{System Overview}\label{sec:model-over}

%We consider a \emph{TV white space network}, where unlicensed white space devices (WSDs) operates on the TV white space in an opportunistic manner.
%%A WSD is an unlicensed device operating on TV white spaces in an opportunistic manner.
%A WSD can be either a personal/portable device performing its own services, or an \emph{infrastructure-based} device (e.g., a base station as shown in Figure \ref{fig:system-mdoel}) serving a set of unlicensed devices called end-users/devices (also called ``slave'' devices in \cite{Ofcom2010geo}).
%In the latter case, WSDs are also called as ``WSD'' devices in \cite{Ofcom2010geo}.\footnote{According to \cite{Ofcom2010geo},  the key difference between a slave device  and a WSD device is that ...}
%In this work, we focus on the operation of a particular WSD (called \emph{WSD}) in a white space network.

We consider a TV white space network where unlicensed
WSDs exploit the under-utilized broadcast television spectrum (called TV white space, or \emph{spectrum} for simplicity) via a
\emph{geo-location} database.
Each WSD is an infrastructure-based device (e.g., a base station), and serves a set of unlicensed end-users/devices called ``slave'' devices.
%In this context, WSDs are also referred to as ``WSD'' devices as in \cite{ofcom2012}, or called WSDs.
We assume that the number of unlicensed WSDs is large enough, so that the spectrum demand of a particular WSD does not affect other WSDs' demand.
This allows us to concentrate on the interaction between the database and each WSD.
%\com{Under this assumption, seems that the database needs to reserve spectrum separately for each WSD, as different WSDs correspond to different locations? This means that the database  can not give the spectrum reserved for one WSD to another WSD? }

%\rev{As mentioned previously, there are two types of different spectrum resources in TV white space networks: (i) the un-registered (un-used) spectrum, and (ii) the registered but under-utilized spectrum. In this work, we focus on the secondary utilization and trading of the second type spectrum resource. Accordingly, we  model a broker-based secondary spectrum market, where TV licensees lease their under-utilized spectrums to the unlicensed WSDs via the geo-location database (spectrum broker).
%Specifically, the database reserves spectrum from TV licensees in advance (Step 0 in Figure \ref{fig:system-mdoel}), and then resells the reserved spectrum to WSDs for secondary utilization (Steps 1-3 in Figure \ref{fig:system-mdoel}). Both the spectrum reservation and trading/sccess processes are performed periodically, but with different time scales. The spectrum reservation process is performed with a relatively large time period (e.g, one day or one week)-called \emph{reservation period} (indexed by $T$), while the spectrum trading/access processes is performed with a relatively small time period (e.g., one hour)-called \emph{access period} (indexed by $t$).}

We focus on the secondary sharing and trading of the under-utilized licensed spectrum of TV licensees.
In particular, we model a broker-based secondary spectrum market, where the geo-location database acts as a spectrum broker, reserving spectrum from TV licensees in advance and then reselling the reserved spectrum to unlicensed WSDs.
%The detailed spectrum reservation and trading/access processes are illustrated in Figure~\ref{fig:system-mdoel}.

%\com{This paragraph still overlaps with both the discussions of Figure 1 on Page 2, and with the discussions of Figure 2 in Section 3.2. I suggest the following simplification:
%(1) Since we have introduced the two types of spectrum in Section 1.1, we do not need to repeat the two types again here. Instead, we should just focus on introducing the 2nd type, the one that we will focus on in this paper.
%(2) We do not repeat Step 0 and Steps 1-3 in Figure 1. Instead, we should move the definitions of reservation period and access period into Section 3.2, when we introduce Figure 2. }

\subsection{Broker-based Spectrum Reservation Market}\label{sec:reservation_market_detailed}

\begin{figure}[t]
%	\vspace{-2mm}
	\centering
	\includegraphics[width=3.5in]{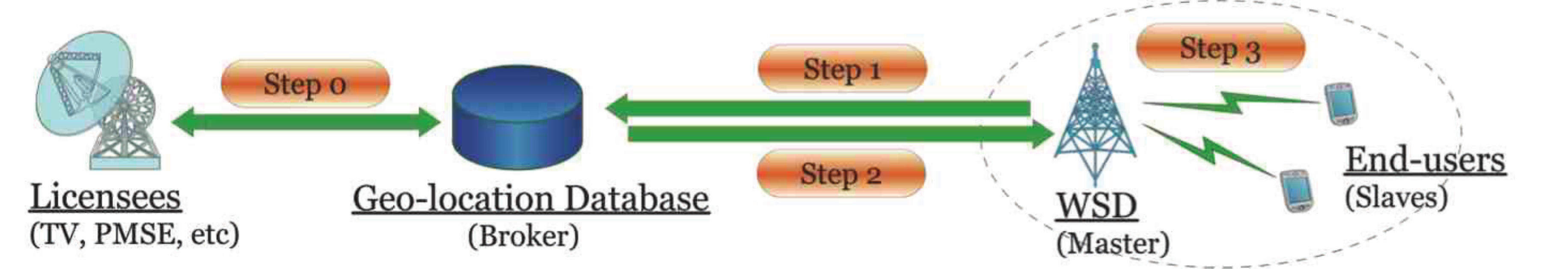}
	\vspace{-2mm}
	\caption{Spectrum reservation and access processes.
		Step 0: the database reserves {\bbd} for every reservation period $T$;
		Step 1: the WSD reports the realized demand in every access period $t$;
		Step 2: the database returns {\bbd} to the WSD in every access period $t$;
		Step 3: the WSD serves end-users in every access period $t$.
	}\label{fig:system-mdoel-process}
%	\vspace{-4mm}
\end{figure}

Now we discuss the proposed spectrum reservation market more detailedly.
Let $c$ denote the unit price (\emph{cost}) at which the database reserves spectrum from TV licensees.
Let $w$ denote the unit price (\emph{wholesale price}) at which the database sells spectrum to the WSD.
Let $r$  and $s$ denote the unit price (\emph{market price}) at which the WSD serves the subscribed and un-subscribed end-users, respectively.\footnote{In Section \ref{sec:model-demand}, We will discuss the two types of users in details.}
In order to concentrate on the {\bbd} reservation problem, we consider a fixed spectrum trading model, that is, the trading prices $c,\ w,\ r,$ and $s$ are fixed system parameters.\footnote{\rev{Our model does allow the possibility of changing the prices over a longer time horizon. Specifically, we can divide the whole time period into multiple frames, each lasting for certain time (say several hours).
		At the beginning of each frame, the WSD can adjust the trading price of $r$ and $s$ according to the congestion level of spectrum.
		Then trading prices remain fix during a frame, and our results and analysis characterize the system within this frame.}}
%\footnote{{Our model does allow the possibility of changing the prices over a longer time horizon.
%Specifically, we can divide the whole time period into multiple frames, each lasting for certain time (say several hours).
%	At the beginning of each frame, the WSD can adjust the trading price of $r$ and $s$ according to the congestion level of spectrum.
%	Then trading prices remain fix during a frame, and our results and analysis characterize the system within this frame.}}
This implies that our proposed {\bbd} reservation framework does not need to alter the {\bbd} trading process, and thus is compatible with many existing spectrum market mechanism designs.
Moreover, to make the trading model meaningful, we  assume that {$\min\{r,s\} > w > c$}, i.e., both the database and the mater will benefit from the trading process.
%\com{This assumption should be moved to the beginning of the subsection, immediately after we introduce the notations of r, s, w, and c.}

We illustrate the detailed spectrum reservation and trading/access processes in Figure \ref{fig:system-mdoel-process} and Algorithm \ref{algo_economic_interaction}.
It is notable that the spectrum reservation process (Step 0) is performed at a relatively large time period (e.g, oncen every day or every week), called the  \emph{reservation period} (denoted by $T$); while the spectrum trading/access processes (Steps 1-3) are performed at a relatively small time period (e.g., once per hour), called the \emph{access period} (denoted by $t$).

\begin{algorithm}[h]
\DontPrintSemicolon
\small
\For{each \textbf{reservation period} $T = 1, 2, \ldots$}
{Step 0: The database reserves  $\K$ unit of {\bbd} from TV licensees at a unit price $c$, for each reservation period;
\For{each \textbf{access period} $t=1,...,T$}
{
Step 1: The WSD collects the realized end-user demand $\D$, and requests $\D$ units of spectrum from the database in each access period; \;
Step 2: The database sells  $\min\{\K, \D\}$;\;
Step 3: The WSD serves end-users using the received {\bbd} at a market price $r$ or $s$ in each access period.
%\For{each operator $m=1,...,M$}{
%\If{$\d_m>\q_m$}{
%Step 4: {\op} $m$ requests $(\d_m - \q_m)^+$ units of shared \tvws;
%}
%Step 5: {\op} $m$ serves end-users using  dedicated {\tvws} $min(\d_m,\q_m)$
%and  shared {\tvws} $(\d_m-\q_m)^+$;
%}
}
}
%\textbf{Reservation Period}\;
%~~{Step 0: The database reserves  $\K$ unit of {\bbd} from TV licensees at a unit price $c$, for each reservation period;} \;
%%~~Step 1: Each {\op} $m\in\M$ determines the order $\q_m$ of dedicated spectrum (initial inventory);\;
%%~~Step 2: Each {\op} $m\in\M$ announces the service price $\p_m$ to end-users; \;
%\caption{The economic interaction of operator $m$ with the database and the end-users}
\caption{Algorithmic statement for the three-stage hierarchical model.}
\label{algo_economic_interaction}
\end{algorithm}

We focus on the following database's {{\bbd} reservation problem}: \emph{how to determine the proper {\bbd} reservation amount $\K$ to maximize the database profit?}
The problem is challenging due to the  demand stochasticity (see Section \ref{sec:model-demand}) as well as the information asymmetry (see Section \ref{sec:model-asym}).
%Furthermore, the reservation decision depends greatly on (i) the information scenario, namely, whether the database and the WSD experience the same degree of uncertainty about
%the end-user demand (see Section \ref{sec:model-asym}), and (ii) the risk-bearing scheme, namely, who will bear the risk of {\bbd} over-reservation (see Section \ref{sec:model-risk}).
% Moreover, there is a risk of {\bbd} over-reservation due to the demand uncertainty. Thus,
Moreover, the {\bbd} reservation decision also depends on the risk-bearing scheme  (see Section \ref{sec:model-risk}), namely, who (i.e., the database or the WSD) will bear the risk of {\bbd} over-reservation. This further complicates the problem.

%We further notice that the above trading process consists of two sub-problems: the \emph{bandwidth reservation problem} (which determines the {\bbd} reservation level $\K$) and the \emph{real spectrum trading problem} (which determines the related trading prices, e.g., $c,w,r,s$).

%Note that stage 0 is done per hour (i.e., reservation period) and stages 1-3 are performed per minute (i.e., access period).  All prices (i.e., $c$, $w$, $r$, and $s$) are pre-fixed. The only decision variable is the {\bbd} reservation amount $K$. Thus, our proposed model does not need to significantly alter the transaction process (i.e., reservation, wholesale, and final-sale), which implies that it is compatible with many practical commercial network models.

\subsection{Demand Stochasticity}\label{sec:model-demand}
%\com{Yuan: as in the Section \ref{sec:contract_multiple}, we also need to use notation to illustrate the aggregated demand of all WSD. Hence, I think we may need to change this a little bit.}

In each access period, a WSD $\n \in \Nset$ uses the purchased spectrum to serve its end-users.
We consider two types of end-users for each WSD: registered end-users (called \emph{subscribers}) and unregistered end-users (called \emph{random access users} or \emph{random users}).
%\footenote{these two types of users have different demand properties, of which the WSD has different knowledge.
%	%In practice, these two kinds of end-users are common and have different properties.
%	The subscribed end-users characterize the residents in the WSD's serving area,
%	and these users can sign a service contract with the WSD in advance. Because of this, the WSD has a good knowledge regarding the demand of these users based on the long-term interactions.
%	%of the WSD's serving area who sign a contract with an operator in advance.
%	%The demand of this kind of end-users is independent of the short-term wireless characteristics and is known by the WSD.
%	The random end-users characterize the travelers to the WSD's serving area,
%	and these users do not have any prior contractual relationship with the WSD. It is difficult for the WSD to predict the demand from these users.}
%%We call the registered end-users as
%%\emph{subscribers}, and the un-registered end-users as \emph{random access users} (or \emph{random users}).
%%\footnote{In the rest of paper, we will call unregistered random access users as random users for simplicity.}.
Let $ \J_{\n}$ and $ \I_{\n}$ denote the sets of WSD $\n$'s subscribers and random users, respectively.

{Specifically, subscribers characterize the residents in the WSD's serving area, and these users can sign a service contract with the WSD in advanced. Because of this, the WSD has a good knowledge regarding the demand of these users based on the long-term interactions. The random end-users characterize the travelers to the WSD's serving area,
	and these users do not have any prior contractual relationship with the WSD. It is difficult for the WSD to predict the demand from these users.}
%	 report their {\bbd} demands in advance via bilateral agreements with the WSD, while random users request {\bbd} in real-time according to their instantaneous demands.}
%, quality of services (QoS) requirements, and wireless characteristics.
Naturally, we assume that subscribers have a higher priority in obtaining service than random users.
That is, when the {\bbd} received by the WSD (from the database) is not enough to meet all end-users' demand,
the WSD will satisfy the subscribers's demand first, and then serve the random users using the remaining spectrum.
Recall that $r$ and $s$ are the unit prices (of \bbd) for serving subscribers and random users, respectively.
Due to the high priority of subscribers, it is reasonable to assume that $r > s$.

Let $\Da_{\n,j} $ and $\Db_{\n,i} $ denote the {\bbd} demands
of a subscriber $j\in\J_{\n}$ and a random user $i\in\I_{\n}$ (to WSD $n$) in one access period, respectively.
We assume that (i) $\Da_{\n,j}  $ keeps unchanged within each  reservation period $T$ (but may vary across $T$), which implies that each contract's validity is larger than one access period;
and (ii) $\Db_{\n,i}$ keeps unchanged within each access period $t$ (but may vary across $t$), which implies that each random user's average QoS and wireless characteristic remain constant in each access period.\footnote{Although the small scale fading coherence time can be much smaller than one access period, we can use proper modulation and coding schemes to combat the impact of fast fading. The assumption on demand $\Db_{\n,i}$ implies that the large scale fading does not change faster than one access period (e.g., users do not move often).}
The total demand (of all subscribers and random users) of WSD $\n$ in one access period~is:
\begin{equation}\label{total_demand}
%\vspace{-1mm}
\textstyle
\D_{\n} = \sum_{j \in \J_{\n}} \Da_{\n,j}  + \sum_{i \in \I_{\n}} \Db_{\n,i} \triangleq  \Da_{\n} +  \Db_{\n},
%\vspace{-2mm}
\end{equation}
where $\Da_{\n} \triangleq  \sum_{j \in \J_{\n}} \Da_{\n,j}$ is total subscriber demand, and
${\Db_{\n}} \triangleq \sum_{i \in \I_{\n}} \Db_{\n,i} \triangleq  \Da_{\n}$ is total random user demand.
For convenience, we refer to $\Da_{\n} $ as the \emph{scheduled demand} of WSD $n$ (as it is known at the beginning of each reservation period, and keeps unchanged during the whole  reservation period), and refer to ${\Db_{\n}}$ as the \emph{bursty demand} of WSD $n$ (as it is known only at the beginning of each access period, and changes randomly in different access periods).\footnote{Note that such a two-fold demand formulation in Eq.~(\ref{total_demand}) is widely used
in economic literature to characterize the asymmetry of demand information (see, e.g., {\cite{larivier1999contract}-\cite{ozer2006}}).
It can represent a lot of practical demand scenarios, such as (i) the two-stage demand used in the electricity market, where $\Da_{\n}$ is the pre-ordered demand and $\Db_{\n}$ is the real-time replenishment, and (ii) the forecast demand with error, where $\Da_{\n}$ is the estimated demand and $\Db_{\n}$
is the forecast error.}

Based on the assumptions mentioned above, the scheduled demand $\Da_{\n}$ is a random variable changing each reservation period $T$, and the bursty demand $\Db_{\n}$ is a random variable changing each access period $t$.
For simplicity, we assume that   $\Da_{\n}$ and $\Db_{\n}$ are independent and identically distributed (i.i.d) in different reservation periods and access periods, respectively.
Let $f(\Da )$ and $F(\Da)$ denote the probability
density function (pdf) and cumulative distribution function
(cdf) of $\Da$, and $g(\Db )$ and $G(\Db)$ denote the pdf and cdf of $\Db$, respectively.
As in many mechanism design literature (see, e.g., \cite{gao2011contract,duan2013contract,sheng2013contract}), we assume that such distribution information are public information to both the database and the WSD.
In practice, they can be obtained through machine learning in a sufficiently long time period.
%\textbf{As WSDs are spatially separated and hence decoupled, we will focus on a typical WSD and omit the subscript $\n$ in $\D_{\n}$, $\Da_{\n}$, and $\Db_{\n}$ in \rev{the following sections} for notational clarity.}
{\textbf{As mentioned previously, the number of WSDs is large enough so that one WSD's strategy is independent of others.
%As each WSD serves a distinct pool of end-users,
Hence, we can concentrate on the interaction between the database and one WSD.}}

Since the total demand $\D$ changes randomly in each access period $t$, while the {\bbd} reservation is performed at the beginning of each reservation period $T$, the database or the WSD faces a {\bbd} reservation problem under \textbf{demand stochasticity}.
Obviously, a higher reservation can serve more demand potentially, but may also lead to a higher risk of {\bbd} over-reservation. A lower reservation, however, may lead to a higher loss due to the spectrum under-reservation.~~~~~~~~~~~
%
%\vspace{2mm}

Next we draw some useful properties of the scheduled demand $\Da$ and the bursty demand $\Db$.
First, we notice that the random users' bursty demand  usually depends on the real-time market price $s$ and end-users' wireless characteristics.
As an example commonly used in the networking literature {(e.g., \cite{yuan2012}-\cite{niyato2009investment})}, a random user $i$'s utility $\pi_i$ can be defined as the difference between the achiavable data rate (e.g., the Shannon capacity assuming high SNR \cite{goldsmith2005wireless}) and the payment, e.g.,\footnote{{This is just an illustrative example. Our analysis applies to more generic utility functions. }}
\begin{equation*}\label{random_user_profit}
\textstyle
\pi_i = \beta \cdot \Db_i \cdot \ln\left(\frac{P_i|h_i |^2 }{ \Db_i n_0 } \right)- s\cdot \Db_i ,
\vspace{-1mm}
\end{equation*}
where  $\left|h_i\right|$ is the channel gain, $P_i$ is the
transmission power, $n_0$ is the noise power per unit bandwidth, and $\beta$ denotes the monetary income per unit of data rate.
Based on the above utility definition, the optimal bursty demand for a random user $i$ that maximizes its payoff $\pi_i$ is
\begin{equation*}
\textstyle
\Db_i  =\frac{P_i\cdot  e^{-{(1+s/\beta)}} \cdot |h_i |^2 }{   n_0 }  .
\end{equation*}
Notice that the channel coefficient $h_i$ satisfies: (i) $h_i \sim \mathcal{C}(0,1 )$, the complex normal distribution (when the channel experiences the Rayleigh fading), and (ii)  $h_i$ is {i.i.d} for different users $i \in \I_{\n}$.
Therefore, both $\Db_i $ and $\Db $ follow the chi-square distribution \cite{goldsmith2005wireless} (with different degrees of freedom).
%First, the random user demand usually depends on the market price $s$ and random users' wireless characteristics.
%As an example commonly used in the networking literature (e.g., \cite{yuan2012}-\cite{niyato2009investment}),
%we assume that both $\Db_i $ and $\Db $ follow chi-square distributions  \cite{goldsmith2005wireless}  (with different degrees of freedom) \footnote{We present the detailed explanation in \cite{techrpt}}.
Note, however,  that our analysis also holds for other demand distributions such as the normal distribution.
% a random user $i$'s utility $\pi_i$ can be defined as the difference between the achievable data rate (e.g., the Shannon capacity assuming high SNR \cite{goldsmith2005wireless}) and the payment, that is,\footnote{{This is just an illustrative example. Our analysis applies to more generic utility functions. }}
%\begin{equation*}\label{random_user_profit}
%\textstyle
%\pi_i = \beta \cdot \Db_i \cdot \ln\left(\frac{P_i|h_i |^2 }{ \Db_i n_0 } \right)- s\cdot \Db_i ,
%\end{equation*}
%where  $\left|h_i\right|$ is the channel gain, $P_i$ is the
%transmission power, $n_0$ is the noise power per unit bandwidth, and $\beta$ denotes the monetary income per unit of data rate.
%Based on the above utility definition, the optimal demand of a random user $i$ that maximizes the payoff $\pi_i$ is
%\begin{equation*}
%\textstyle
%\Db_i  =\frac{P_i\cdot  e^{-{(1+s/\beta)}} \cdot |h_i |^2 }{   n_0 }  .
%\end{equation*}
%Notice that the channel coefficient $h_i$ satisfies: (i) $h_i \sim \mathcal{C}(0,1 )$, the complex normal distribution (when the channel experiences the Rayleigh fading), and (ii)  {i.i.d} for different users.
%Therefore, both $\Db_i $ and $\Db $ follow chi-square distributions  \cite{goldsmith2005wireless}  (with different degrees of freedom). {Note  that our analysis also holds for other demand distributions such as the normal distribution.}

Second, the subscribers' scheduled demand $\Da$ is a long-term  average demand (changing every reservation period, e.g., one day), and usually independent of the short-term wireless characteristics.
Our analysis holds for arbitrary $\Da$ distribution with the increasing failure rate (IFR), i.e., $\frac{f(\Da)}{1- F(\Da)}$ is increasing in $\Da$.\footnote{Such an IFR constraint is widely used in the mechanism design literature (e.g., \cite{cachon2001contract,ozer2006}).
Many commonly used distributions, such as
the uniform distribution, exponential distribution, and normal distribution, satisfy the IFR constraint.}

%
%The total users' demand in each minute $t$ is written as:
%\begin{align}
%\label{total_demand} D &= \sum_{j \in \mathcal {J}}{b_j^c} +
%\sum_{i \in \mathcal {I}}{b_i}^{r*} \nonumber \nonumber \\
%&= \sum_{j \in \mathcal {J}}{b_j^c} +
%\sum_{i \in \mathcal {I}}{\theta_i}e^{-{(1+s)}} \nonumber \nonumber \\
%&= \Da + {\Db}e^{-(1+s)}
%\end{align}
%where $ \mathcal {J}$ and $ \mathcal {I}$ are the set of contract
%users and demand users, respectively, $\Da = \sum_{j \in \mathcal
%{J}}{b_j^c}$ is the total demand of contract users and
%${\Db} = \sum_{i \in \mathcal {I}}{\theta_i}$ is random
%users' aggregate wireless characteristics.
%
%Since the amount of accessable {\bbd} is reserved prior to
%realization of total users' demand, hence the demand is uncertain
%for both WSD and database when they make reserved bandwidth
%decision.
%
%%Also noted that the amount of accessable {\bbd} is reserved prior
%%to realization of total users' demand, hence ${\Db}$ is
%%random variable for both WSD and database when they make reserved
%%bandwidth decision.
%
%
%

%\begin{table}[t]
%\centering
%\caption{Database's and WSD's Knowledge about Demand}
%\label{table:information}
%\small
%\begin{tabular}{|c||c|c|}
%\hline
%& \textbf{Scheduled Demand - $\Da$}
%& \textbf{Bursty Demand - $\Db$}
%\\
%\hline
%\hline
% Database
% & distribution $f(\Da)$
% & distribution $g(\Db)$
%   \\
%\hline
%WSD
% & value $\Da$
% & distribution $g(\Db)$
%   \\
%   \hline
%\end{tabular}
%\end{table}

\subsection{Information Asymmetry}\label{sec:model-asym}

%According to (\ref{total_demand}), the total demand $\D$ contains two parts: (i) the scheduled demand $\Da$, and (ii) the bursty demand $\Db$.

Due to the different proximities to end-users, the database and the WSD usually have different knowledge about the scheduled demand $\Da$ and the bursty demand $\Db$.
Table \ref{table:information} illustrates the difference between the database's knowledge and the WSD's knowledge regarding the end-user demand \emph{at the beginning of each reservation period $T$ (when making the reservation decision)}. Specifically,
% which will affect the reservation decision eventially.
\begin{itemize}
\item
\emph{Bursty demand $\Db$ of random users:} Notice that $\Db$ changes randomly every access period.
Thus,
neither the WSD nor the database knows the exact value of $\Db$ at the beginning of the reservation period.
That is, both the WSD and the database only know the distribution of ${\Db}$.
\item
\emph{Scheduled demand $\Da$ of subscribers:}
Notice that $\Da$ keeps unchanged within each reservation period.
Thus, the WSD is able to know the exact value of $\Da$ (e.g., through bilateral agreements signed with subscribers) at the beginning of the reservation period.
The database, however, does not know the exact value of $\Da$ unless the WSD shares such information.
That is, the database only knows the distribution of ${\Da}$.
\end{itemize}

%\begin{table}[t]
%\centering
%\caption{Database's and WSD's Knowledge about Demand}
%\label{table:information}
%\small
%\begin{tabular}{|c||c|c|}
%\hline
%\multirow{2}{*}
%& \textbf{Scheduled Demand}
%& \textbf{Bursty Demand}
%\\
%&\textbf{$\Da$}
%&\textbf{$\Db$}
%\\
%\hline
%\hline
% Database
% & $f(\Da)$
% &  $g(\Db)$
%   \\
%\hline
%WSD
% & $\Da$
% & $g(\Db)$
%   \\
%   \hline
%\end{tabular}
%\end{table}

We refer to the difference between the database's knowledge  and the WSD's knowledge regarding demand information as \textbf{information asymmetry}.
The co-existence of these two types of end-users and the information asymmetry provide incentives for the WSD to misreport its private information.
Without the existence of random users, the WSD would request the database to reserve spectrum equal to the demand of subscribed users.
With the existence of random end-users, the WSD would reserve the amount of spectrum larger than the demand of subscribed end-users, in order to gain more revenue by serving both the subscribed end-users and the  random end-users.
However, the exact value of the random end-users demand is unknown by the WSD at the beginning of the reservation period.
Hence,
to maximize its own profit, the WSD would optimize the value to be reported to the database, instead of truthfully revealing his information of the certain demand of subscribed users. Such strategic misreporting will make it difficult for the database to make the optimal reservation decision to maximize the database's payoff.
To hedge information asymmetry, it is important to design an incentive compatible mechanism to elicit the WSD's private demand information (i.e., $\Da$).
In this work, we will propose contract-theoretic {\bbd} reservation mechanisms to achieve this goal.
%to ensure the WSD shares its private information $\Da$ with the database credibly.
%Note that we will also study the {\bbd} reservation under  {information symmetry} (where the database also knows the exact value of $\Da$) as benchmark.

\subsection{Risk-Bearing Scheme}\label{sec:model-risk}

Due to the  demand stochasticity, there is a risk of {\bbd} over-reservation.\footnote{Note that spectrum under-reservation will hurt the profits of both the database and the WSD directly, and thus there is no need to discuss the risk sharing under spectrum under-reservation.
Under spectrum over-reservation, however, the database and the WSD must decide who will pay for the over-reserved spectrum.}
Thus, the {\bbd} reservation decision depends greatly on the risk-bearing scheme.
Namely, who will bear the risk of {\bbd} over-reservation, i.e.,  the database or WSDs?
We refer to the former scheme as DB-bearing-risk (Scheme I) and the latter scheme as WSD-bearing-risk (Scheme II).
Specifically,
\begin{itemize}
\item
\emph{DB-bearing-risk (Scheme I):} In this case, the WSD only pays for the spectrum it actually purchases in each access period, and thus the database bears all the risk of {\bbd} over-reservation.
That is, in each access period, the WSD will only pay for $\min\{\K,\D\}$ units of spectrum that it consumes.
\item
\emph{WSD-bearing-risk (Scheme II):} In this case, the WSD pays for all the spectrum reserved, and thus the WSD bears all the risk of {\bbd} over-reservation. That is, in each access period, the WSD will pay for all $\K$ units of reserved spectrum, even if the total demand $\D$ is smaller than $\K$.
\end{itemize}

In this paper, we will study the {\bbd} reservation problem under both risk-bearing schemes systematically.
In the following sections, we first study the centralized/integrated {\bbd} reservation solution as a (centralized) benchmark (Section \ref{sec:integrated}).
Then we study the decentralized  reservation solution without information sharing as another (decentralized) benchmark (Section \ref{sec:asymm-no}), and show that it may lead to a poor performance (in terms of database profit and network profit) due to the asymmetry of information.
To this end, we study the decentralized  reservation solution with contract-based credible information sharing (Section \ref{sec:asymm-contract}).
{To facilitate the understanding, we have listed the key results of this work  in Table \ref{table:key-result}.
%Then we compare two different risk-bearing schemes with the centralized solution and show that a critical price affects the performance of two different risk-bearing schemes.}
%For comparison, the database's demand information is $D = \Da +
%{\Db}e^{-(1+s)}$, where $s$ is deterministically known and
%$\Da$ has c.d.f. $F(\cdot)$ and $\Db$ has c.d.f. $G(\cdot)$.
%On the other hand, the WSD's demand information is $D = \Da +
%{\Db}e^{-(1+s)}$, where where $s$ and $\Da$ is
%deterministically known $\Db$ has c.d.f. $G(\cdot)$. Hence
%the information is asymmetric. The database learns the uncertain
%private information $\Da$ after this information is shared credibly
%by the WSD. The uncertainty random users' demand
%${\Db}e^{-(1+s)}$ is realized each minute $t$. Assume
%$F(\cdot)$ and $G(\cdot)$ are twice differentiable, and $f(\cdot)$
%and $g(\cdot)$ are strictly positive over their supports.

%\input{Section_symmetry}
%!TEX root = main_CR_Database.tex
%SourceDoc main_CR_Database.tex

 %\section{Information Symmetry: A Benchmark Solution}\label{sec:symm}
%\section{Information Symmetry: A Benchmark Soltuion}\label{sec:symm}
\section{Integrated {\bbd} Reservation Solution}\label{sec:integrated}

In this section, we consider an \emph{integrated} system, where the database and the WSD act as an integrated decision maker to maximize their aggregate profit (called  {\emph{network profit}}, denoted by $\Utot$).
We will study this integrated/centralized optimal {\bbd} reservation  {as the centralized benchmark}.

%In this section, we consider the {\bbd} reservation under information symmetry,
%where both the database and the WSD know the precise value of $\Da$ and the distribution of ${\Db} $.
%We will study the \emph{centralized/integrated optimal reservation} solution (as a benchmark).
%That is, the database and the WSD act as an integrated decision maker to maximize their aggregate profit (also called \emph{social welfare}, denoted by $\Utot$).

%In this case, there is \emph{no} information asymmetry as both parties know the precise value of $\Da$ and the distribution of  ${\Db} $.
Obviously, in this case the integrated player (database and WSD) knows the precise value of $\Da$ and the distribution of ${\Db} $.
Moreover, there is no difference between the DB-bearing-risk scheme and the WSD-bearing-risk scheme.
Specifically, given any {\bbd} reservation $\K$,
the expected {network profit} is
\begin{equation}\label{centralized_profit}
\begin{aligned}
\textstyle
\Utot
(\K,\Da) = & r \cdot {\min{ \{\K,\Da \}}} + s \cdot \EX_{\Db} \big[ \min\big\{{\Db} , (\K-{\Da})^{+} \big\} \big]
- c\cdot {\K} ,
\end{aligned}
\end{equation}
where $(x)^+= \max\{x,0\}$.
This formula implies that the WSD will satisfy the subscribers' scheduled demand first (1st term), and then satisfy the random users' bursty demand using the remaining spectrum (2nd term).

%The above aggregate profit is differentiable everywhere except the point $\K = \Da$.
Next we study the centralized optimal reservation $\Kso$ that  maximizes the {network profit} defined in (\ref{centralized_profit}).
Notice that when $\K \leq \Da$, we have $\frac{\partial{\Utot(\K,\Da)}}{\partial{\K}} = r-c >0 $, which implies that the optimal $\K$ cannot be smaller than $\Da$; when $\K \geq \Da$, we have (i) $\frac{\partial{\Utot(\K,\Da)}}{\partial{\K}} = s\big[1-G(\K-\Da)\big] - c$,
and (ii) $\frac{\partial^2{\Utot(\K,\Da)}}{\partial{\K}^2} =-s \cdot g\big( \K -\Da \big) \leq 0$.
Thus, the centralized optimal reservation $\Kso$ is given by the first-order condition $\frac{\partial{\Utot(\K,\Da)}}{\partial{\K}} = 0$, and more formally,
\vspace{-1mm}
\begin{equation}
\label{centralized_bandwidth}
\textstyle
\Kso =  \Da + {G^{-1}\Big( \frac{s-c}{s}\Big)}.
\vspace{-1mm}
\end{equation}
Intuitively, $\Kso$ consists of two parts: (i) the scheduled demand $\Da$, and (ii) the best response to the bursty demand $\Db$.
Note that the centralized optimal reservation $\Kso$ is a function of $\Da$, but not a function of $\Db$.
This is because the integrated player knows the precise value of $\Da$, but not the value of $\Db$.

%\input{Section_asymmetry_wholesale}
%!TEX root = main_CR_Database.tex
%SourceDoc main_CR_Database.tex

%\section{Information Asymmetry: Wholesale Price Contract}\label{sec:asymm}
%Now we study the optimal {\bbd} reservation under information asymmetry. It is common that the WSD and the database establish a transaction through a wholesale price agreement. Under this contract, the WSD pays a wholesale price $w$ to the database for each unit ordered. We study two types of wholesale price purchasing arrangement:
%\begin{inparaenum}[\itshape 1\upshape)]
%\item when the wholesale price is exogenous; and
%\item when the database sets the wholesale price.
%\end{inparaenum}
%
%The first arrangement (exogenous wholesale price) models a situation in which the price is set by the external market. The latter arrangement (database sets the price) models a situation in which the database has power that allows it to take the leadership that allows it to take the leadership position in contract negotiations.
%
%Similar to Section \ref{sec:symm}, we consider two cases:
%\begin{inparaenum}[\itshape 1\upshape)]
%\item database selects the reserved {\bbd} and bears the risk of the {\bbd} reservation;
%\item database selects a reservation level, whereas the WSD selects an order level and bears the risk of shortages and/or excess inventory.
%\end{inparaenum}

\section{Decentralized {\bbd} Reservation --  No Information Sharing}\label{sec:asymm-no}

Now we consider a general \emph{decentralized} system, where the database and the WSD make decisions independently, aiming at maximizing their individual profits.
In this section, we will study the decentralized {\bbd} reservation solution \emph{under information symmetry} and \emph{under information asymmetry without information sharing} as the decentralized benchmarks.
%In what follows, we study the reservation solutions under two risk-bearing schemes one by one.

%In this section,  we consider the {\bbd} reservation under information asymmetry, where the WSD knows the precise value of $\Da$ but the database does not know (see Section \ref{sec:model-asym}).
%We will study the {\bbd} reservation without information sharing between the database and the WSD in this section.
%In the next section, we will propose contract-based {\bbd} reservation to promote information sharing.

%It is common that the WSD and the database establish a transaction through a wholesale price agreement.
%Under this contract, the WSD pays a wholesale price $w$ to the database for each unit ordered.
%%We assume that the wholesale price is set by the external market.
%We first discuss the Scenario where database selects the reserved {\bbd} and bears the risk of the {\bbd} reservation, then the Scenario where WSD selects an reservation level and bears the risk of shortages and/or excess inventory is studied.
%Similar to Section \ref{sec:symm}, we consider two cases:
%\begin{inparaenum}[\itshape 1\upshape)]
%\item database selects the reserved {\bbd} and bears the risk of the {\bbd} reservation;
%\item database selects a reservation level, whereas the WSD selects an order level and bears the risk of shortages and/or excess inventory.
%\end{inparaenum}

%\subsection{Exogenous Wholesale Price Contract}
%In this contract, the database charges a wholesale price $w$ per unit {\bbd} to the WSD, where $w$ is determined by the external market.
\vspace{-1mm}
\subsection{Scheme I: DB-Bearing-Risk}\label{sec:asymm_database}

%Under the DB-bearing-risk scheme, the database determines the {\bbd} reservation $\K$, and bears the risk of {\bbd} over-reservation.
%In other words, at each access period, the WSD will pay for at most $\D$ units (i.e., the realized end-user demand) of bandwidth.

Under the DB-bearing-risk scheme, the WSD only pays for the spectrum it actually uses, and thus the database bears all the risk of {\bbd} over-reservation.
That is, in each access period, the WSD will only purchase $\min\{\K,\D\}$ units of \bbd.
%, while not pay for the over-reserved {\bbd} $(\K-\D)$ (when $\K>\D$).

%\begin{figure*}
%\vspace{-2mm}
%  \centering
%   \includegraphics[width=3in]{Figure/K_vs_xi_without_contract_xx2}
%  ~~~~
%  \includegraphics[width=3in]{Figure/total_profit_dif_w_asy_diff_Vxi_xx2}
%  \vspace{-2mm}
%  \caption{ (a) Spectrum Reservation vs Scheduled Demand $\Da$, and (b) Network Profit vs Wholesale Price $w$. Here, $\sigma_{\Da}$ denotes the variance of the scheduled demand $\Da$.}\label{fig:K_vs_xi}.
%%  \com{Both figures: increase the marker size of the circle}
%  \vspace{-5mm}
% \end{figure*}

 \begin{figure*}
% 	\vspace{-2mm}
 	\centering
 	\includegraphics[width=2.8in]{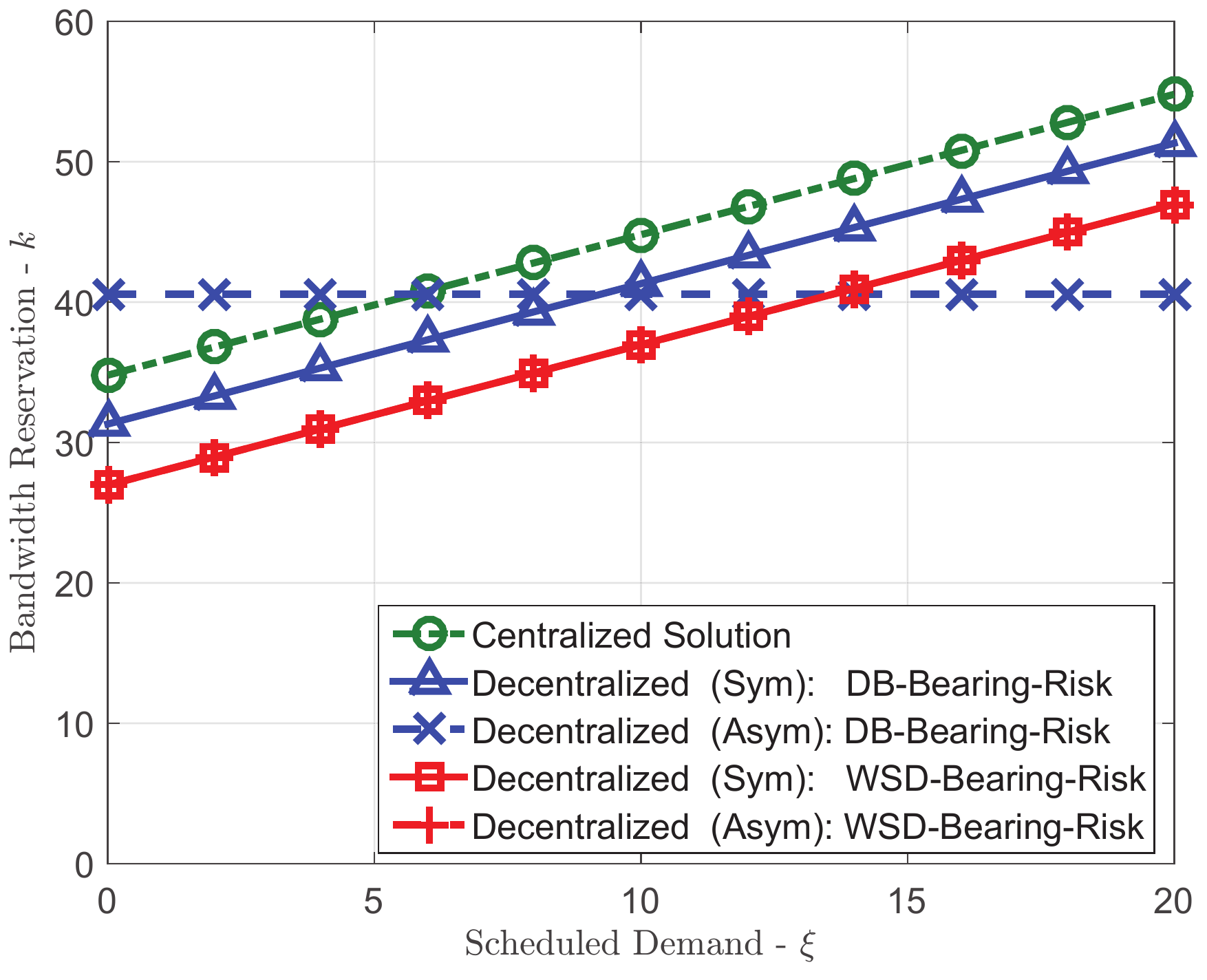}
 	~~~~~~~~
 	\includegraphics[width=2.8in]{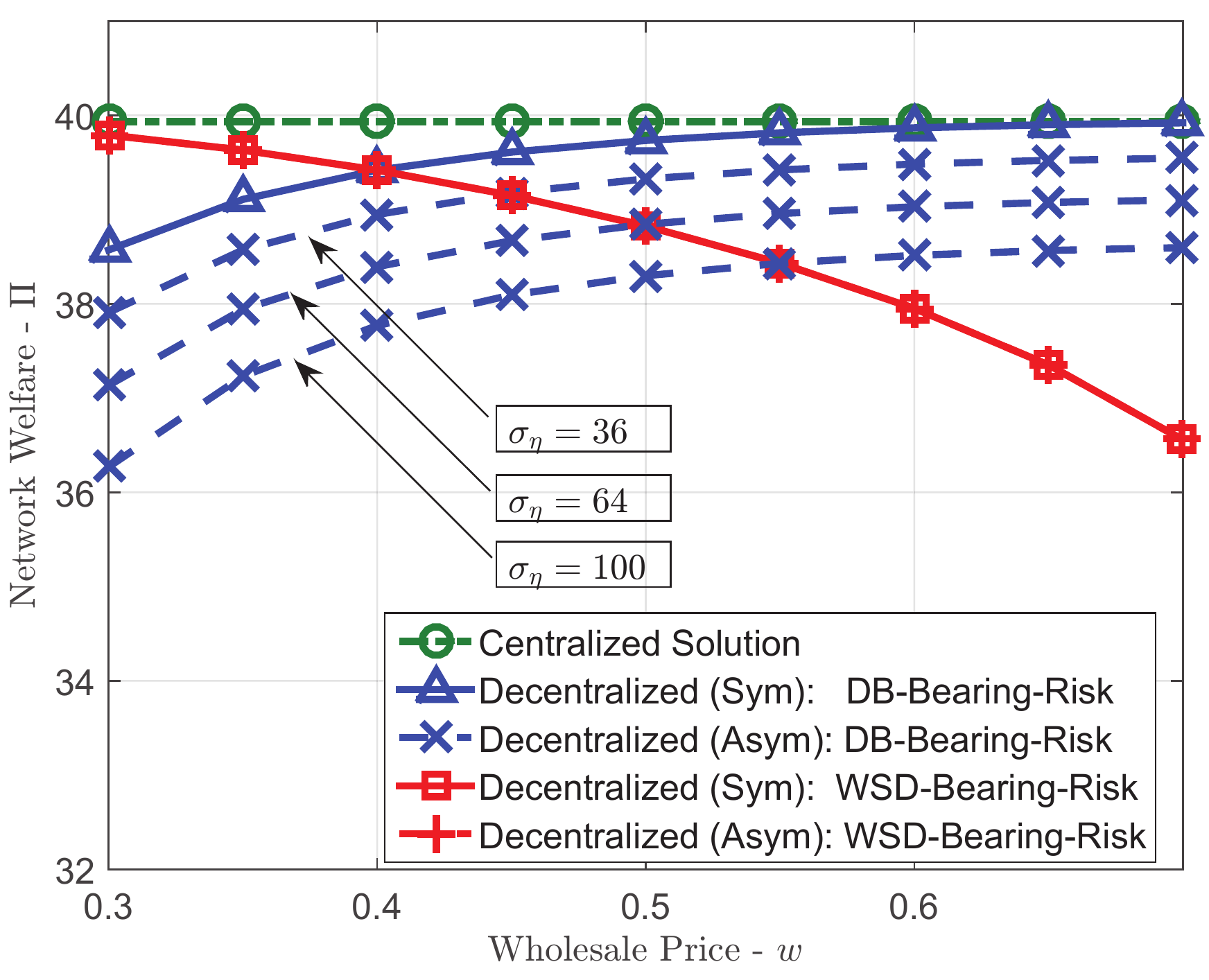}
 	\vspace{-2mm}
 	\caption{ (a) Spectrum Reservation vs Scheduled Demand $\Da$, and (b) Network Profit vs Wholesale Price $w$. Here, $\sigma_{\Da}$ denotes the variance of the scheduled demand $\Da$.}\label{fig:K_vs_xi}.
 	%  \com{Both figures: increase the marker size of the circle}
% 	\vspace{-5mm}
 \end{figure*}

\subsubsection{\textbf{Information Symmetry}}\label{sec:asymm_database-sym}

We first study the database's optimal {\bbd} reservation solution under \emph{information symmetry}, where the database is assumed to know the precise value of $\Da$.
Specifically, for any reservation $\K$, the WSD's and the database's  (ex-ante) expected profits are, respectively,~~~~~~
\begin{equation}\label{asym_decentral_WSD_profit}
\begin{aligned}
\textstyle
\Ums (\K,\Da ) = &  (r-w) \cdot \min  \{\K, \Da \}  \\
& \quad{} + (s-w)\cdot \EX_{\Db} \big[ \min\big\{{\Db} ,  (\K-\Da )^{+}\big\} \big],
\end{aligned}
\end{equation}
\begin{equation}\label{sym_dencentral_DB_profit}
\textstyle
\Udb (\K, \Da )  =
w \cdot \EX_{\Db} \big[  \min \{ \Db  + \Da, \K \} \big]  - c\cdot {\K}.   ~~~~~~~~~~~~~~~~~~~~~~~~~~~~~~~
\end{equation}
The optimal reservation for the database
 (i.e., that maximizes its profit defined in (\ref{sym_dencentral_DB_profit})) is
\begin{equation}\label{sym_dencentral_bandwidth}
\textstyle
\Kdbsym  = \Da +
 {G^{-1} \Big( \frac{w-c}{w}  \Big)}.
\end{equation}
Similar to the centralized optimal reservation $\Kso $, the above decentralized optimal reservation $\Kdbsym $ under information symmetry is also a function of $\Da$.

\subsubsection{\textbf{Information Asymmetry}}\label{sec:asymm_database-asym}

In practice, the demand information is asymmetric between the database and the WSD as discussed in Section \ref{sec:model-asym}.
Now we study the database's optimal {\bbd} reservation solution under \emph{information asymmetry}, where the database does not know the precise value of $\Da$.

We first show that the reservation solution $\Kdbsym $ in (\ref{sym_dencentral_bandwidth}) under information symmetry may not be the database's optimal solution in this case, as it
cannot ensure that the WSD shares its private information $\Da$ with the database credibly.
Notice that (i) the WSD profit $\Ums (\K,\Da )$   in (\ref{asym_decentral_WSD_profit})   increases with the {\bbd} reservation $\K$, and (ii) the database's optimal  {\bbd} reservation  $\Kdbsym$   in (\ref{sym_dencentral_bandwidth}) is linear to $\Da$.
This implies that the WSD has an incentive to inflate its private information $\Da$.
%This incentive arises because (i) the WSD's profit $\pi^{\mathrm{m}} (K,\Da )$ is increasing in the database's {\bbd} reservation choice $K$, and (ii) the database's optimal  {\bbd} reservation  $K^{\mathrm{d}}_{\mathrm{sym}}$ is linear to $\Da$.
The key reason behind this phenomenon is that \emph{the database bears all the risk of over-reservation}.

As a consequence, the database will \emph{not} trust the information (i.e., the value of $\Da$) informed by the WSD, and therefore will act based on its own prior distribution information of $\Da$ and $\Db$. That is, it will maximize the following expected profit:
\begin{equation} \label{asym_decentral_DB_profit}
\Udba(\K) \triangleq \EX_{\Da} \big[ \Udb (\K, \Da )\big]  =
w \cdot \EX_{\Da,\Db} \big[  \min \{ \Db  + \Da, \K \} \big]  - c\cdot {\K},
\end{equation}
where the expectation $\EX_{\Da,\Db}$ is taken over the distribution of $\Da$ and $\Db$.
The optimal reservation for the database that maximizes its expected profit defined in (\ref{asym_decentral_DB_profit}) is
\begin{equation}
\label{asym_decentral_bandwidth}
\textstyle
\Kdbasy=  (F \times G
 )^{-1} \Big( \frac{w-c}{w}  \Big),
\end{equation}
where
$F \times G$ is the joint c.d.f. of $\Da +
{\Db} $.
%\begin{equation}
%\label{asym_decentral_bandwidth}
%K_1=  (F \circ G
% )^{-1} \big( (w-c)/{w}  \big),
%\end{equation}
%where
%$F \circ G$ is the joint c.d.f. distribution of $X + Y $.
%
%\begin{equation}\label{asym_decentral_bandwidth2}
%K_2  = x +
% {G^{-1} \big( (s-w)/{s}  \big)}
%\end{equation}
%
%Suppose $s,c$ are fixed and $c \leq w \leq s$. We want to derive under what condition that $K_1 > K_2$. From the simulation result, we can tell that there exist a $w^{*} \in (c,r)$ such that when:
%\begin{enumerate}
%\item  $w < w^{*}$, then $K_2 > K_1$;
%\item  $w > w^{*}$, then $K_1 > K_2$;.
%\end{enumerate}
%and $w^{*}$ is increasing in $V(X)/V(Y)$, where $V(X)$ and $V(Y)$ are the variance of distributions. I wonder if it is possible to get the mathematics proof? If not, if we suppose that $X$ and $Y$ follow specific distributions such as normal, then can we proof it?

Note that $\Kdbasy$ is not a function of $\Da$, which is different from (\ref{centralized_bandwidth}) and (\ref{sym_dencentral_bandwidth}). This implies that the database cannot adjust its {\bbd} reservation decision to account for the WSD's private information. Therefore, both parties's profits may reduce due to the ignorance of information $\Da$ (that WSD has) in the {\bbd} reservation.
To solve this problem, we will propose  a \emph{spectrum reservation contract} to achieve the credible information sharing between the database and the WSD in  {Section \ref{sec:contract-DB}}.

\subsection{Scheme II: WSD-Bearing-Risk}\label{sec:asymm_WSD}

Under the WSD-bearing-risk scheme, the WSD pays for all the spectrum reserved, and thus the WSD  bears all the risk of {\bbd} over-reservation. That is, in each access period, the WSD will pay for all $\K$ units of reserved spectrum, even if the total demand $\D $ is smaller than $\K$.

\subsubsection{\textbf{Information Symmetry}}\label{sec:asymm_WSD-sym}

%Under the WSD-bearing-risk scheme, the WSD determines the {\bbd} reservation $\K$, and bears the risk of {\bbd} over-reservation.
%In other words, at each access period, the WSD will pay for all $\K$ units of reserved {\bbd} even if the realized total demand $\D$ is smaller than $\K$.

Similarly, we first study the WSD's optimal {\bbd} reservation decision under \emph{information symmetry}.
%If the decision right of the database can be transferred to the WSD, then
Specifically, for any reservation $\K$, the WSD's and the database's (ex-ante) expected profits are, respectively,
\begin{equation}\label{asym_decentral_WSD_profit2}
\begin{aligned}
\textstyle
%\Ums (\K,\Da ) = &  r \cdot \min  \{\K, \Da \}  - w \cdot \K \\
%& + s\cdot \EX_{\Db} \big[ \min\big\{{\Db} ,  (\K-\Da )^{+}\big\} \big],
\Ums (\K,\Da ) = &  r \cdot \min  \{\K, \Da \} + s\cdot \EX_{\Db} \big[ \min\big\{{\Db} ,  (\K-\Da )^{+}\big\} \big] - w \K,
\end{aligned}
\end{equation}
\begin{equation}\label{sym_dencentral_DB_profit2}
\begin{aligned}
\textstyle
\Udb (\K, \Da )  =
(w - c ) \cdot {\K} .~~~~~~~~~~~~~~~~~~~~~~~~~~~~~~
\end{aligned}
\end{equation}

{Note that if the WSD bears the risk, then the WSD will determine the  {\bbd} reservation amount. Otherwise, the database will always choose a very large reservation as it does not bear the risk of over-reservation.}
Accordingly, the optimal reservation for the WSD (i.e., that maximizes its profit defined in (\ref{asym_decentral_WSD_profit2})) is
\begin{equation}\label{sym_dencentral_bandwidth2}
%\textstyle
%\Kmssym
%= \Da +
% {G^{-1} \Big( \frac{s-w}{s}  \Big)},
\Kmssym= \Da +
 {G^{-1} \Big( \frac{s-w}{s}  \Big)},
\end{equation}
which is also a function of $\Da$.

%Eq.(\ref{sym_dencentral_DB_profit2}) and Eq.(\ref{sym_dencentral_bandwidth2}) show that the database has incentive to adjust wholesale price $c<w<s$ to maximize its own profit. This incentive arises because there exits a unique wholesale price $w$ that maximizes database profit $\pi^{\mathrm{d}} (K, \Da )$.

\subsubsection{\textbf{Information Asymmetry}}\label{sec:asymm_WSD-asym}

Since the WSD itself holds the private information under information asymmetry, the WSD's expected profit under information asymmetry is exactly same as (\ref{asym_decentral_WSD_profit2}).
Thus, the optimal reservation for the WSD under information asymmetry is same as that under information symmetry, i.e.,
\begin{equation}\label{asym_decentral_bandwidth2}
\Kmsasy  =
\Kmssym= \Da +
 {G^{-1} \Big( \frac{s-w}{s}  \Big)}.
\end{equation}
%\com{why the fonts in (11) and (12) are different?}
%\com{do they look like different}

Notice that the database profit $\Udb(\K, \Da)$ defined in  (\ref{sym_dencentral_DB_profit2}) is increasing in the {\bbd} reservation $\K$.
This implies that it is possible for the database to improve its profit by incentivizing the WSD to increase the {\bbd} reservation $\K$.
In  {Section \ref{sec:contract-WSD} }, we will propose a \emph{spectrum reservation contract} to maximize the database profit under the WSD-bearing-risk scheme.

\subsection{Comparison}

Now we compare the above decentralized optimal reservations (without information sharing).
It is easy to see that these decentralized solutions  deviate from the integrated optimal solution (\ref{centralized_bandwidth}), due to the ``double marginalization'' effect as well as the lack of information on the database side under information asymmetry.

%\com{Overall, 5.3 reads like a collection of observations without key insights.
%Need to focus on explaining the insights.
%Also, Lemmas 2 and 3 are very similar in format. It seems to be quite redundant of presenting both lemmas in the current format. }

\subsubsection{\textbf{Performance under Information Symmetry}}

%Because each player (WSD or database) charges a price that is greater than their marginal cost ("double marginalization"), this two cases does not achieve system-wide optimal expected profit.
%However, it is an open question as to which mechanism has better performance in terms of achieving system-wide efficiency.
We first compare two {\bbd} reservation solutions under information symmetry, i.e.,  $ \Kdbsym$ and $\Kmssym $.
%Compare the social welfares under the integrated optimal solution (i.e., $\Kso$ in (\ref{centralized_bandwidth})) and  the decentralized optimal solutions (i.e., $\Kdbsym$ in (\ref{sym_dencentral_bandwidth}) and $\Kmssym$ in (\ref{sym_dencentral_bandwidth2})), we have
%Note that in Eq.(\ref{centralized_profit}), only $K$ affects the result of aggregate profit. Compared Eq.(\ref{sym_dencentral_bandwidth}) and Eq.(\ref{sym_dencentral_bandwidth2}) with Eq.(\ref{centralized_bandwidth}), we have:
\begin{lemma}\label{theorem:optimality_sym}
There exists a critical wholesale price $w^{*} = \sqrt{sc}$ such that \begin{enumerate}
\item when $w < w^{*}$, then $\Kso > \Kmssym > \Kdbsym$;
\item when $w > w^{*}$, then $\Kso  > \Kdbsym> \Kmssym$.
\end{enumerate}
\end{lemma}

We illustrate the {\bbd} reservation solutions vs scheduled demand $\Da$ in {Figure \ref{fig:K_vs_xi}.a}, where $s=0.8$, $w=0.5$, $c=0.2$, and obviously, $w >\sqrt{sc}=0.4$.
It is easy to see that $\Kdbsym$ under DB-bearing-risk  (the blue triangle curve) is always larger than $\Kmssym$  under WSD-bearing-risk ({the red square curve}). This is because with a large wholesale price (e.g., $w>\sqrt{sc}$), the risk of over-reservation that the WSD bears under WSD-bearing-risk is higher than that the database bears under DB-bearing-risk, and thus the WSD will reserve less spectrum than the database.
We can further see that $\Kdbsym$ and $\Kmssym$ are smaller than $\Kso$ in the integrated system (the green circle curve) .
The gap between $\Kdbsym$ (or $\Kmssym$) and $\Kso$ is caused by the double marginalization effect.

\begin{lemma}\label{theorem:optimality_sym_sw}
Under information symmetry, there exists a critical wholesale price $w^{*} = \sqrt{sc}$ such that \begin{enumerate}
\item when $w < w^{*}$,
the optimal network profit under   WSD-bearing-risk   (i.e., under $\Kmssym$) is larger  than that under   DB-bearing-risk   (i.e., under $\Kdbsym$);
\item when $w > w^{*}$, the optimal network profit under   WSD-bearing-risk   (i.e., under $\Kmssym$) is smaller than that under   DB-bearing-risk   (i.e., under $\Kdbsym$)
\end{enumerate}
\end{lemma}

Lemma
\ref{theorem:optimality_sym_sw} can be obtained by Lemma
\ref{theorem:optimality_sym}, together with the fact that the  {network profit} increases with $k$ when $k \leq \Kso $.
%further implies that (i) when $w < w^{*}$, then the WSD-bearing-risk scheme (Scenario II) can address a better social welfare (because $\Kmssym > \Kdbsym$); and (ii) when $w > w^{*}$, then the DB-bearing-risk scheme (Scenario I) can address a better social welfare (because $ \Kdbsym> \Kmssym$).
%, where $\EDa$ and $\EDb$ are the mean of $\Da$ and $\Db$, respectively.
For clarity, we illustrate the network profit under different reservation solutions vs wholesale price $w$ in  {Figure \ref{fig:K_vs_xi}.b}.
We can see that  (i) the centralized optimal network profit ({the green circle curve}) does not depend on the wholesale price $w$, and (ii) the decentralized optimal network profit under   DB-bearing-risk   (the blue triangle curve) increases with the wholesale price $w$, while the decentralized optimal network profit under   WSD-bearing-risk   (the red square curve) decreases with the wholesale price.
This is because with a larger wholesale price, the database will reserve more spectrum  under DB-bearing-risk (hence, the network profit increases), while the WSD will reserve less spectrum   under WSD-bearing-risk (hence, the network profit decreases).

%A brief explanation is as follows.
%If we rewrite Eq.(\ref{sym_dencentral_DB_profit}) as:
%\begin{equation}\label{sym_dencentral_DB_profit_rewrite}
%\begin{aligned}
%\pi^{\mathrm{d}} (K, \Da )  &=
%(w - c)\cdot K - w\cdot \EX_{\Db} \big[ K - \Da - \Db \big]^{+} \\
%& =  (w - c)\cdot K - w\cdot \EX_{\Db} \big[ K - D \big]^{+} ~~~~~
%\end{aligned}
%\end{equation}
%The first term is the profit margin on the unit of reserved {\bbd} and the last term is the unrealized revenues on unsold units.
%and Eq.(\ref{asym_decentral_WSD_profit2}) as:
%\begin{equation}\label{asym_decentral_WSD_profit2_rewrite}
%\begin{aligned}
%\pi^{\mathrm{m}} (K,\Da ) = &  (r-s) \cdot \Da    \\
%& + ( s - w ) \cdot K  - s \cdot \EX_{\Db} \big[ K - D \big]^{+}
%\end{aligned}
%\end{equation}
%
%When $w < w^{*}$, i.e. $(s-w)/s > (w-c)/w $, $\pi^{\mathrm{m}} (K,\Da )$ in Eq. (\ref{asym_decentral_WSD_profit2}) is larger than $\pi^{\mathrm{d}} (K, \Da )$ in Eq. (\ref{sym_dencentral_DB_profit}) given the same reserved {\bbd} $K$. In other words, WSD has more incentive to reserve more {\bbd} to maximize its own profit. Hence, it is optimal for choosing WSD to decide reserved bandwidth.
%%the WSD's ratio of profit margin per units to unsold cost per unit, $(w-c)/w$, is less than the database's ratio $(s-w)/s$

\subsubsection{\textbf{Performance under Information Asymmetry}}

We now compare two {\bbd} reservation solutions under information asymmetry, i.e.,  $ \Kdbasy$ and $\Kmsasy $.

%We now consider the {\bbd} reservation solutions and the resultant network profit under information asymmetry.

%As shown in (\ref{asym_decentral_bandwidth}) and (\ref{asym_decentral_bandwidth2}),
From {Figure \ref{fig:K_vs_xi}.a}, we can see that
$\Kdbasy$ (blue dashed curve with mark ``x'') under DB-bearing-risk is independent of $\Da$, while $\Kmsasy$ (red dashed curve with mark ``$+$'') under WSD-bearing-risk increases linearly with $\Da$.
Obviously, $\Kdbasy > \Kmsasy $ when $\Da$ is small (e.g., $\Da < 14$), while $\Kdbasy < \Kmsasy $ when  $\Da$ is large (e.g., $\Da > 14$).
This is because the database makes the reservation decision $\Kdbasy$ without knowing the exact value of $\Da$, and thus it is likely to over-reserve spectrum  when $\Da$ is small, while under-reserve spectrum when $\Da$ is large.
Similarly, from Figure \ref{fig:K_vs_xi}.b, we can see that (i) the decentralized optimal  {network profits} under DB-bearing-risk (the blue dash curves with mark ``x'') increases with the wholesale price $w$, while the decentralized optimal network profit under WSD-bearing-risk (the red dash curve with mark ``$+$'', overlapping with the
red square curve) decreases with $w$.
The reason is similar to that under information symmetry, i.e., a larger wholesale price will increase the database's  reservation $\Kdbasy $ under DB-bearing-risk, but reduce the WSD's reservation $\Kmsasy $ under WSD-bearing-risk.
%We can see that the decentralized optimal network profit under the DB-bearing-risk scheme increases with the wholesale price, while the decentralized optimal network profit under the WSD-bearing-risk scheme decreases with the wholesale price.
Moreover, we can see that the decentralized optimal network profit under DB-bearing-risk (the blue dash curves with mark ``x'') decreases with the variance of scheduled demand $\Da$ (denoted by  $\VDa$).
This is because the database's spectrum reservation $\Kdbasy $ under DB-bearing-risk does not consider the exact value of $\Da$; hence, a larger variance of $\Da$ will lead to a larger network profit loss.

%\com{Can we add a concise paragraph, concluding subsection C with strong insights? Maybe we can move the first paragraph of the Section VI here. }

\subsection{Observation}
By the above comparison, we can see that performances of the decentralized optimal solution under information asymmetry (i.e., $\Kdbasy$ in (\ref{asym_decentral_bandwidth}) and $\Kmsasy$ in (\ref{asym_decentral_bandwidth2})) depend on the wholesale price $w$ and the variance of scheduled demand $\Da$. Moreover, both of these solutions
may lead to low profits for both the database and the WSD (comparing with the centralized benchmark),
due to the lack of information and/or the double marginalization effect.

%\input{Section_asymmetry}

%!TEX root = main_CR_Database.tex
%SourceDoc main_CR_Database.tex

 \section{Decentralized {\bbd} Reservation -- Contract-Theoretic Approach}
 \label{sec:asymm-contract}
 In the previous section, we have shown that lacking of information and/or the double marginalization effect may result in profit losses for both the database and the WSD. In this section, we will propose a contract-theoretic approach to achieve credible information sharing and hedge double marginalization in {\bbd} reservation.
%In the previous section, we have shown that the decentralized optimal solution under information asymmetry (i.e., $\Kdbasy$ in (\ref{asym_decentral_bandwidth}) and $\Kmsasy$ in (\ref{asym_decentral_bandwidth2}))
%may lead to low profits for both the database and the WSD (comparing with the centralized benchmark),
%due to the lack of information and/or the double marginalization effect.
%In this section, We will propose a contract-theoretic approach to achieve credible information sharing and hedge double marginalization in {\bbd} reservation.

\subsection{Contract under DB-Bearing-Risk}\label{sec:contract-DB}

As shown in (\ref{asym_decentral_bandwidth}), under the DB-bearing-risk scheme, the profit loss under information asymmetry is mainly due to the lack of information $\Da$ (when the database makes  the {\bbd} reservation decision).
% To solve the inefficiency problem of first case, i.e., database bears the risk of excess bandwidth,
Therefore, we propose a \emph{Spectrum Reservation Contract} to achieve the credible information sharing between the database and the WSD.
We derive the optimal contract that maximizes the database profit under information asymmetry analytically.
Simulations demonstrate that with the optimal contract, the total network profit can also be improved, comparing with that (under information asymmetry) without credible information sharing.

\subsubsection{\textbf{Contract Design}}

The key idea of a spectrum reservation contract is as follows.
To motivate the WSD credibly reveal its private information $\Da$,
the database put an additional charge on the WSD for spectrum reservation (on top of the wholesale charge of $w \cdot \min{[ \K, \Da ]}$).
This \textbf{forces the WSD to share the cost of over-reservation}, such that the WSD has no incentive to inflate the value of $\Da$.

Based on this idea, we design the following \textbf{contract}: $\CTRdb \triangleq \{ \langle \K (\Da ), \P (\Da ) \rangle \}_{\forall \Da}$, which consists of a menu of contract items, $\langle \K(\Da), \P(\Da)\rangle$, each intending for a possible scheduled demand $\Da$. Here, $\K(\Da)$ and $\P(\Da)$ denote the {\bbd} reservation and the WSD's payment to the database, respectively, when the scheduled demand is $\Da$.\footnote{Note that $\P(\Da)$ is the WSD's payment for reserving spectrum via the database, and is not the total cost of using spectrum.}
The detailed {\bbd} reservation process is as follows.
\begin{enumerate}
  \item Before reserving spectrum, the database announces the contract $\CTRdb =  \{ \langle \K (\Da ), \P (\Da ) \rangle  \}_{\forall \Da}$;
 %, which consists of a menu of $\langle \K(\Da), \P(\Da)\rangle$ for all $\Da$;
  \item The WSD selects the contract item $\langle \K(\hDa), \P(\hDa)\rangle$ that maximizes its expected  profit, based on its private information $\Da$;
       %(by which he impliedly informs the database its private information $\Da$);
  \item The database reserves {\bbd} $\K(\hDa)$ for one reservation period, and charges the WSD a reservation fee $\P(\hDa)$ (Step 0 in Figure \ref{fig:system-mdoel-process});
  \item The database sells spectrum to the WSD in each access period (Steps 1-3 in Figure \ref{fig:system-mdoel-process}).
      %(which is independent to the wholesale fee $w$).
\end{enumerate}

When the WSD with information $\Da$ chooses a contract item $\langle \K(\hDa), \P(\hDa)\rangle$ (i.e., that intended for information $\hDa$), the WSD profit, the database profit, and the aggregate profits (network profit) are, respectively,
\begin{equation}
\label{asy_contract_WSD_profit_new}
\begin{aligned}
\textstyle
& \Ums (\K(\hDa), \P(\hDa),\Da ) = (r-w) \cdot  \min  \{ \K(\hDa), \Da \}  \\
& \qquad{}   + (s-w) \cdot \EX_{\Db} \big[ \min\big\{{\Db} ,  (\K(\hDa)-\Da )^{+}\big\} \big] - \P(\hDa) ,
\end{aligned}
\end{equation}
\begin{equation}
\label{asy_contract_database_profit}
\begin{aligned}
\Udb (\K(\hDa), \P(\hDa), \Da )  = & ~
w \cdot \EX_{\Db} \big[  \min \{ \Db  + \Da, \K(\hDa)\} \big]  \\
& ~ - c \cdot \K(\hDa) + \P(\hDa),~~~~~~~~~~
\end{aligned}
\end{equation}
\begin{equation}
\label{asy_contract_total_profit_new}
\begin{aligned}
 & \Utot (\K(\hDa), \P(\hDa), \Da) = r\cdot {\min{ \{\K(\hDa), \Da \}}} \\
 & \qquad{} + s \cdot \EX_{\Db} \big[ \min\big\{{\Db} , (\K(\hDa) -{\Da})^{+} \big\} \big] - c\cdot \K(\hDa).~~~~~~~~~
\end{aligned}
\end{equation}

%We say that a contract is \emph{feasible}, if and only if (i) the WSD with any information $\Da$ prefers the contract item $\langle \K( {\Da}), \P( {\Da})\rangle$ (i.e., that intended for $\Da$) than all other contract items $\langle \K( {\hDa}), \P( {\hDa})\rangle, \forall \hDa \neq \Da$, and (ii) the WSD can achieve a minimum acceptance profit $\Umsmin$ when choosing $\langle \K( {\Da}), \P( {\Da})\rangle$. The former condition is referred to as \emph{Incentive Compatibility} (IC), and the latter condition is referred to as \emph{Individual Rationality} (IR). Formally, we have the following definitions.
%\com{We can consider using definition environment for "feasible" and "optimal" contracts, as they are very important. }

We define a \emph{feasible} contract as follows.
\begin{definition}[Feasible Contract]\label{def:feasible_contract}
A contract is feasible, if and only if
\begin{itemize}

\item
\emph{Incentive Compatibility (IC)}:
The WSD with any information $\Da$ prefers the contract item $\langle \K( {\Da}), \P( {\Da})\rangle$ (that is intended for $\Da$) than all other contract items $\langle \K( {\hDa}), \P( {\hDa})\rangle, \forall \hDa \neq \Da$. Formally, we have
\begin{equation}\label{IC_requirement}
\begin{aligned}
\Ums (\K( {\Da}), \P( {\Da}),\Da ) \geq \Ums (\K(\hDa), \P(\hDa),\Da ),\ \forall \hat{\Da}, \Da.
\end{aligned}
\end{equation}

\item
\emph{Individual Rationality (IR)}:
The WSD can achieve a minimum acceptance profit $\Umsmin$ when choosing $\langle \K( {\Da}), \P( {\Da})\rangle$. Formally, we have
\begin{equation}
\label{AC_requirement}
\Ums (\K( {\Da}), \P( {\Da}),\Da ) \geq  \Umsmin,\ \forall \Da.~~~~~~~~~~~~~~~~~~
\end{equation}

\end{itemize}
\end{definition}

Moreover, we define an optimal contract, denoted by $\CTRdb^* =  \{ \langle \Kdbctr (\Da ), \Pdbctr (\Da ) \rangle \}_{\forall \Da}$, as follows.
\begin{definition}[Optimal Contract]\label{def:optimal_contract}
The contract $\CTRdb^* =  \{ \langle \Kdbctr (\Da ), \Pdbctr (\Da ) \rangle \}_{\forall \Da}$ is optimal if this contract is  feasible and maximizes the database expected profit.
Formally, the optimal contract is given by
\begin{equation}
\label{optimal_contract_asy}
\begin{aligned}
\max_{\langle \K (\Da ), \P (\Da ) \rangle, \forall \Da } & \EX_{\Da} \big[ \Udb (\K( {\Da}), \P( {\Da}), \Da ) \big],\\
\mathrm{subject~to:~~} & \mathrm{IC~and~IR~in~(\ref{IC_requirement})~and~(\ref{AC_requirement}).}
\end{aligned}
\end{equation}

\end{definition}

In the following, we first provide the necessary and sufficient conditions for a feasible contract.
Then, we derive the optimal contract systematically.
For clarity, we present all of the detailed proofs in \cite{techrpt}.
%We denote the {\bbd} reservation and reservation fee in the optimal contract by $ \langle\Kdbctr(\Da),
%{\Pdbctr}(\Da)\rangle$ for each $\Da$.

\subsubsection{\textbf{Feasibility}}\label{sec:contract_feasibility_BR}
Suppose that a contract $\CTRdb = \{ \langle \K (\Da ), \P (\Da ) \rangle \}_{\forall \Da}$ is feasible.
Then, the following   necessary conditions hold.

\begin{proposition}[Necessary Condition I for Feasibility]\label{lemma:nece1}
$$ \K (\Da_1) > \K (\Da_2), \mathrm{~~if~and~only~if~~} \P (\Da_1) > \P (\Da_2).$$
\end{proposition}

%For detailed proof, please refer to Appendix-\ref{lemma:nece1-proof}.

\begin{proposition}[Necessary Condition II for Feasibility]\label{lemma:nece2}
$$\K (\Da_1) \geq \K (\Da_2),\quad \forall \Da_1 > \Da_2.$$
\end{proposition}

%For detailed proof, please refer to Appendix-\ref{lemma:nece2-proof}.

Proposition \ref{lemma:nece1} implies that in a feasible contract,
a larger {\bbd} reservation $\K(\cdot)$ must correspond to a larger reservation fee $\P(\cdot)$.
This is quite intuitive, as the WSD's profit is increasing in $\K(\cdot)$ but decreasing in $\P(\cdot)$.
Proposition \ref{lemma:nece2} implies that the {\bbd} reservation $\K(\cdot)$ increases with the value of scheduled demand $\Da$.

For convenience, we denote $\Ums ( \Da ) \triangleq \Ums (\K( {\Da}), \P( {\Da}),\Da )$ as the WSD profit when  choosing the contract item intended for its true private information $\Da$.
Given any feasible $\K(\Da)$ (i.e., those non-decreasing with $\Da$), we have the following necessary conditions for the feasible $\P(\Da)$, or equivalently, for the WSD profit $\Ums ( \Da )$.

\begin{proposition}[Necessary Condition III for Feasibility]\label{lemma:nece3}
$$\Ums (\Da_1) \geq \Ums (\Da_2),\quad \forall \Da_1 > \Da_2.$$
\end{proposition}

%For detailed proof, please refer to Appendix-\ref{lemma:nece3-proof}.

\begin{proposition}[Necessary Condition IV for Feasibility]\label{lemma:nece4}
\begin{equation*}\label{eq:nece4}
\begin{aligned}
\Ums (\Da) = & \Ums (\underline{\Da}) + (r-s)\cdot (\Da-\underline{\Da}) \\
&~ + \int_{\underline{\Da}}^{\Da}
 (s-w ) \cdot G\big(  \K(x) - x \big)\,\mathrm{d}{x} .
\end{aligned}
\end{equation*}
\end{proposition}

%For detailed proof, please refer to Appendix-\ref{lemma:nece4-proof}.
%This quantity is referred to as the mater's expected information rent under database risk-bearing scenario.

Proposition  \ref{lemma:nece3} implies that in a feasible contract, the WSD profit increases with the value of $\Da$.
Proposition \ref{lemma:nece4} further gives the detailed form of the WSD profit in a feasible contract, given any feasible $\K(\Da)$.
Note that the third term on the r
Here, $\underline{\Da}$ is the minimum achievable value of scheduled demand $\Da$, i.e., $g(\Da) = 0$ if $\Da < \underline{\Da}$.

By Proposition \ref{lemma:nece4}, we can get the following feasible reservation fee $\P(\Da)$ directly:
\begin{equation}
\label{eq:opt-p}
\begin{aligned}
\P (\Da)  = & - { \Ums (  \Da )} + (r-w) \cdot  \min  \{\K( {\Da}), \Da \} \\
& + (s-w) \cdot  \EX_{\Db} \big[ \min\big\{{\Db} ,  (
\K( {\Da})-\Da )^{+}\big\} \big],
\end{aligned}
\end{equation}
where $ { \Ums (  \Da )}$ is given in Proposition \ref{lemma:nece4}.

We have shown the necessary conditions for a feasible contract through Propositions \ref{lemma:nece1}-\ref{lemma:nece4}.
Next we show that these conditions are also sufficient for a contract to be feasible.
%for a contract to be feasible.

%\com{How are the three conditions in Prop 5 correspond to necessary conditions? }

\begin{proposition}[Sufficient Conditions for Feasibility]\label{lemma:sufficent}
A contract $\CTRdb =  \{ \langle \K (\Da ), \P (\Da ) \rangle   \}_{\forall \Da}$ is feasible, if the following conditions hold:
\begin{itemize}
  \item $\K (\Da )$ is non-decreasing in $\Da$ {(i.e., Necessary Condition II in Proposition \ref{lemma:nece2}),}
  %\item  $P (\Da)$ is given by the WSD's expected profit $\Ums  (\Da)$ defined in Lemma \ref{lemma:nece4} with $\Ums  (\underline{\Da}) \geq \Umsmin $.
  \item  $\P (\Da )$ is given by (\ref{eq:opt-p}) {(i.e., Necessary Condition IV in Proposition \ref{lemma:nece4}),}
  \item  $\Ums (\underline{\Da}) \geq \Umsmin $ {(i.e., IR Condition).}
  \end{itemize}
\end{proposition}

Intuitively, the first two conditions guarantee the IC condition for the contract, and the last condition guarantees the IR condition for the contract. Therefore, the conditions in Proposition \ref{lemma:sufficent} are sufficient.
%For detailed proof, please refer to Appendix-\ref{lemma:sufficent-proof}.

\subsubsection{\textbf{Optimality}}\label{sec:contract_optimality_BR}

Now we study the database's optimal contract characterized by (\ref{optimal_contract_asy}).
By (\ref{asy_contract_WSD_profit_new}) and (\ref{asy_contract_database_profit}), we notice that the total profit can be freely transferred between the database and the WSD through the reservation fee $\P(\Da)$.
Therefore, to maximize the database profit, we need to shrink the WSD's profit as much as possible.
This leads to the following optimality condition immediately.

\begin{proposition}[Optimality Condition I]\label{lemma:OCI}
\begin{equation*}\label{eq:opt1}
\Ums  (\underline{\Da} ) = \Umsmin.
\end{equation*}
\end{proposition}

%For detailed proof, please refer to Appendix-\ref{lemma:OCI-proof}.

Proposition \ref{lemma:OCI} implies that in the optimal contract, the database will assign the minimal acceptable profit to the WSD.
Intuitively, if the WSD profit $\Ums  (\underline{\Da} ) = X > \Umsmin$, then the database can immediately improve its profit by increasing the reservation fee $\P(\Da)$ by a constant $(X-\Umsmin)$ for all $\Da$.

Denote $\Udb ( \Da ) \triangleq \Udb ( \K( {\Da}), \P( {\Da}),\Da )$ and  $\Utot ( \Da ) \triangleq \Utot ( \K( {\Da}), \P( {\Da}),\Da )$.
By (\ref{asy_contract_WSD_profit_new})-(\ref{asy_contract_total_profit_new}), we can write the database's   profit as $\Udb  ( {\Da} ) = \Utot  ( {\Da} ) - \Ums  ({\Da} )$.
Together with Proposition \ref{eq:nece4} and Proposition \ref{lemma:OCI}, we can rewrite the database profit maximization problem  (\ref{optimal_contract_asy}) as follows.
\begin{equation*}
\begin{aligned}
\max_{  \K (\Da ) , \forall \Da } ~~
\EX_{\Da} \big[ \Udb ( \Da ) \big] & \triangleq  \textstyle \int_{\underline{\Da}}^{\bar{\Da}} \phidb \big(\K(\Da),\Da \big) \cdot f (\Da )   \mathrm{d}{\Da} -  \Umsmin, \end{aligned}
\end{equation*}
\begin{equation}
\label{asy_optimal_contract_formulation}
\begin{aligned}
\mathrm{subject~to:~~} &  \K(\Da) \mathrm{~is~non\mbox{-}decreasing~in~} \Da,
\end{aligned}
\end{equation}
where
$$
\textstyle
\phidb \big(\K(\Da),\Da \big) \triangleq \Utot  ( {\Da} ) - \frac{1-F (\Da )}{f (\Da )}\big[ r-s + (s-w) \cdot G\big(\K(\Da) - \Da\big)\big].
$$

We first notice that $\phidb (\K(\Da),\Da ) $ is related to a particular $\Da$ only, and is independent of other $\hDa \neq \Da$. Thus, the optimal solution of (\ref{asy_optimal_contract_formulation})
%(i.e., the optimal {\bbd} reservation $\Kdbctr(\Da)$)
can be obtained by maximizing $\phidb\big(\K(\Da),\Da\big)$ for each $\Da$ independently (as long as the non-decreasing condition is not violated).
However,
due to the non-convexity of $G(\cdot)$, $\phidb (\K(\Da),\Da ) $ is non-convex in $\K(\Da)$, and thus the classic Karush-Kuhn-Tucker (KKT) analysis cannot be directly applied here.\footnote{As an example mentioned in Section \ref{sec:model-demand}, the bursty demand $\Db$'s distribution $G(\cdot)$ is the chi-square distribution, which is non-convex.}

Next we can show that $\phidb (\K(\Da),\Da ) $ has the nice property of  piecewise convexity.
Based on this, the maximizer of $\phidb (\K(\Da),\Da ) $ is unique, and it satisfies the first-order condition: $\frac{\partial {\phidb (\K, \Da )}}{\partial{\K } }=0$.
%\footnote{We leave the detailed discussion in our technical report \cite{xx}.}
Formally, the optimal $\K(\Da), \forall \Da$, is given by
\begin{equation}
\label{asy_optimal_K_in_contract}
\begin{aligned}
\textstyle
 \frac{\partial {\phidb (\K, \Da )}}{\partial{\K } } = & s \cdot [1-G ( \K(\Da) -\Da  )  ]  - c
  \\
  & - \frac{1-F(\Da)}{f(\Da)}\cdot ( s-w  )\cdot g( \K(\Da) - \Da  )  = 0 .
\end{aligned}
\end{equation}
We can further check that optimal $\K(\Da)$  given by (\ref{asy_optimal_K_in_contract}) is indeed non-decreasing in $\Da$, due to the IFR assumption for $F(\cdot)$, i.e., $ \frac{1-F(\Da)}{f(\Da)} $ decreases with $\Da$.
Therefore, we have the following optimal contract under DB-bearing-risk.

\begin{theorem}\label{thrm:optimal-contract}
Under DB-bearing-risk, the database's optimal contract  $ \CTRdb^* = \{ \langle \Kdbctr(\Da),{\Pdbctr}(\Da) \rangle \}_{\forall \Da}$ is given by: $\forall \Da \in [\underline{\Da}, \bar{\Da}]$,
\begin{itemize}
  \item  $\Kdbctr(\Da)$  is given by (\ref{asy_optimal_K_in_contract}), and
  \item  ${\Pdbctr}(\Da)$ is given by (\ref{eq:opt-p}) with $\Ums  (\underline{\Da}) = \Umsmin $.
\end{itemize}
\end{theorem}

%This Theorem can be simply proved by using Lemma \ref{lemma:sufficent} and Lemma \ref{lemma:OCI}.

Now we provide some useful properties for the optimal contract $  \CTRdb^* = \{ \langle \Kdbctr(\Da),{\Pdbctr}(\Da) \rangle \}_{\forall \Da}$. Specifically,
\begin{equation}\label{eq:marginal-wholesale-price-BR}
\begin{aligned}
\textstyle
\frac{\mathrm{d} \Pdbctr }
    {\mathrm{d} \Kdbctr }
=
\frac{ \mathrm{d} \Pdbctr  / \mathrm{d}{\Da} }
     { \mathrm{d}  \Kdbctr / \mathrm{d}{\Da}}
=
 ( s-w  ) \cdot \big[ 1 - G( \Kdbctr-\Da )  \big] \geq 0,~~~
\end{aligned}
\end{equation}
\begin{equation}\label{eq:marginal-wholesale-price-SOC-BR}
\begin{aligned}
\textstyle
\frac{\mathrm{d}^2 \Pdbctr~}
    { \mathrm{d}~{\Kdbctr}^2 }
& \textstyle
=
\frac{ \mathrm{d}
\big(\frac{\mathrm{d} \Pdbctr }
{\mathrm{d}  \Kdbctr }\big) / \mathrm{d}{\Da}}
{\mathrm{d}  \Kdbctr  / \mathrm{d}{\Da}}
=
\frac{ -( s-w  ) \cdot g( \Kdbctr-\Da ) \cdot \left( {\mathrm{d}  \Kdbctr / \mathrm{d}{\Da}} - 1 \right)}{\mathrm{d}  \Kdbctr / \mathrm{d}{\Da}} \leq 0.
\end{aligned}
\end{equation}
%which shows that $\Pdbctr$ is concavely increasing in $K^{\mathrm{cr}}_{\mathrm{asy}}$. %Eq.(\ref{eq:marginal-wholesale-price-SOC-BR}) and Eq.(\ref{eq:marginal-wholesale-price-SOC-AP}) holds when
%$\left( {\mathrm{d}}/{\mathrm{d} \Da} \right) g( K(\Da) - \Da  )  > 0$.
The above properties show that $\Pdbctr$ is concavely increasing in $\Kdbctr$ (which can be seen from Figure \ref{fig:k_vs_p_WSD}.a).
This implies that the database's reservation fee charge for each additional unit of {\bbd} reservation will decrease with the total amount of {\bbd} reservation.

\subsection{Contract under WSD-Bearing-Risk}\label{sec:contract-WSD}
Comparing (\ref{centralized_bandwidth}) and (\ref{asym_decentral_bandwidth2}), we can see that under WSD-bearing-risk, the gap between the centralized optimal reservation $\Kso$ and the decentralized optimal reservation $\Kmsasy $ (under information asymmetry without information sharing) is mainly due to the double marginalization effect, which further leads to some loss in both the database profit and the total network profit.
%\com{This has not been carefully explained yet. This at equation (12)}
The perfect coordination of the WSD's optimal solution (\ref{asym_decentral_bandwidth2}) and the centralized optimal solution (\ref{centralized_bandwidth}) requires the wholesale price to be as low as the cost (i.e., $w = c$). This is obviously undesirable for a profit-maximizing database.
To this end, we propose a \emph{Spectrum Reservation Contract} to mitigate the double marginalization effect in this case.
%Our simulations show that such a contract can improve both the database's and the WSD profit comparing with \rev{???}.
Similarly, we analytically derive the optimal contract that maximizes the database profit under information asymmetry. Simulations demonstrate that with the optimal contract, the total network profit can also be improved, comparing with that (under information asymmetry) without credible information sharing.

\subsubsection{\textbf{Contract Design}}

The detailed contract formulation under   WSD-bearing-risk   is   similar to that under   DB-bearing-risk   (in Section \ref{sec:contract-DB}).
Specifically, to motivate the WSD to order {\bbd} according to the database's profit-maximizing objective, the database charges
the WSD for the spectrum reservation (in addition of the wholesale charge of $w \cdot \K$).\footnote{Note that this wholesale charge is different from that under DB-bearing-risk. The latter is $w \cdot \min{[ \K, \Da ]}$, as the WSD only needs to pay for the spectrum it actually purchases.}
This \textbf{forces the database to share the cost of over-reservation}, such that the WSD operates as the database desired.

%To motivate the WSD to credibly report its private information $\Da$,
%the database can require a quantity commitment from the WSD before reserving bandwidth. To do so, database reserves {\bbd} which is equal to the WSD's order quantity and charges corresponding payment from WSD. The advance purchase could be costly to the WSD if the realized demand turns out to be smaller than the advance purchase quantity. Intuitively, this commitment limits a WSD with a low $\Da$ to place a large order.
%The essence behind this idea is to \textbf{share part of the reservation cost $cK$ with the WSD}, such that the WSD has no incentive to request a reservation $K$ larger than needed.

Similarly, we design the following  \textbf{contract}: $\CTRms \triangleq \{ \langle \K (\Da ), \P (\Da ) \rangle \}_{\forall \Da}$, where each contract item $\langle \K(\Da), \P(\Da)\rangle$ specifies a {\bbd} reservation level $\K(\Da)$ and the corresponding WSD's payment $\P(\Da)$.
The detailed {\bbd} reservation process is the same as that in Section \ref{sec:contract-DB}.
However, the definitions for the database's and the WSD profits are different, due to the different risk-bearing schemes.

%This can be achieved through a \emph{contract}. The key steps is the same as the {\bbd} reservation contract (\ref{sec:contract-DB}). However, the expected profits are different.
%\begin{enumerate}
%  \item Before reserving bandwidth, the database offers a contract $\Psi^{ad} = \left\{ \langle K^{ad} (\Da ), P^{ad} (\Da ) \rangle,\ \forall \Da \right\}$, which consists of a menu of $\langle K^{ad}(\Da), P^{ad}(\Da)\rangle$ for all $\Da$;
%  \item The WSD selects the contract item $\langle K^{ad}(\hat{\Da}), P^{ad}(\hat{\Da})\rangle$ that maximizes its expected profit, based on its private information $\Da$;
%       %(by which he impliedly informs the database its private information $\Da$);
%  \item The database then reserves {\bbd} $K^{ad}(\hat{\Da})$ and charges the WSD $P^{ad}(\hat{\Da})$.
%      %(which is independent to the wholesale fee $w$).
%\end{enumerate}

Specifically, when the WSD with information $\Da$ chooses a contract item $\langle \K(\hDa), \P(\hDa)\rangle$ (i.e., that intended for $\hDa$), the WSD's profit, the database profit, and the aggregate profits ({network profit}) are, respectively,
\begin{equation}
\label{asy_contract_WSD_profit_ad}
\begin{aligned}
& \Ums (\K(\hDa), \P(\hDa),\Da ) = r \cdot \min  \{ \K(\hDa), \Da \} - w \cdot \K(\hDa) \\
& \qquad{} \qquad{} + s \cdot \EX_{\Db} \big[ \min\big\{{\Db} ,  (\K(\hDa)-\Da )^{+}\big\} \big] - \P(\hDa),
\end{aligned}
\end{equation}
\begin{equation}
\label{asy_contract_database_profit_ad}
\begin{aligned}
\Udb (\K(\hDa), \P(\hDa), \Da )  = &
(w - c) \cdot \K(\hDa)   + \P(\hDa),~~~~~~~~
\end{aligned}
\end{equation}
\begin{equation}
\label{asy_contract_total_profit_new_ad}
\begin{aligned}
&  \Utot (\K(\hDa), \P(\hDa), \Da) =  r\cdot {\min{ \{\K(\hDa), \Da \}}}  \\
& \qquad{} \qquad{} + s \cdot \EX_{\Db} \big[ \min\big\{{\Db} , (\K(\hDa) -{\Da})^{+} \big\} \big] - c\cdot \K(\hDa).
\end{aligned}
\end{equation}
Obviously, the aggregate profit in (\ref{asy_contract_total_profit_new_ad}) is same as that in (\ref{asy_contract_total_profit_new}), that is,
 the {network profit} does not depend on the choice of the risk-bearing scheme.

{Similar as in Definition \ref{def:feasible_contract} and \ref{def:optimal_contract}}, we first define the contract feasibility and optimality.

%We define a \emph{feasible} contract as follows.
\begin{definition}[Feasible Contract under WSD-risk-bearing]\label{def:feasible_contract_wsd}
The contract $\CTRms = \{ \langle \K (\Da ), \P (\Da ) \rangle \}_{\forall \Da}$ is feasible, if and only if it satisfies the following conditions.
\begin{equation}
\label{IC_requirement_ad}
\mathrm{IC:}~~\Ums (\K( {\Da}), \P( {\Da}),\Da ) \geq \Ums (\K(\hDa), \P(\hDa),\Da ),\ \forall \hat{\Da},  \Da;
\end{equation}
\begin{equation}
\label{AC_requirement_ad}
\mathrm{IR:}~~\Ums (\K( {\Da}), \P( {\Da}),\Da ) \geq  \Umsmin,\ \forall \Da.~~~~~~~~~~~~~~~~~~
\end{equation}

%\begin{itemize}
%
%\item
%\emph{Incentive Compatibility (IC)}:
%%The WSD with any information $\Da$ prefers the contract item $\langle \K( {\Da}), \P( {\Da})\rangle$ (i.e., that intended for $\Da$) than all other contract items $\langle \K( {\hDa}), \P( {\hDa})\rangle, \forall \hDa \neq \Da$. Formally, we have
%\begin{equation}
%\label{IC_requirement_ad}
%\mathrm{IC:}~~\Ums (\K( {\Da}), \P( {\Da}),\Da ) \geq \Ums (\K(\hDa), \P(\hDa),\Da ),\ \forall \hat{\Da},  \Da;
%\end{equation}
%
%
%\item
%\emph{Individual Rationality (IR)}:
%%The WSD can achieve a minimum acceptance profit $\Umsmin$ when choosing $\langle \K( {\Da}), \P( {\Da})\rangle$. Formally, we have
%\begin{equation}
%\label{AC_requirement_ad}
%\mathrm{IR:}~~\Ums (\K( {\Da}), \P( {\Da}),\Da ) \geq  \Umsmin,\ \forall \Da.~~~~~~~~~~~~~~~~~~
%\end{equation}
%
%\end{itemize}
\end{definition}

%Moreover, we judge whether the contract $\CTRms = \{ \langle \K (\Da ), \P (\Da ) \rangle \}_{\forall \Da}$ is \emph{optimal} based on the following definition.
We denote the optimal contract by  $\CTRms^* = \{ \langle \Kmsctr (\Da ), \Pmsctr (\Da ) \rangle \}_{\forall \Da}$, which is defined below.
\begin{definition}[Optimal Contract]\label{def:optimal_contract_wsd}
The contract $\CTRms^* =  \{ \langle \Kmsctr (\Da ), \Pmsctr (\Da ) \rangle \}_{\forall \Da}$ is optimal if this contract is  feasible   and   maximizes the database expected profit. Formally, the optimal contract is given by
\begin{equation}
\label{optimal_contract_asy_ad}
\begin{aligned}
\max_{\langle \K (\Da ), \P (\Da ) \rangle, \forall \Da } & \EX_{\Da} \big[ \Udb (\K( {\Da}), \P( {\Da}), \Da ) \big],\\
\mathrm{subject~to:~} & \mathrm{IC~and~IR~in~(\ref{IC_requirement_ad})~and~(\ref{AC_requirement_ad}).}
\end{aligned}
\end{equation}
\end{definition}

\subsubsection{\textbf{Feasibility}}
It is easy to check that the necessary conditions II and III in Propositions \ref{lemma:nece2}-\ref{lemma:nece3} also hold for the feasible contract under WSD-bearing-risk.
%However, the detailed form of the WSD's profit is a bit different to that under the DB-bearing-risk scheme (given in Proposition \ref{lemma:nece4}).
However, the necessary condition IV in Proposition \ref{lemma:nece4} is a bit different.
Specifically,
%for a feasible advance purchase contract is similar to the condition for a {\bbd} reservation contract (\ref{sec:contract_feasibility_BR}), except the following condition:
\begin{proposition}[Necessary Condition IV for Feasibility under WSD-bearing-risk]\label{lemma:nece4_ad}
%\rev{When WSD bears the risk, the expected profit of WSD can be obtained as:}
Given a feasible $\K( {\Da})$, the WSD's expected profit is
\begin{equation*}\label{eq:nece4_ad}
\begin{aligned}
\Ums  (\Da) = & \Ums  (\underline{\Da}) + (r-s)\cdot (\Da-\underline{\Da}) + \int_{\underline{\Da}}^{\Da}
 s \cdot G\big( \K(x) - x \big)\,\mathrm{d}{x} .
\end{aligned}
\end{equation*}
\end{proposition}
%This quantity is the mater's expected information rent under WSD risk-bearing scenario.

\noindent
Accordingly, the feasible reservation fee $\P (\Da)$ is
\begin{equation}
\label{eq:opt-p-ad}
\begin{aligned}
\P (\Da)  = & -  {\Ums  (  \Da )} + r \cdot  \min  \{\K( {\Da}), \Da \} \\
&  + s \cdot \EX_{\Db} \big[ \min\big\{{\Db} ,  (\K( {\Da})-\Da )^{+}\big\} \big] - w \cdot \K ( {\Da}),
\end{aligned}
\end{equation}
where $\Ums  (  \Da )$ is given in Proposition \ref{eq:nece4_ad}.

\subsubsection{\textbf{Optimality}}\label{sec:contract_optimality_AD}
Notice that the optimality condition in Proposition \ref{lemma:OCI} also holds for the WSD-bearing-risk scheme.
%Use similar method in section \ref{sec:contract_optimality_BR}, we can derive the optimal contract maximizing the database's expected profit.
%\begin{lemma}[Optimality Condition I]\label{lemma:OCI_AD}
%\begin{equation}\label{eq:opt1_AD}
%\Ums  (\underline{\Da} ) = \Umsmin.
%\end{equation}
%\end{lemma}
%For detailed proof, please refer to Appendix-\ref{lemma:OCI-proof}.
%Intuitively, the database assigns the minimal acceptable profit to the WSD.
Thus, we can similarly rewrite the database profit
maximization problem (\ref{optimal_contract_asy_ad}) as
\begin{equation*}
\begin{aligned}
\max_{  \K (\Da ) , \forall \Da } & \textstyle  ~~ \EX_{\Da} \big[ \Udb ( \Da ) \big] \triangleq  \int_{\underline{\Da}}^{\bar{\Da}} \phims \big(\K(\Da),\Da \big) \cdot f (\Da )   \mathrm{d}{\Da} -  \Umsmin,
\end{aligned}
\end{equation*}
\begin{equation}
\label{asy_optimal_contract_formulation_AD}
\begin{aligned}
\mathrm{subject~to:~~} &  \K(\Da) \mathrm{~is~non\mbox{-}decreasing~in~} \Da,
\end{aligned}
\end{equation}
where
$$
\textstyle
\phims \big(\K(\Da),\Da \big) \triangleq \Utot  ( {\Da} ) - \frac{1-F (\Da )}{f (\Da )} \cdot \big[ r-s + s \cdot G\big(\K(\Da) - \Da\big)\big].
$$

%Since each $\phidb (K(\Da),\Da ) $ is related to $\Da$ only, the optimal reservation $\Kdbctr(\Da)$ can be obtained by maximizing $\phidb\big(K(\Da),\Da\big)$ for each $\Da$ independently, if   $\Kdbctr(\Da)$ satisfies the non-decreasing condition.

Using a similar analysis as in Section   \ref{sec:contract-DB}, we can show that the optimal solution
of (\ref{asy_optimal_contract_formulation_AD}) can be obtained by maximizing $\phims \big(\K(\Da),\Da \big)$ for each $\Da$ independently.
Moreover, the optimal $\K(\Da)$ satisfies the first-order condition: $ \frac{\partial {\phims (\K, \Da )}}{\partial{\K } } =0$. Formally,
\begin{equation}
\label{asy_optimal_K_in_contract_AD}
\begin{aligned}
\textstyle
 \frac{\partial {\phims (\K(\Da), \Da )}}{\partial{\K } } = & s \cdot [1-G ( \K(\Da) -\Da  )  ]  - c \\
  & - \frac{1-F(\Da)}{f(\Da)}\cdot s \cdot g( \K(\Da) - \Da  )  = 0 .
\end{aligned}
\end{equation}
Therefore, the optimal contract under the WSD-bearing-risk scheme is given in the following theorem.

\begin{theorem}\label{thrm:optimal-contract-AD}
Under   WSD-bearing-risk, the optimal contract  $ \CTRms^* = \{ \langle \Kmsctr(\Da),{\Pmsctr}(\Da) \rangle \}_{\forall \Da}$ is given by: $\forall \Da \in [\underline{\Da}, \bar{\Da}]$,
\begin{itemize}
  \item  $\Kmsctr(\Da)$  is given by (\ref{asy_optimal_K_in_contract_AD}), and
  \item  ${\Pmsctr}(\Da)$ is given by (\ref{eq:opt-p-ad}) with $\Ums  (\underline{\Da}) = \Umsmin $.
\end{itemize}
\end{theorem}

%\begin{theorem}[\textbf{Optimality}]\label{thrm:optimal-contract-AD}
%\begin{itemize}
%    \item[~]
%  \item  ${K^{\mathrm{ap}}_{\mathrm{asy}}}(\Da)$  is given by (\ref{asy_optimal_K_in_contract_AD}),
%  \item  ${P^{\mathrm{ap}}_{\mathrm{asy}}}(\Da)$ is given in (\ref{eq:opt-p-ad}) with $\Ums  (\underline{\Da}) = \Umsmin $.
%\end{itemize}
%\end{theorem}

We provide some useful properties for the optimal
contract $\CTRms^* = \{ \langle \Kmsctr (\Da ),  \Pmsctr (\Da ) \rangle \}_{\forall \Da}$. Specifically,
\begin{equation}\label{eq:marginal-wholesale-price-AP}
\begin{aligned}
\textstyle
\frac{\mathrm{d} \Pmsctr }
    {\mathrm{d} \Kmsctr }
=
\frac{ \mathrm{d} \Pmsctr  / \mathrm{d}{\Da} }
     { \mathrm{d}  \Kmsctr / \mathrm{d}{\Da}}
=
s\cdot \big[ 1 - G( \Kmsctr-\Da )  \big] - w,~~~~~~~~~
\end{aligned}
\end{equation}
\begin{equation}\label{eq:marginal-wholesale-price-SOC-AP}
\begin{aligned}
\textstyle
\frac{\mathrm{d}^2 \Pmsctr~}
    { \mathrm{d}~{\Kmsctr}^2 }
& \textstyle
=
\frac{ \mathrm{d}
\big(\frac{\mathrm{d} \Pmsctr }
{\mathrm{d}  \Kmsctr }\big) / \mathrm{d}{\Da}}
{\mathrm{d}  \Kmsctr  / \mathrm{d}{\Da}}
=
\frac{ -s \cdot g( \Kmsctr-\Da ) \cdot \left( {\mathrm{d}  \Kmsctr / \mathrm{d}{\Da}} - 1 \right)}{\mathrm{d}  \Kmsctr / \mathrm{d}{\Da}} \leq 0.
\end{aligned}
\end{equation}
The second property shows that $\Pmsctr$ is concave in $\Kmsctr$, and the first property shows that $\Pmsctr$ is non-monotonous in $\Kmsctr$. More precisely, $\Pmsctr$ first increases with $\Kmsctr$ and then decreases with $\Kmsctr$, as illustrated in Figure { \ref{fig:k_vs_p_WSD}.a}.

%\com{The first order equation \ref{eq:marginal-wholesale-price-AP} should be $s\cdot \big[ 1 - G( \Kmsctr-\Da )  \big] - w$, in such form that we can explain why the figure \ref{fig:k_vs_p_WSD} has such form. }

%We can also seek the relationship of $K$ and $P$ in an optimal contracts $ \Psi^{**} = \{ \langle {K^{\mathrm{ap}}_{\mathrm{asy}}}(\Da),{P^{\mathrm{ap}}_{\mathrm{asy}}}(\Da) \rangle \}$ under WSD bears risk scenario. We have
%\begin{equation}\label{eq:marginal-wholesale-price-AP}
%\begin{aligned}
%\textstyle
%\frac{\mathrm{d} P^{\mathrm{ap}}_{\mathrm{asy}} }
%    {\mathrm{d} K^{\mathrm{ap}}_{\mathrm{asy}} }
%=
%\frac{ \mathrm{d} P^{\mathrm{ap}}_{\mathrm{asy}} (\Da) / \mathrm{d}{\Da} }
%     { \mathrm{d} K^{\mathrm{ap}}_{\mathrm{asy}}(\Da)/ \mathrm{d}{\Da}}
%=
%  s \cdot \big[ 1 - G( K-\Da )  \big]  - w ,
%\end{aligned}
%\end{equation}
%\begin{equation}\label{eq:marginal-wholesale-price-SOC-AP}
%\begin{aligned}
%\textstyle
%\frac{\mathrm{d}^2 P^{\mathrm{ap}}_{\mathrm{asy}} }
%    { \mathrm{d} {K^{\mathrm{ap}}_{\mathrm{asy}}}^2 }
%&=
%\frac{ \mathrm{d}
%\big(\frac{\mathrm{d} P^{\mathrm{ap}}_{\mathrm{asy}} }
%{\mathrm{d}  K^{\mathrm{ap}}_{\mathrm{asy}} }\big) / \mathrm{d}{\Da}}
%{\mathrm{d}  K^{\mathrm{ap}}_{\mathrm{asy}} (\Da) / \mathrm{d}{\Da}} \\
%&=
%\frac{ - s \cdot g( K-\Da )\left( {\mathrm{d}  K^{\mathrm{cr}}_{\mathrm{asy}} (\Da) / \mathrm{d}{\Da}} - 1 \right)}{\mathrm{d}  K^{\mathrm{cr}}_{\mathrm{asy}} (\Da) / \mathrm{d}{\Da}}  \\
%& \leq 0,
%\end{aligned}
%\end{equation}
%which shows that $P^{\mathrm{ap}}_{\mathrm{asy}}$ is concavely increasing in $K^{\mathrm{ap}}_{\mathrm{asy}}$.

\vspace{-2mm}
\subsection{Comparison}

Now we compare the optimal contract $\CTRdb^* = \{ \langle \Kdbctr (\Da ),  \Pdbctr (\Da ) \rangle \}_{\forall \Da}$ under the DB-bearing-risk scheme (in Theorem \ref{thrm:optimal-contract}) and  the optimal contract $\CTRms^* = \{ \langle \Kmsctr (\Da ),  \Pmsctr (\Da ) \rangle \}_{\forall \Da}$ under the WSD-bearing-risk scheme (in Theorem \ref{thrm:optimal-contract-AD}).

%\com{since we have added the P-K relationship with different variance, do you think we should add a Lemma to illustrate that? }

Figure {\ref{fig:k_vs_p_WSD}.a} compares
the structures of both contracts, by showing the relationships of reservation and reservation fee under both optimal contracts.
\begin{itemize}
\item For the optimal contract $\CTRdb^*$ under DB-bearing-risk, we can see that the reservation fee $p^*$ monotonically increases with the {\bbd} reservation $k^*$. This is because the WSD always benefits from a larger spectrum reservation level (as it does not need to bear the risk); hence, the database can charge a higher reservation fee for a higher reservation level.

\item For the optimal contract $\CTRms^*$ under WSD-bearing-risk, we can see that the reservation fee $p^*$ first increases and then decreases with the {\bbd} reservation $k^*$. This is because the WSD's profit first increases with the reservation level, and then decreases with the reservation level (due to the high risk of over-reservation); hence, the reservation fee first increases with the reservation level, and then decreases with the reservation level.

\end{itemize}
%We can see that {in the optimal contract $\CTRdb^*$ under DB-bearing-risk, the reservation fee $p^*$ increases with the {\bbd} reservation $k^*$, while in the optimal contract $\CTRms^*$ under
%WSD-bearing-risk, the reservation fee $p^*$ first increases and then decreases with the {\bbd} reservation $k^*$.}
%The reasons are as follows. Under DB-bearing-risk, the WSD always benefits from a larger spectrum reservation level (as it does not need to bear the risk); hence, the database can charge a higher reservation fee for a higher reservation level. Under WSD-bearing-risk, the WSD's profit first increases with the reservation level, and then decreases with the reservation level (due to the high risk of over-reservation); hence, the reservation fee first increases with the reservation level, and then decreases with the reservation level.
We can further see that under the same reservation level $k^*$, the reservation fee under DB-Bear-Risk is larger than that under WSD-Bear-Risk, hence charges
%\rev{This is because under DB-Bear-Risk, we need to transfer risk from the database to the WSD, while under WSD-Bear-Risk, we need to transfer risk from the WSD to the database.}\com{this does not seem to explain either I or II. }
a higher reservation fee to compensate its expected cost due to over-reservation.

%\begin{figure*}
%\vspace{-4mm}
%  \centering
%   \includegraphics[width=3in]{Figure/P_vs_K_diff_variance}
%  ~~~~
%  \includegraphics[width=3in]{Figure/K_vs_xi_contract}
%  \vspace{-2mm}
%  \caption{ (a) Illustration of Optimal Contracts, (b) Contract-based Spectrum Reservations vs Scheduled Demand $\Da$. Here, $\sigma_{\Da}$ denotes the variance of scheduled demand $\Da$.}\label{fig:k_vs_p_WSD}
%  \vspace{-4mm}
% \end{figure*}
%
 \begin{figure*}
% 	\vspace{-4mm}
 	\centering
 	\includegraphics[width=2.8in]{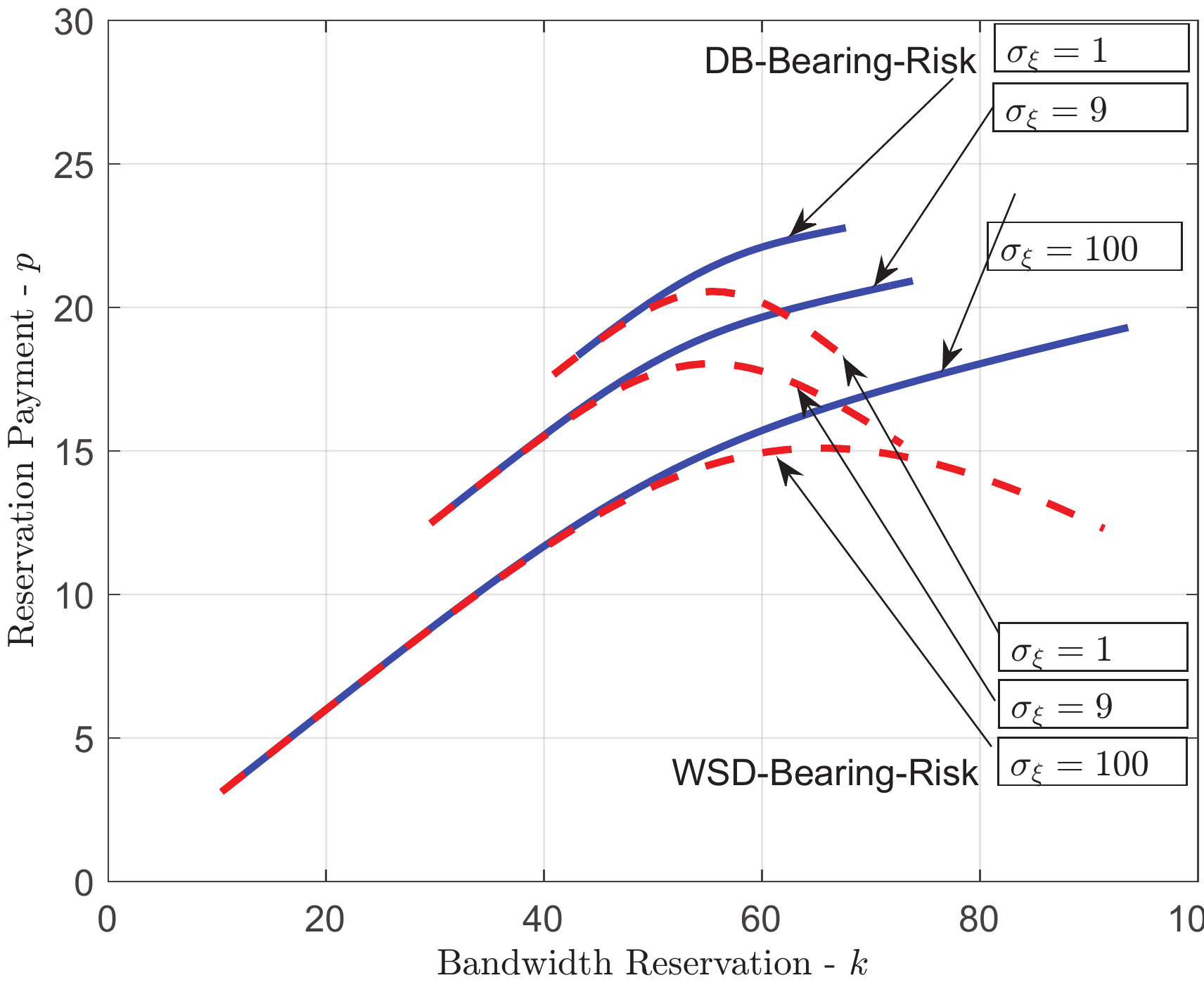}
 	~~~~~~~~
 	\includegraphics[width=2.8in]{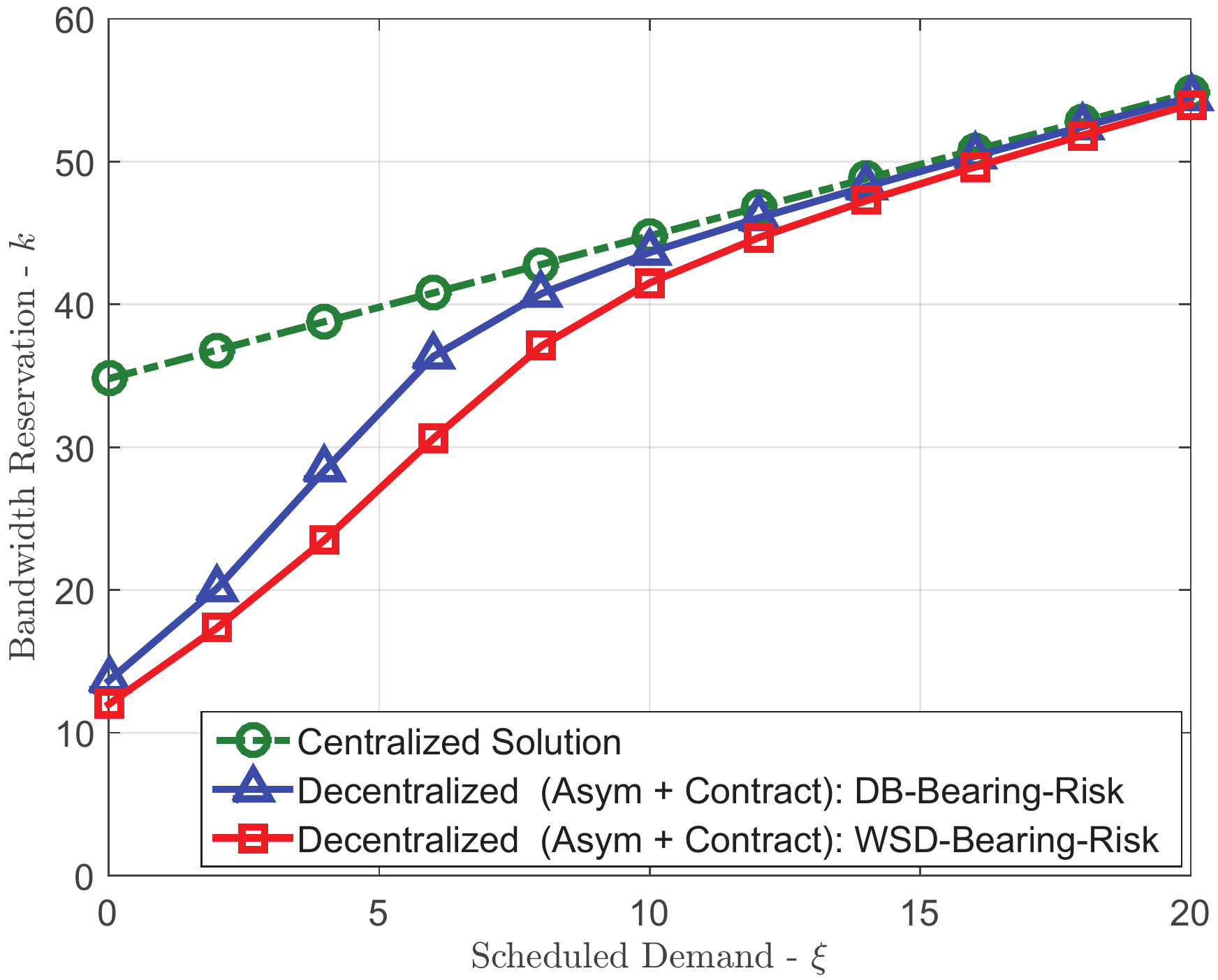}
 	\vspace{-2mm}
 	\caption{ (a) Illustration of Optimal Contracts, (b) Contract-based Spectrum Reservations vs Scheduled Demand $\Da$. Here, $\sigma_{\Da}$ denotes the variance of scheduled demand $\Da$.}\label{fig:k_vs_p_WSD}
% 	\vspace{-4mm}
 \end{figure*}

Then we compare the {\bbd} reservations under both contracts.
By Proposition \ref{lemma:nece2}, both $\Kmsctr (\Da)$ and $\Kdbctr (\Da)$ are increasing in $\Da$.
By  (\ref{asy_optimal_K_in_contract}) and (\ref{asy_optimal_K_in_contract_AD}), we further have the following observation.
\vspace{-2mm}
\begin{lemma}[Contract-based {\bbd} reservation]\label{lemma:contract_compare}
$$
\Kmsctr (\Da) \leq \Kdbctr (\Da) \leq \Kso(\Da),\ \ \forall \Da \in [\underline{\Da}, \bar{\Da}],
$$
and $\Kmsctr({\Da}) = \Kdbctr (\Da)
= \Kso( {\Da})$ only when $\Da = \bar{\Da}$.
\end{lemma}
\noindent That is, only when the realized scheduled demand $\Da$ reaches its maximum value (i.e., $\Da = \bar{\Da}$), the {\bbd} reservations under both optimal contracts are identical, and are equal the integrated optimal {\bbd} reservation.
Under other values of $\Da$, the {\bbd} reservation in the contract  $\CTRms^*$ (under WSD-bearing-risk) is smaller than that in the contract  $\CTRdb^*$ (under  DB-bearing-risk), which is further smaller than the integrated optimal {\bbd} reservation.
%\com{explain the intuition behind this observation. Besides, the following explanation is not clear.}
%The difference in capacity is the system inefficiency because of the information asymmetry. This is because the database distorts the quantities of {\bbd} downward to limit the information rents earned by the WSD.
%This is illustrated by Figure~\ref{fig:k_vs_p_WSD}.b.

We illustrate the result of Lemma \ref{lemma:contract_compare} in Figure~\ref{fig:k_vs_p_WSD}.b. Intuitively,
When the database bears the risk, it has an incentive to charge a high reservation fee in order to force the WSD to shoulder some of the potential cost.
When the WSD bears the risk, however, the database has less incentive to charge a high reservation fee.
Hence, for the same $\Da$, we find that $\Pdbctr(\Da) > \Pmsctr(\Da)$.
Combined with Proposition \ref{lemma:nece1}, we have $\Kmsctr(\Da) < \Kdbctr(\Da)$.

\vspace{-2mm}
\section{Numerical Results}\label{sec:simu}
In this section, we provide numerical results to compare the performances of the proposed contract-based {\bbd} reservation mechanisms.
%We will use these numerical results, together with our previous analysis, \rev{to outline a promising choice of \emph{contract selection} strategy under different risk-bearing schemes.}
%\com{A bit confusing. We we have characterized the optimal contracts. What do we mean "contract selection"?}
Practically speaking, the database's contract choice depends on many factors, among which the {\bbd} reservation decision and the resulting (expected) profit are the most important ones.
Hence, we will present the expected profits (of the database, WSD, and the aggregated one) under different contracts associated with different risk-bearing schemes.
Unless specified otherwise,  we assume the following spectrum trading parameters: $r=1$,
$s=0.8$, $w=0.5$, and $c=0.2$.
We further assume that the scheduled demand $\Da$ follows the normal distribution, and the bursty demand $\Db$ follows the chi-square distribution.\footnote{The parameter setting is for an illustrative purpose; similar insights can be obtained using other parameter settings.}

\begin{figure*}
  \centering
%  \vspace{-5mm}
  \includegraphics[width=2.8in]{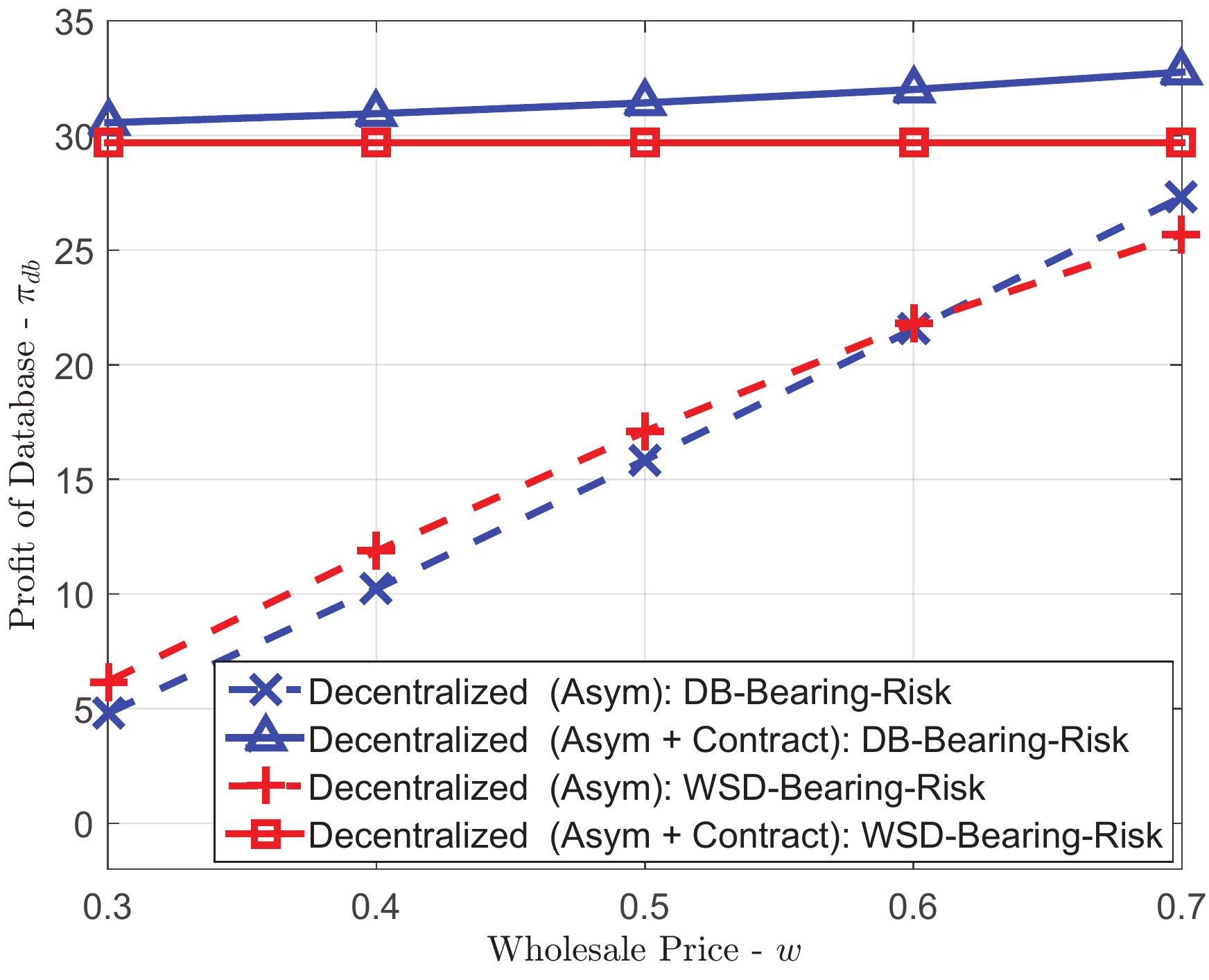}
  ~~~~~~~~
  \includegraphics[width=2.8in]{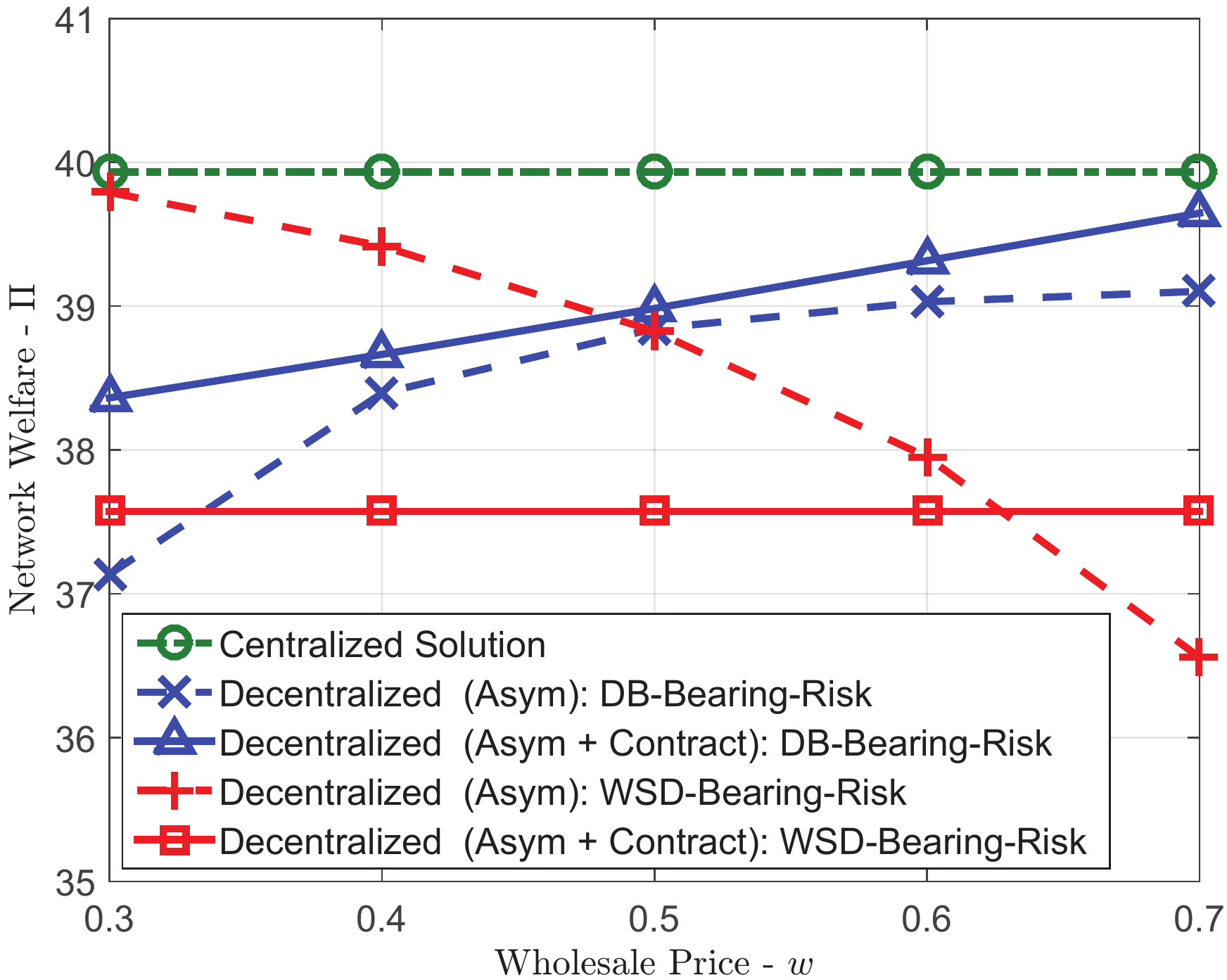}
  \vspace{-2mm}
  \caption{ (a)  Database Profit vs Wholesale Price, and  (b) {Network Profit} vs Wholesale Price.}\label{fig:total_vs_w}
%  \vspace{-2mm}
\end{figure*}

%\begin{figure*}
%	\centering
%	\vspace{-5mm}
%	\includegraphics[width=3.2in]{Figure/database_profit_vs_w_all_without_exte}
%	~~~~
%	\includegraphics[width=3.2in]{Figure/total_profit_vs_w_all_without_exterm}
%	\vspace{-2mm}
%	\caption{ (a)  Database Profit vs Wholesale Price, and  (b) {Network Profit} vs Wholesale Price.}\label{fig:total_vs_w}
%	\vspace{-2mm}
%\end{figure*}

%\vspace{-2mm}
\subsection{Profit vs Wholesale Price}

Figure \ref{fig:total_vs_w} illustrates (a) the database profit and (b) the network profit (aggregate profit)  achieved in different {\bbd} reservation solutions (associated with information asymmetric under different wholesale prices $w$.)
In this simulation, we assume that $\Da$ follows the normal distribution with mean $\EDa = 30$ and variance $\VDa^2 = 64$, and $\Db$ follows the chi-square distribution with mean $\EDb = 30$ and variance $\VDb^2 = 60$.

%\vspace{-1mm}
%\com{use itemize}
From Figure \ref{fig:total_vs_w}.a, we have the following observations regarding the database profit.
\begin{itemize}
\item Under both risking bear-schemes, the contract-based {\bbd} reservation leads to a much higher profit for the database, compared to the reservation solution without information sharing.
\item The database can achieve a higher profit with the optimal {\bbd} reservation contract under DB-bearing-risk (the blue triangle curve) than that under WSD-bearing-risk (the red square curve).
\end{itemize}
This is quite counter-intuitive. The reason is that the WSD is more risk-averse than the database.

From Figure \ref{fig:total_vs_w}.b, we have the following observations.
\begin{itemize}
\item
\emph{Centralized Optimal Network Profit:}
The green circle curve denotes the optimal network profit achieved in the centralized reservation solution $\Kso$ given in (\ref{centralized_bandwidth}), which is independent of the wholesale price $w$, and serves as an upper-bound of the network profit under any other reservation solution.
\item
\emph{Network Profit under DB-Bear-Risk:}
The blue ``x'' (dash) curve and blue triangle (solid) curve denote the network profit achieved under DB-Bearing-Risk, without and with contract, respectively. Specifically, the former one is achieved from the reservation solution without information sharing, i.e., $\Kdbasy $ given in (\ref{asym_decentral_bandwidth}). The latter one is achieved from the optimal {\bbd} reservation contract $\CTRdb^*$ given in Theorem \ref{thrm:optimal-contract}.
Obviously, information sharing based on the optimal {\bbd} reservation contract proposed in this paper improves the total network profit up to $5\%$.

\item
\emph{Network Profit under WSD-Bear-Risk:}
The red ``$+$'' (dash) curve and red square (solid) curve denote the network profit achieved under WSD-Bearing-Risk, with and without contract, respectively.
Specifically, the former one is achieved from the reservation solution without information sharing, i.e., $\Kmsasy $ given in (\ref{asym_decentral_bandwidth2}). The latter one is achieved from the optimal {\bbd} reservation contract $\CTRms^*$ given in Theorem \ref{thrm:optimal-contract-AD}.
Different with the DB-Bearing-Risk scheme, we can see that only when the wholesale price $w$ is large (e.g., $w>0.62$ in this example), the performance under the optimal {\bbd} reservation contract is better than that without information sharing.
This is because the purpose of contract under the WSD-bearing-risk is to reduce the double marginalization effect. Hence, the network profit under WSD-Bearing-Risk contract is independent of the wholesale price.
However, as the objective of contract is maximizing the database profit, the database would charge an equivalent high ``wholesale price" from the WSD. As shown by the Figure \ref{fig:total_vs_w}.b, such equivalent ``wholesale price" lies between $0.6$ and $0.7$. This high equivalent high wholesale price decreases the performance of social welfare.
%can increase the network profit; while when  the wholesale price $w$ is small, the optimal contract will decrease the network profit.
%This is because when $w$ is small, the WSD only needs to pay a low purchasing fee. Moreover, in this case, the WSD's best reservation decision without contract is already close to the centralized optimal reservation (comparing $\Kmsasy $ and $\Kso$). Thus, the database will tend to charge a high registration fee through the reservation contract, which may drive the WSD's reservation decision away from the centralized optimal reservation.
%%When $w$ is large, the observation is just the opposite.
%\com{the discussion in this paragraph are not very clear}

\end{itemize}

%\com{It would be intuitive to talk about database profit first, and then the network profit. This means that we should exchange the locations of Figure 6.a and Figures 6.b and their corresponding discussions.
%For the same reason, we should exchange the positions of Figures 7.a and 7.b and their corresponding discussions. }

{Our results provide the following important insight for a general reservation problem: it is not only individually better, but also socially better to leave the over-reservation risk to the less risk-averse decision maker.}

%\com{A separate paragraph, as this conclusion draws from both Figure 6.a and Figure 6.b? }

%\com{emphasize that this is a counter-intuititve message and explain the intuitition. }

%When we compare the result under DB-Bearing-Risk case, we can see that the optimal contract with credible information sharing can improve the total profit (\rev{network profit}) compared to that without information sharing. The performance differences of two mechanism under information symmetry and information asymmetry have been explained in the previous section.

%Concerning the result under WSD-Bearing-Risk case, only when the value of $w$ is high that the optimal contract with credible information sharing can improve the total profit (\rev{network profit}) compared to that without information sharing. The reason is that.... In terms of database's profit, we can see the performance of two kinds of contracts are much better than that of any other schemes. The reason is that each item in contract is the unique solution that maximize the profit of database given any realized $\xi$. We can also see that the performance of contract with DB-Bearing-Risk scheme is better than that of the contract with WSD-Bearing-Risk scheme. The reason is that the WSD extracts less information rent from the database with a {\bbd} reservation contract (Eq.(\ref{lemma:nece4}) and Eq.(\ref{lemma:nece4_ad})).

%\com{Do we need to put the figure with extreme point, i.e., $w=c$ and $w = s$. I have draw the picture, but the result seems ugly, and we need to explain more. }

\subsection{Profit vs Scheduled Demand Variance}
%\com{Try to put Figure 6 and section 7.2 on the same page. }

Figure \ref{fig:total_vs_vxi} illustrates (a) the database profit and (b) the network profit achieved in different {\bbd} reservation solutions (associated with information asymmetry), under different scheduled demand variance $\VDa^2$.
Notice that $\VDa^2$ reflects the degree of information asymmetry. That is, a higher $\VDa^2$ implies a larger variance of $\Da$, and thus a higher uncertainty of the database regarding $\Da$.
%\rev{ Figure \ref{fig:total_vs_vxi} show the relation of total aggregate profit, and database's profit with $\VDa^2/\VDb$.}
In this simulation, we assume that $\Da$ follows the normal distribution with mean $\EDa = 30$ (and with different variances), and $\Db$ follows the chi-square distribution with mean $\EDb = 30$ and variance $\VDb^2 = 60$.

%\begin{figure*}
%  \centering
%  \includegraphics[width=3.2in]{Figure/database_profit_diff_vxi_all}
%  ~~~~
%  \includegraphics[width=3.2in]{Figure/total_profit_diff_vxi_all}
%  \vspace{-2mm}
%  \caption{ (a) Database Profit vs Scheduled Demand Variance, and (b) {Network Profit} vs Scheduled Demand Variance.}\label{fig:total_vs_vxi}
%  \vspace{-2mm}
%\end{figure*}

\begin{figure*}
	\centering
	\includegraphics[width=2.8in]{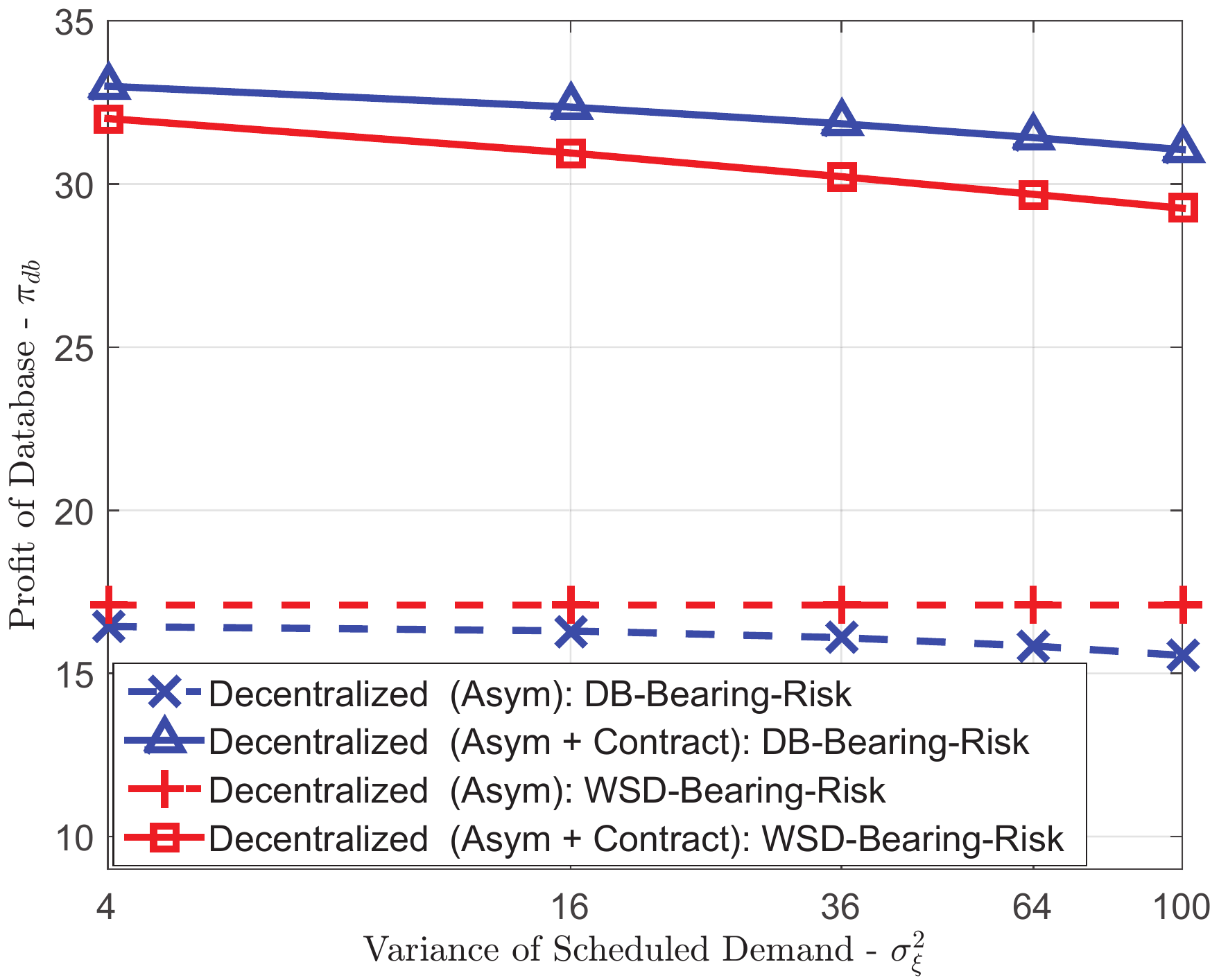}
	~~~~~~~~
	\includegraphics[width=2.8in]{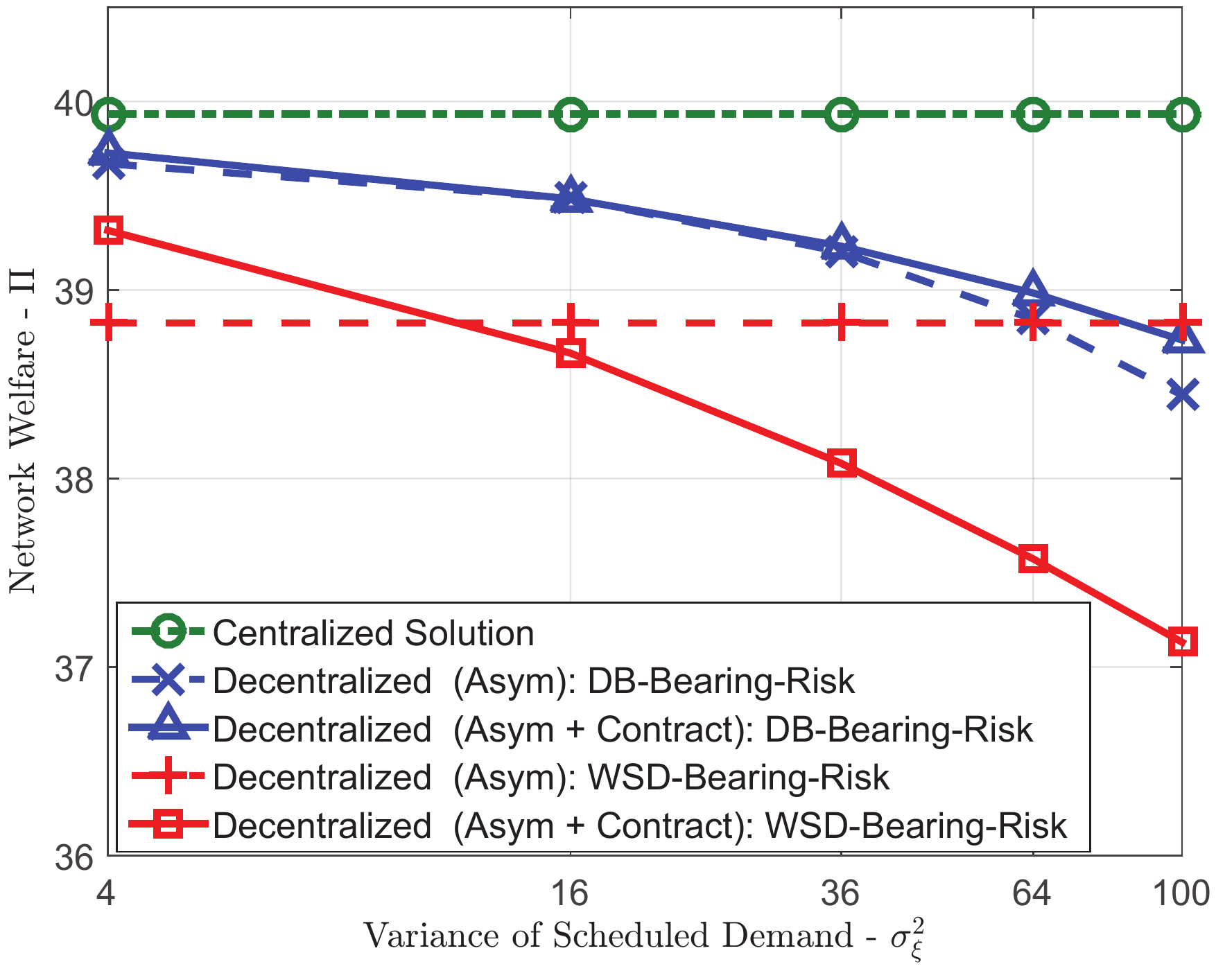}
	\vspace{-2mm}
	\caption{ (a) Database Profit vs Scheduled Demand Variance, and (b) {Network Profit} vs Scheduled Demand Variance.}\label{fig:total_vs_vxi}
	\vspace{-2mm}
\end{figure*}

From Figure \ref{fig:total_vs_vxi}.a, we can further see that under both risk-bearing schemes, the optimal contracts ($\CTRdb^*$ and $\CTRms^*$) can greatly improve the database profit. Moreover, the database can achieve a slightly higher profit with the optimal contract $\CTRdb^*$ under DB-bearing-risk, than the optimal contract $\CTRms^*$ under WSD-bearing-risk.

Figure \ref{fig:total_vs_vxi}.b leads to a similar observation as Figure \ref{fig:total_vs_w}.b. Specifically, under DB-bearing-risk, the optimal contract $\CTRdb^*$ can always increase with the total network profit; while under WSD-bearing-risk, the optimal contract $\CTRms^*$ can only increase the total network profit when $\VDa^2$ is small (i.e., when the degree of information asymmetry is low).
We can further see that the profits under both optimal contracts decrease with $\VDa^2$.
This is because with a larger $\VDa^2$, the scheduled demand $\Da$ varies more dramatically. As the scheduled demand $\Da$ is the private information of the WSD, the larger variance of $\Da$ means that the database needs to pay a higher information rent to the WSDs.

%\vspace{-3.0mm}
%\vspace{3mm}

\section{Conclusion}\label{sec:conclusion}
We  propose a broker-based spectrum reservation market model for TV white space network, under {stochastic demand} and information {asymmetry}.
%where the database acts as a {spectrum broker} reserving the {\bbd} from TV licensees and then resells the {\bbd} to unlicensed WSDs.
%We study the {optimal {\bbd} reservation} for the database under {stochastic nature of demand} and information {asymmetry}.
%Under {information symmetry}, we derive the optimal {\bbd} reservation in a centralized/integrated operational manners (as a benchmark).
To solve the problem, we propose a {contract-based} {\bbd} reservation framework, which ensures WSDs share their private information credibly.
We analyze the incentive compatibility of contracts, and further derive the optimal contracts under different risk-bearing schemes.
Our analysis and extensive simulations indicate that the optimal contract under DB-bearing-risk
%\com{Huang: can? or always?}
leads to a higher database profit and higher network profit than that under WSD-bearing-risk.
\rev{As there is no large-scale commercializatin of TV white space network with detailed spectrum reservation scheme, our work can serve as a first step to give theoretical insights into the problem of risk-bearing between the database and the WSD, and promote the economic study of such a network.}
%It is important to note that our proposed contract-based {\bbd} reservation framework is compatible with (and hence can be applied to) many existing spectrum market scenarios, as it does not need to alter the real spectrum trading process.
%%It is important to note that our proposed contract-based {\bbd} reservation framework focuses only on the reservation decision and the necessary economic incentives, while does not need to alter the real spectrum trading process.
%%This implies that our results are compatible with (and thus applicable to) many existing spectrum market scenarios.

\rev{In this work, we have focused on the TV white space network, where the primary users are the TV broadcasters. As the TV towers have fixed locations and TV programs have well planned schedules, the database has full information regarding the primary usage of TV spectrum ahead of time. This allows us to focus on the demand uncertainty from unlicensed users in this paper.
	On the other hand, the issue of primary usage uncertainly becomes much more important,
	if we consider the Licensed Shared Access (LSA) and Authorised Shared Access (ASA) models, where unlicensed users may access specific non-TV band (\emph{e.g.,} $3.5$ GHz band in the United States and $2.3$ GHz band in Europe).
	This is because these bands are used for ship- and air-borne radar systems which are critical to the operation of the national defense.
	Our model can be directly extended to analyze the LSA/ASA systems, if
	there is no penalty to the database and the WSD for not being able to serve all demands.
	However, when the expected payoffs of the database and the WSD depend on both the demand randomness and the available spectrum randomness, it would be much more challenging to obtain theoretical results by solving the contract design problem.
	We will consider the issue of two-sided uncertainty and the interaction among the licensee, the database, and the WSDs in our future work.}

\vspace{-0.5mm}
%\input{reference2}
%!TEX root = main_CR_Database.tex
%SourceDoc main_CR_Database.tex

\bibliographystyle{IEEEtran}

\begin{thebibliography}{10}

\bibitem{yuan2012}
Y.~Luo, L.~Gao, and J.~Huang, ``Spectrum broker by geo-location database,'' in
  \emph{IEEE GLOBECOM}, 2012, pp. 5427-5432.

\bibitem{federal2010second}
%FCC Tech. Rep., ``In the matter of unlicensed operation in the TV broadcast
%  bands: Third memorandum opinion and order,''
%%  [online]~\url{https://apps.fcc.gov/edocs_public/attachmatch/FCC-12-36A1.pdf},
%  April 2012.
FCC 12-36, \emph{Third Memorandum Opinion and Order}, USA, Federal Communications Commission, 2012.

\bibitem{ofcom2012}
Ofcom, Technical Report 08-260, ``Implementing TV white spaces,'' London, U.k., Feb. 2015.

%\bibitem{Saifan2011sensing}
%A.~E.~K. Saifan, R. and Y.~Guan, ``Efficient Spectrum Searching and Monitoring
%  in Cognitive Radio Network,'' in \emph{IEEE MASS},  2011.
%
%\bibitem{mishra2006cooperative}
%S.~Mishra, A.~Sahai, and R.~Brodersen, ``Cooperative sensing among cognitive
%  radios,'' in \emph{IEEE ICC}, June 2006.

%\bibitem{ECC2010dractreport}
%CEPT-ECC, Tech. Rep., ``Ecc report 159: Technical and operational requirements for the
%  possible operation of cognitive radio systems in the ?white spaces? of the
%  frequency band 470-790 mhz,''  January 2011.

%\bibitem{ECC2010dractreport}
%ETSI, Tech. Rep., ``ETSI EN 301 598: White Space Devices: Wireless Access Systems operating in the 470 MHz to 790 MHz TV broadcast band; Harmonized EN covering the essential requirements of article 3.2 of the R$\&$TTE Directive,''  April 2014.


\bibitem{FCC2007initial}
  S.~Jones and T.~Phillips, ``Initial Evaluation of the Performance of Prototype
  Tv-Band White Space Devices,'' Federal Communication Commission, Tech. Rep.,
  2007.

%\bibitem{spectrumbridge}
%Spectrum Bridge, ``Spectrum Bridge TV white space,''
%  [online]~\url{http://www.spectrumbridge.com}.

\bibitem{IEEE80222}
IEEE, ``Ieee 802.22 working group on wireless regional area networks,''
  [online]~\url{http://www.ieee802.org/22/}.

 \bibitem{ECC2010dractreport}
 CEPT-ECC, Tech. Rep., ``Ecc report 185: Further definition of technical and operational requirements for the
 operation of the operation of white space devices in the band 470-790 Mhz,''  January 2013.

 \bibitem{ETSIreport}
 ETSI, Tech. Rep., ``ETSI harmonized European standard for white spaces devies V 1.1.1,''  April 2014.

%\bibitem{COGEU2015}
%European COGEU project,
%  [online]~\url{http://www.ict-cogeu.eu/}.

%\bibitem{niyato2009}
%D.~Niyato, E.~Hossain, and Z.~Han, ``Dynamic spectrum access in ieee 802.22-
%  based cognitive wireless networks: a game theoretic model for competitive
%  spectrum bidding and pricing,'' \emph{IEEE Wireless Communications},
%  vol.~16, no.~2, pp. 16--23, April 2009.
%
%\bibitem{Qian2011}
%L.~Qian, F.~Ye, L.~Gao, X.~Gan, T.~Chu, X.~Tian, X.~Wang, and M.~Guizani,
%  ``Spectrum trading in cognitive radio networks: an agent-based model under
%  demand uncertainty,'' \emph{IEEE Transactions on Communications}, vol.~59,
%  no.~11, pp. 3192--3203, 2011.

\bibitem{bogucka}
H.~Bogucka, M.~Parzy, P.~Marques, J.~Mwangoka, and T.~Forde, ``Secondary
  spectrum trading in TV white spaces,'' \emph{IEEE Communications Magazine},
  vol.~50, no.~11, pp. 121--129, November 2012.

%\bibitem{larivier1999contract}
%M.~A. Lariviere, ``Supply chain contracting and coordination with stochastic
%  demand,'' \emph{Quantitative Models for Supply Chain Management}, vol.~17,
%  no.~2, pp. 235--268, 1999.

\bibitem{cachon2001contract}
G.~P. Cachon and M.~A. Lariviere, ``Contracting to assure supply: How to share
  demand forecasts in a supply chain,'' \emph{Management Science}, vol.~47,
  no.~5, pp. 629--646, 2001.

\bibitem{ozer2006}
{\"O}.~{\"O}zer and W.~Wei, ``Strategic commitments for an optimal capacity
  decision under asymmetric forecast information,'' \emph{Management Science},
  vol.~52, no.~8, pp. 1238--1257, 2006.

\bibitem{niyato2008}
D.~Niyato and E.~Hossain, ``Spectrum trading in cognitive radio networks: A
  market-equilibrium-based approach,'' \emph{IEEE Wireless Communications},
  vol.~15, no.~6, pp. 71--80, December 2008.

%\bibitem{bahl2009}
%P.~Bahl, R.~Chandra, T.~Moscibroda, R.~Murty, and M.~Welsh, ``White space
%  networking with Wi-Fi like connectivity,'' in \emph{ACM SIGCOMM}, 2009, pp. 27-38.
%
%
%\bibitem{gurney2008geo}
%D.~Gurney, G.~Buchwald, L.~Ecklund, S.~Kuffner, and J.~Grosspietsch,
%  ``Geo-location database techniques for incumbent protection in the TV white
%  space,'' in \emph{IEEE DySPAN}, Oct 2008, pp. 1-9.
%
%%\bibitem{goncalves2011value}
%%V.~Goncalves and S.~Pollin, ``The value of sensing for TV white spaces,'' in
%%  \emph{IEEE DySPAN}, May 2011.
%
%\bibitem{feng2011database}
%X.~Feng, J.~Zhang, and Q.~Zhang, ``Database-assisted multi-AP network on TV
%  white spaces: Architecture, spectrum allocation and AP discovery,'' in
%  \emph{IEEE DySPAN}, 2013, pp. 265-276.

\bibitem{chen2012}
X.~Chen and J.~Huang, ``Game theoretic analysis of distributed spectrum sharing
  with database,'' in \emph{IEEE ICDCS}, 2012, pp. 255-264.

\bibitem{feng2013database}
X.~Feng, Q.~Zhang, and J.~Zhang, ``Hybrid pricing for TV white space
  database,'' in \emph{IEEE INFOCOM}, 2013, pp. 1995-2003.

\bibitem{luo2013}
Y.~Luo, L.~Gao, and J.~Huang, ``Price and inventory competition in oligopoly TV
  white space markets,'' \emph{IEEE Journal
  on Selected Areas in Communications}, vol.~33, no.~5, pp. 1002--1013, May 2015.

\bibitem{luo2014wiopt}
Y.~Luo, L.~Gao, and J.~Huang, ``Trade Information, Not Spectrum: A Novel TV White Space Information Market Model'', \emph{IEEE WiOpt}, 2014, pp. 405-412.

\bibitem{luo2014SDP}
Y.~Luo, L.~Gao, and J.~Huang, ``Information Market for TV White Space'', \emph{IEEE INFOCOM Workshop on SDP}, 2014.

\bibitem{luo2015information}
Y.~Luo, L.~Gao, and J.~Huang, ``MINE GOLD to Deliver Green Cognitive Communications,'' \emph{IEEE Journal
	on Selected Areas in Communications}, vol.~PP, no.~99, 2015.

\bibitem{luo2015magazine}
Y.~Luo, L.~Gao, and J.~Huang, ``Business Modeling for TV White Space Networks'', \emph{IEEE Commun. Mag.}, vol.~53, no.~5, pp.~82-88, May~2015.

\bibitem{luo2015INFOCOM}
Y.~Luo, L.~Gao, and J.~Huang, ``HySIM: A Hybrid Spectrum and Information Market for TV White Space Networks'', \emph{IEEE INFOCOM}, 2015, pp. 604-609.


%\bibitem{kalathil2011contract}
%D.~M. Kalathil and R.~Jain, ``Spectrum sharing through contracts for cognitive
%  radios,'' \emph{IEEE Transactions on Mobile Computing}, vol.~12, no.~10, pp.
%  1999--2011, 2013.

\bibitem{gao2011contract}
L.~Gao, X.~Wang, Y.~Xu, and Q.~Zhang, ``Spectrum trading in cognitive radio
  networks: A contract-theoretic modeling approach,'' \emph{IEEE Journal on Selected Areas in
  Communications}, vol.~29, no.~4, pp. 843--855, 2011.

\bibitem{duan2013contract}
L.~Duan, L.~Gao, and J.~Huang, ``Cooperative spectrum sharing: a contract-based
  approach,'' \emph{IEEE Transactions on Mobile Computing}, vol.~13, no.~1,
  pp. 174--187, 2014.

\bibitem{sheng2013contract}
S.~P. Sheng and M.~Liu, ``Profit incentive in a secondary spectrum market: A
  contract design approach,'' in \emph{IEEE INFOCOM}, 2013, pp. 836-844.

\bibitem{niyato2009investment}
D.~Niyato, E.~Hossain, and Z.~Han, ``Dynamics of multiple-seller and
  multiple-buyer spectrum trading in cognitive radio networks: A game-theoretic
  modeling approach,'' \emph{IEEE Transactions on Mobile Computing}, vol.~8,
  no.~8, pp. 1009--1022, 2009.

\bibitem{goldsmith2005wireless}
A.~Goldsmith, \emph{Wireless communications}, Cambridge, U.K., Cambridge Univ. Press, 2005.

\bibitem{techrpt}
Y. Luo, L. Gao, and J. Huang, Tech. Report, online at
    http://jianwei.ie.cuhk.edu.hk/publication/WSN-Contract.pdf

\end{thebibliography}


\appendix
%\input{Section_Appendix_Proofs_TCCN}

%!TEX root = main_CR_Database.tex
%SourceDoc main_CR_Database.tex

%\newpage
%
%\noindent
%\textbf{Technical Report} for ``\emph{Spectrum Reservation Contract Design in TV White Space Networks}''
%
%\appendix
%\section{Appendix}\label{sec:appendix}

\section{Appendix}\label{sec:appendix}
%\com{Huang: The main text does not mention appendix at all. This is strange. Where in the main text should we mention Appendix A? }

\subsection{Total Spectrum Reservation}
\label{sec:contract_multiple}

%% \com{explain that this is a different model, or a generalization of the previous model?}
%In Section \ref{sec:asymm-contract}, we have derived the optimal reservation contract for each WSD.
%Through the optimal reservation contract, the database can maximize its expected profit from each WSD. 
%Note that when multiple WSDs are located at the same location, they can potentially use the same spectrum resource in different access periods.
%That is, the spectrum reserved for a particular WSD may be assigned to another co-located WSD (as long as it is not used by the former WSD).
%Hence, the database can take advantage of such multiplexing to reduce the total reservation amount, hence reduce its reservation cost. This is especially useful under the DB-bearing-risk scheme, where the database bears the risk of over-reservation.
%
%It is important to note that such a reservation reduction is transparent to WSDs. 
%Namely, when WSDs' realized demands are equal or higher than their claimed reservation amounts, the database needs to replenish spectrum (with a cost higher than the reservation cost) to satisfy the WSDs' demands. 
%This implies that reducing the aggregate reservation amount may lead to a high spectrum replenishing cost for the database. 
%Therefore, there is a tradeoff between the reservation cost reduction and the replenishing cost.
%%As we consider the number of co-located WSDs is large enough so that each WSD's spectrum reservation decision is dependent of other WSDs. Hence, 
In this section, we will show how the database determines the optimal aggregate reservation after knowing WSDs' contract item choices. 

%%In this section, we will study the database's optimal aggregate  reservation under both risk-bearing schemes.
%Note that exploring the multiplexing gain makes the contract design even more challenging, especially when considering the choice of multiple WSDs are correlated. 
%Hence, we consider a sub-optimal problem where the database does the optimal aggregate reservation after receiving contract items selected by all WSDs. In such case, the optimal contract designing problem is the same as that in Section \ref{sec:asymm-contract}. 

\subsubsection{Aggregate Reservation under DB-Bearing-Risk}\label{sec:total_db}

Under the DB-bearing-risk scheme, a WSD only purchases the spectrum that it actually needs in each access period, and hence the database can sell the over-reserved spectrum to other co-located  WSDs in each access period.

%Let $\CTRdb^* =  \{ \langle \Kdbctr (\Da ), \Pdbctr (\Da ) \rangle \}_{\forall \Da}$ denote the optimal contract (defined in Theorem \ref{thrm:optimal-contract}) under the DB-bearing-risk scheme.
Consider a set $\Nset $ of co-located WSDs.
Suppose that in a particular reservation period, the scheduled demand of each WSD $n \in \Nset$ is $\Da_{\n}$.
By the incentive compatibility of the optimal contract $\CTRdb^* =  \{ \langle \Kdbctr (\Da ), \Pdbctr (\Da ) \rangle \}_{\forall \Da}$ (defined in Theorem \ref{thrm:optimal-contract}), each WSD $n $ will choose the reservation amount $\Kdbctr (\Da_n)$ intended for its demand type.
Without the further optimization on the aggregate reservation, the database will simply reserve an amount $\Kdbctr (\Da_n)$ for each WSD $n$, hence the total reservation is $\TKdb = \sum_{n\in\Nset}\Kdbctr (\Da_n)$.
Next we study how to further optimize the aggregate reservation for the database.

%Also note that the database knows all WSDs private information $\Da_{\n}$, $\n \in \Nset$ by offering the optimal contract $\CTRdb$ defined in Theorem \ref{thrm:optimal-contract}.

Let $\TDa = \sum_{\n \in \Nset} \Da_{\n}$ denote the aggregated scheduled demand of all WSDs in $\Nset$,
and let $\TDb = \sum_{\n \in \Nset} \Db_{\n}$ denote the aggregated bursty demand of all WSDs in $\Nset$.
Note that $\TDa$ does not change during the whole reservation period, while $\TDb$ changes every access period.
Moreover, the database can deduce the previous value of the scheduled demand $\Da_{\n}$ of each WSD $n$ (hence the total scheduled demand  $\TDa$) from  the WSDs' selections.
However, neither the database nor the WSDs can obtain the precise value of  $\TDb$, as it changes every access period.
Let $h(\TDb)$ and $H(\TDb)$ denote the p.d.f and c.d.f. of $\TDb$, respectively.

It is easy to see that when the realized total demand $\TD = \TDa + \TDb$ is smaller than the total reservation $\TKdb$, the database can reduce the reservation cost by reducing the reservation amount.
With a reduced aggregate reservation, on the other hand, the database may need to replenish the spectrum reservation (with a higher  replenishment cost) when the realized total demand $\TD $ is larger than the aggregate reservation.
Let $\OTKdb$ denote that the reduced  aggregate reservation  of the database (for WSDs $\Nset$).
Let $\TKdb = \sum_{\n \in \Nset} \Kdbctr(\Da_{\n})$ denote the total reservation requests of all WSDs, and we have $\OTKdb \leq \TKdb $ as each WSD will not purchase an amount larger than its request.

The database profit depends on the actual realization of total demand $\TD$.
Let $c$ be the database's reservation cost, and let $ c^{\textsc{ex}}>c$ be the replenishment cost.
Then, we have,
%According to different realizations of the total demand $\TD$, the database can obtain different revenue.
(a) when $\TD \leq \OTKdb $, the database can reduce the reservation cost by $c \cdot ( \TKdb - \OTKdb  )$;
(b) when $\OTKdb \leq \TD \leq \TKdb$, the database needs to replenish a  reservation amount $\TD - \OTKdb$  to meet the requirements of WSDs, which will introduce  a total replenishment cost of $\ce \cdot (\TD - \OTKdb) $; (c) when $  \TKdb \leq \TD$, the database needs to replenish a  reservation amount $\TKdb - \OTKdb$  to meet the requirements of WSDs, which will introduce  a total replenishment cost of $\ce \cdot (\TKdb - \OTKdb) $.
Based on the above discussion, the expected increasing profit of the database is
\begin{equation}\label{eq:db_revenue_add}
\begin{aligned}
%\AUdb = & \EX_{\TDb}\bigg[ c \cdot ( \TKdb - \OTKdb  ) \cdot \Pr( \TD \leq \OTKdb  \leq \TKdb  ) \\
%		& \qquad{}- \ce \cdot (\TD - \OTKdb) \cdot \Pr(  \OTKdb \leq \TD \leq \TKdb  ) \\
%		& \qquad{} -  \ce\cdot ( \TKdb - \OTKdb ) \cdot \Pr(  \OTKdb \leq \TKdb\leq \TD ) \\
%		& \qquad{} - \c \cdot(  \OTKdb - \TKdb   )\cdot \Pr(   \TD \leq \TKdb \leq \OTKdb   ) \\
%		& \qquad{} + \w \cdot (\TD - \TKdb  )\cdot \Pr(  \TKdb \leq \TD \leq \OTKdb  )   \\
%		&\qquad{} - \c \cdot( \OTKdb - \TD )  \cdot \Pr(  \TKdb \leq \TD \leq \OTKdb  )   \\
%		& \qquad{} + \w \cdot ( \OTKdb - \TKdb  ) \cdot \Pr(  \TKdb \leq \OTKdb \leq \TD   )
%		\bigg]
%		\AUdb = & \EX_{\TDb}\bigg[ c \cdot ( \TKdb - \OTKdb  ) \cdot \Pr( \TD \leq \OTKdb     ) - \ce \cdot (\TD - \OTKdb) \cdot \Pr(  \OTKdb \leq \TD \leq \TKdb  ) \\
%		& \qquad{} -  \ce\cdot ( \TKdb - \OTKdb ) \cdot \Pr(    \TKdb\leq \TD )
%		 - \c \cdot(  \OTKdb - \TKdb   )\cdot \Pr(   \TD \leq \TKdb \leq \OTKdb   ) \\
%		& \qquad{} + \w \cdot (\TD - \TKdb  )\cdot \Pr(  \TKdb \leq \TD \leq \OTKdb  )    - \c \cdot( \OTKdb - \TD )  \cdot \Pr(  \TKdb \leq \TD \leq \OTKdb  )   \\
%		& \qquad{} + \w \cdot ( \OTKdb - \TKdb  ) \cdot \Pr(  \TKdb \leq \OTKdb \leq \TD   )
\AUdb = & \EX_{\TDb}\bigg[ c \cdot ( \TKdb - \OTKdb  ) \cdot \Pr( \TD \leq \OTKdb \leq \TKdb    ) \\
& \qquad{} - \ce \cdot (\TD - \OTKdb) \cdot \Pr(  \OTKdb \leq \TD \leq \TKdb  ) \\
& \qquad{} -  \ce\cdot ( \TKdb - \OTKdb ) \cdot \Pr(    \OTKdb \leq \TKdb\leq \TD )
\bigg]
\end{aligned}
\end{equation}
%\com{Lin: Do we need to consider the case of $\TKdb < \OTKdb$?}
%where $c$ is the database's reservation cost, and $ c^{\textsc{ex}}>c$ is the replenishment cost.
%The first term of (\ref{eq:db_revenue_add}) denotes the benefit by saving the waste reservation,
%the second and the third terms are the replenishment cost for the database in order to satisfy the requirements of WSDs.

The database's optimal aggregate reservation $\OTKdb^*$ that maximizes (\ref{eq:db_revenue_add}) satisfies
\begin{equation}\label{eq:db_total_optimal}
\begin{aligned}
%		&\ce + \w + \c \cdot ( \TKdb - \OTKdb  ) \cdot h(  \OTKdb - \TDa ) \\
%		& \qquad{} \qquad{} - (  \c + \ce + \w  ) \cdot H(  \OTKdb - \TDa ) = 0
%		&\ce + \w + \c \cdot ( \TKdb - \OTKdb  ) \cdot h(  \OTKdb - \TDa )  - (  \c + \ce + \w  ) \cdot H(  \OTKdb - \TDa ) = 0.
&\ce +  \c  \cdot ( \TKdb - \OTKdb^{*}  ) \cdot h(  \OTKdb^{*} - \TDa )  \\
& \qquad{} \quad{} - (  \c + \ce  ) \cdot H(  \OTKdb^{*} - \TDa ) = 0.
\end{aligned}
\end{equation}
%\begin{equation}\label{eq:db_total_optimal}
%\begin{aligned}
%\OTKdb^{*} = \TDa + {H^{-1}} \Big( \frac{ c^{\textsc{ex}}}{c+ c^{\textsc{ex}}}  \Big).
%\end{aligned}
%\end{equation}
Obviously, $\OTKdb^{*} $ is  a function of $\TDa$.

%Figure \ref{fig:total_vs_WSD} illustrates the database profit with and without aggregate reservation optimization under different variance $\VDa^2$ of scheduled demand  $\Da$.
%The blue-square and the red-circle curves denote the database profit with and without further optimization on the aggregate reservation, respectively.
%Notice that $\VDa^2$ reflects the degree of information asymmetry.
%That is, a higher $\VDa^2$ implies a larger variance of $\Da$, and thus a higher uncertainty of the database regarding $\Da$.
%In this simulation, we assume that different WSDs' scheduled demand $\Da_{\n}$, $\n \in \Nset$ are i.i.d., and different WSDs' bursty demand $\Db_{\n}$, $\n \in \Nset$  are also i.i.d.
%The scheduled demand $\Da_{\n}$ of each WSD $n$ follows a  normal distribution with mean $\EDa = 30$ (and with different variances), and the bursty demand $\Db_{\n}$ of each WSD $n$  follows a normal distribution with mean $\EDb = 30$ and variance $\VDb^2 = 60$.

Figure \ref{fig:total_vs_WSD} illustrates the database profit with and without aggregate reservation optimization under different numbers of WSDs.
The blue-square and the red-circle curves denote the database profit with and without further optimization on the aggregate reservation, respectively.
In this simulation, we assume that different WSDs' scheduled demands $\Da_{\n}$, $\n \in \Nset$ are i.i.d., and different WSDs' bursty demands $\Db_{\n}$, $\n \in \Nset$  are also i.i.d.
%We fixed the distribution of the scheduled demand $\Da_{\n}$ of each WSD $n$ while changing the distribution of each WSD's bursty demand. 
The scheduled demand $\Da_{\n}$ of each WSD $n$ follows a  normal distribution with mean $\EDa = 9$ and variance $\VDa^2 = 3$.
The bursty demand $\Db_{\n}$ of each WSD $n$ follows a chi-square distribution with different values of the degrees of freedom.\footnote{{When the degrees of freedom of a chi-square distribution changes, the mean and the variance of this chi-square distribution change accordingly. Specifically, the value of mean is equal to the value of the degrees of freedom, while the value of variance is two times of the value of the degrees of freedom. }}
%The aggregated bursty demand $\sum_{\n \in \Nset} \Db_{\n}$ follows a chi-square distribution with different values of {the degrees of freedom}. 
%Let $h(\TDb)$ and $H(\TDb)$ denote the p.d.f and c.d.f. of $\TDb$, which are the joint p.d.f. and c.d.f. of $\sum_{\n \in \Nset} \Db_{\n}$.
%and the bursty demand $\Db_{\n}$ of each WSD $n$  follows a normal distribution with mean $\EDb = 30$ (and with different variances) .

\begin{figure}
	\vspace{-4mm}
	\centering
	\includegraphics[width=3.2in]{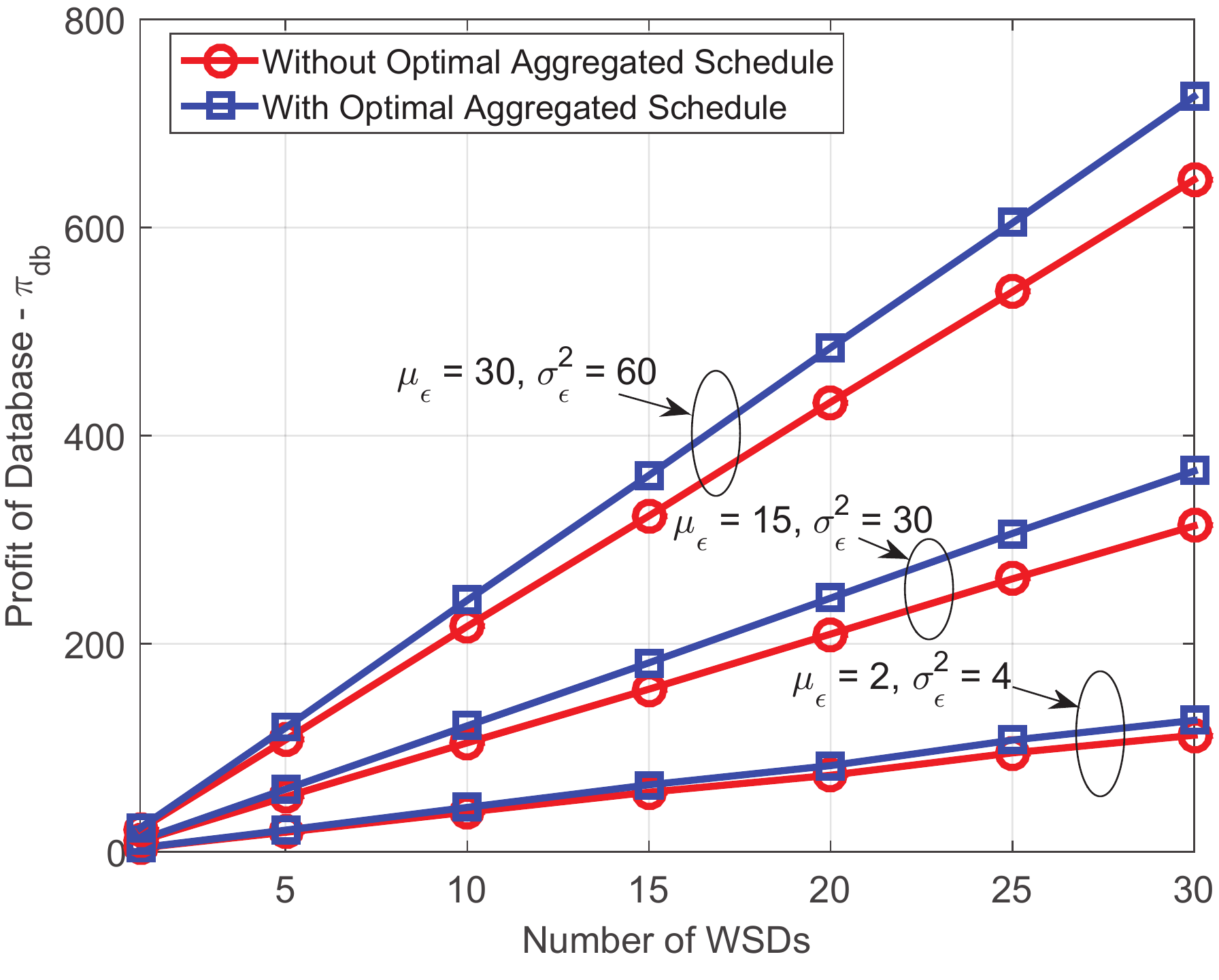}
	\vspace{-2mm}
	%~~~~~~
	%  \includegraphics[width=2.8in]{fig_multiple_WSDs}
	% \caption{ (a)  {Database Profit} vs Wholesale Price, and (b) Database Profit vs Scheduled Demand Variance under different number of WSDs.}\label{fig:total_vs_WSD}
	\caption{ \rev{Database profit vs. numbers of WSDs, where $\EDb$ and $\VDb^2$ denote the mean and variance of the bursty demand of each WSD $n$, respectively.} }\label{fig:total_vs_WSD}
	\vspace{-4mm}
\end{figure}

Figure \ref{fig:total_vs_WSD} shows that with the further optimization, the database can increase its profit up to $12\%$.
This benefit increases with the number of WSDs, as more WSDs submitting their spectrum reservation requirements, more freedom for the database to assign over-reserved spectrum of one WSD to other WSDs in need.
As the mean $\EDb$ and variance $\VDb^2$ of the bursty demand increase, 
%We can also see that 
the difference between the database profit with and without aggregated reservation schedule also increases. 
%with the variance of bursty demand $\VDb^2$.
%$\sum_{\n \in \Nset} \Db_{\n}$. 
% This is because the higher value of variance, the more freedom for the database to do the optimal scheduling.  
%Figure \ref{fig:total_vs_WSD} also shows that the database profit increases with the mean of bursty demand $\EDb$.
%%$\sum_{\n \in \Nset} \Db_{\n}$. 

%We can also see that the  database profit increases with the mean of $\sum_{\n \in \Nset} \Db_{\n}$. 
%
%
%variance $\VDb^2$ of $\Db$. This is because.....

%decreases with the variance $\VDa^2$ of $\Da$ in both cases.
%This is because with a larger variance $\VDa^2$, the scheduled demand $\Da$ varies more dramatically; hence,
%\rev{the inherent cost of information sharing also becomes larger.}
%\com{not clear}
%\com{global change "database's revnue" to "database revenue"}

\subsubsection{Aggregate Reservation under WSD-Bearing-Risk}\label{sec:contract_multiple_WSD}

Under the DB-bearing-risk scheme, a WSD purchases all the spectrum it requests in each access period, even if the realized demand is smaller than the requested reservation. Hence,  the database cannot sell the over-reserved spectrum to other co-located  WSDs in each access period.  In this case, the database does not need to make the further optimization on the aggregate reservation.
Namely, the database's optimal aggregate reservation $\OTKms^{*}$ is\footnote{According to existing regulations \cite{federal2010second, ofcom2012}, WSDs cannot directly communicate with each other to sell their extra spectrum. Hence, we assume that the over-reservation spectrum under the WSD-bearing-risk scheme is wasted. How to let different WSDs trade their over-reservation spectrum through the database will be our future work.},
\begin{equation}\label{eq:ms_total_optimal}
\OTKms^{*} =  \sum_{\n \in \Nset} \Kmsctr(\Da_{\n}),
\end{equation}
which is exactly  the total requested reservation of all WSDs.

%Note that under WSD-bearing-risk, the database needs to return the amount of $\Kmsctr(\Da_{\n})$ to WSD $\n \in \Nset$ based on WSD $\n$'s contract selection. Hence, the database needs to reserve the total amount of $\sum_{\n \in \Nset} \Kmsctr(\Da_{\n})$, and the optimal total reservation for the database is\footnote{According to the regulator's rule \cite{Ofcom2010geo, spectrumbridge, IEEE80222}, WSDs cannot directly communicate to sell their extra spectrum. Hence, we just assume that the over-reservation spectrum under WSD-bearing-risk is wasted. How to let different WSDs trade their over-reservation spectrum through the database will be our future work.}:
%\begin{equation}\label{eq:ms_total_optimal}
%\OTKms^{*} =  \sum_{\n \in \Nset} \Kmsctr(\Da_{\n})
%\end{equation}

\subsection{Proof for Lemma \ref{theorem:optimality_sym}}\label{theorem:optimality_sym-proof}
\begin{proof}
Since $s > w > c$, we have $\Kso$ is always larger than $\Kdbsym$ and $\Kmssym$. Note that $\Kdbsym  = \Da +
 {G^{-1} \Big( {(w-c)}/{w}  \Big)}$ is monotonic increasing with wholesale price $w$ while $\Kmssym
= \Da + {G^{-1} \Big( {(s-w)}/{s}  \Big)}$ decreases monotonically with $w$. We can easily get the conclusion.
\end{proof}

\subsection{Proof for Lemma \ref{theorem:optimality_sym_sw}}\label{theorem:optimality_sym_sw-proof}
\begin{proof}
Note that (\ref{sym_dencentral_bandwidth}) and (\ref{sym_dencentral_bandwidth2}) show that the optimal bandwidth $\K \geq \Da$ for any realized $\Da$. Hence, we can write the expected network profit in (\ref{centralized_profit}) with respect to $\Da$ as:
\begin{equation}\label{centralized_profit_appendix}
\begin{aligned}
%\EX_{\Da}[\Utot] &= ( r - c) E_{\Da}[\K] - (r - s) ( \EX_{\Da}[\K] - \mu_{\Da} ) \\
%& \quad{} - s \cdot \int_0^{\EX_{\Da}[\K] - \mu_{\Da}} G(u) \mathrm{d}u
\EX_{\Da}[\Utot] = & ( r - c) E_{\Da}[\K] \\
& - (r - s) ( \EX_{\Da}[\K] - \mu_{\Da} )  - s \cdot \int_0^{\EX_{\Da}[\K] - \mu_{\Da}} G(u) \mathrm{d}u
\end{aligned}
\end{equation}
where $\mu_{\Da} = \EX[\Da]$ is the mean of the scheduled demand. Moreover, the first derivative with respect to expected bandwidth $\EX_{\Da}[\K]$ is:
\begin{equation}
\begin{aligned}
\frac{\partial{\EX_{\Da}[\Utot]}}{\partial{\EX_{\Da}[\K]}} &= ( s - c) - s \cdot G( \EX_{\Da}[\K] - \mu_{\Da} )
\end{aligned}
\end{equation}
and the second derivative is $-s \cdot g( \EX_{\Da}[K] - \mu_{\Da} ) \leq 0$.
Hence, the maximum solution of (\ref{centralized_profit_appendix}) is $\EX_{\Da}[\Kso]$ and the network profit increases with $\EX_{\Da}[\K]$ when $\EX_{\Da}[\K] < \EX_{\Da}[\K]$. By applying Lemma \ref{theorem:optimality_sym}, we can get the conclusion.
%Since $s > w > c$, we have the centralized optimal bandwidth decision $K^{c}_{sym}$ is larger than the database's optimal decision $K^{d}_{sym}$ in Eq.(\ref{sym_dencentral_bandwidth}), and master's optimal decision $K^{m}_{sym}$ in Eq.(\ref{sym_dencentral_bandwidth2}). Hence the total expected aggregate profit increase with the expected bandwidth. Note that $E_{\xi}(K^{d}_{sym}) = \mu_{\xi} + G^{-1}( (w-c)/w)$ is monotonic increasing with wholesale price $w$ while $E_{\xi}(K^{m}_{sym}) = \mu_{\xi} + G^{-1}( (s-w)/s)$ decreases monotonically with $w$. Then we can conclude that the expected total profit achieved by database's decision is monotonic increasing with $w$ and the expected profit achieved by master's decision is monotonic decreasing with $w$. And the only intersection of these two achieving expected aggregated profit is only when $E_{\xi}(K^{d}_{sym}) = E_{\xi}(K^{m}_{sym})$, i.e., $w = \sqrt{sc}$
\end{proof}

\subsection{Proof for Proposition \ref{lemma:nece1}}\label{lemma:nece1-proof}
\begin{proof}
We prove it by contradiction.
$\leftarrow$: When $\xi = \xi_1$, the following IC constraint must be satisfied:
\begin{equation}
\begin{aligned}
%&(s-w) \EX_{\Db} \big[ \min\big\{{\Db} ,  (\K(\Da_2)-\Da_1 )^{+}\big\} \big] \\
%& -(s-w)\EX_{\Db} \big[ \min\big\{{\Db} ,  (\K({\Da_1})-\Da_1 )\big\} \big] \geq  \P(\Da_2)- \P(\Da_1)
&(s-w) \EX_{\Db} \big[ \min\big\{{\Db} ,  (\K(\Da_2)-\Da_1 )^{+}\big\} \big] \\
&~~ -(s-w)\EX_{\Db} \big[ \min\big\{{\Db} ,  (\K({\Da_1})-\Da_1 )\big\} \big] \geq  \P(\Da_2)- \P(\Da_1)
\end{aligned}
\end{equation}
from which we can find that if $\P(\Da_1) >  \P(\Da_2)$, then $\K (\Da_1) >  \K (\Da_2)$. For otherwise, the IC cannot be satisfied.
$\rightarrow$: We first differentiate (\ref{eq:opt-p}) with respect to $\Da$ and get:
\begin{equation}
\begin{aligned}
\frac{\dt{\P (\Da)}}{\dt{\Da}}  = (s-w) \cdot [ 1 - G(\K(\Da) - \Da)] \cdot \frac{ \dt{\K(\Da)} }{ \dt{\Da} }
\end{aligned}
\end{equation}
Then we can conclude that if $K (\xi_1) > K (\xi_2)$, then $P (\xi_1) > P (\xi_2)$
\end{proof}

\subsection{Proof for Proposition \ref{lemma:nece2}}\label{lemma:nece2-proof}

\begin{proof} We prove it by contradiction. Note that:
\begin{equation}
\begin{aligned}
\frac{\partial^2{ \Ums (\K,\P,\Da )}}{\partial{\K}\partial{\Da}} = (s-w) g \big( \K -\Da \big) &> 0 , \mathrm{~~and~~}
\end{aligned}
\end{equation}
\begin{equation}
\begin{aligned}
\frac{\partial^2{\Ums (\K,\P,\Da )}}{\partial{\K}^2} = -(s-w  ) g \big( \K -\Da \big) < 0.
\end{aligned}
\end{equation}

Hence, for any $\Da_1 > \Da_2$, if $\K(\Da_1) < \K(\Da_2)$. That we have:
\begin{equation}
\begin{aligned}
0 &=\left.\frac{\partial{\Ums\left(\K,\P,\Da_1 \right)}}{\partial{\K}}\right|_{\K=\K(\Da_1)} \nonumber\\
&> \left.\frac{\partial{\Ums\left(\K,\P,\Da_1 \right)}}{\partial{\K}}\right|_{\K=\K(\Da_2)} \nonumber\\
&> \left.\frac{\partial{\Ums\left(\K,\P,\Da_2 \right)}}{\partial{\K}}\right|_{\K=\K(\Da_2)},
\end{aligned}
\end{equation}
where the equality is because of IC and the inequalities are because of the sign of the second-order derivatives. But this contradicts the optimality of $\K\left(\Da_2\right)$. \end{proof}

\subsection{Proof for Proposition \ref{lemma:nece3} } \label{lemma:nece3-proof}

\begin{proof} The IC constraint implies that $\Ums ({\Da}) = \max_{\hDa}  \Ums \big( \K(\hDa), \P(\hDa),\Da  \big)$. The envelope theorem further shows that:
\begin{equation}
\begin{aligned}
\frac{\dt{\Ums} (\Da )}{ \dt{\Da}} & = \left.\frac{\partial{\Ums (\K({\hDa}),\P(\hDa),\Da_i )}}{\partial{\Da}}\right|_{\hDa = \Da} \\
& =  (r-s  ) + (s-w  ) G \big( \K( {\Da})-\Da \big) > 0.
\end{aligned}
\end{equation}
Then we have $\Ums(\Da)$ is increasing in $\Da $. \end{proof}

\subsection{Proof for Proposition \ref{lemma:nece4} } \label{lemma:nece4-proof}
\begin{proof} By using IC constraint and the envelope theorem, we have:
\begin{equation}\label{asy_contract_master_profit_proof}
\begin{aligned}
\frac{\dt{\Ums} (\Da )}{\dt{\Da}} & = \left.\frac{\partial{\Ums (\K({\hDa}),\P(\hDa),\Da_i )}}{\partial{\Da}}\right|_{\hDa = \Da} \\
& =  (r-s  ) + (s-w  ) G \big( \K( {\Da})-\Da \big).
\end{aligned}
\end{equation}
By integrating both sides, we get the Proposition \ref{lemma:nece4}. \end{proof}

\subsection{Proof for Proposition \ref{lemma:sufficent}}\label{lemma:sufficent-proof}

\begin{proof} We only have to show that these three conditions imply IC and IR. Noted that (\ref{eq:opt-p}), $\Ums(\Da) $ is obtained by Lemma (\ref{lemma:nece4}). We therefore have:
\begin{equation}
\begin{aligned}
%&\Ums \big( \K(\hDa),\P(\hDa),\Da  \big) \nonumber\\
%& \quad{} = {\int_{\underline{\Da}}^{\Da}}\frac{\partial{\Ums \big( \K(\hDa),\P(\hDa),x  \big)}}{\partial{x}} \dt{x}+ \Ums \big( \K(\hDa),\P(\hDa),\underline{\Da}  \big) \nonumber\\
%& \quad{} = \Ums \big( \K(\hDa),\P(\hDa),\hDa \big)+ {\int_{\hDa}^{\Da}}\big[(r-s) +(s-w)\\
%& \qquad{} \qquad{} \qquad{} \qquad{} \qquad{} \qquad{} \quad{}  \cdot G\big( \K(\Da) - x \big)  \big]\mathrm{d}{x} \nonumber\\
%& \quad{}= \Ums \big( \K(\Da),\P(\Da),\Da \big) + {\int_{\hDa}^{\Da}}(s-w)\big[G\big( \K(\hDa)- x \big) \\
%& \qquad{} \qquad{} \qquad{} \qquad{} \qquad{} \qquad{} \qquad{} -G\big( \K(x)-x \big)\big] \mathrm{d}{x}
&\Ums \big( \K(\hDa),\P(\hDa),\Da  \big) \nonumber\\
& \quad{} = {\int_{\underline{\Da}}^{\Da}}\frac{\partial{\Ums \big( \K(\hDa),\P(\hDa),x  \big)}}{\partial{x}} \dt{x}+ \Ums \big( \K(\hDa),\P(\hDa),\underline{\Da}  \big) \nonumber\\
& \quad{} = \Ums \big( \K(\hDa),\P(\hDa),\hDa \big) \nonumber\\
& \qquad{} + {\int_{\hDa}^{\Da}}\big[(r-s) +(s-w)  \cdot G\big( \K(\Da) - x \big)  \big]\mathrm{d}{x} \nonumber\\
& \quad{}= \Ums \big( \K(\Da),\P(\Da),\Da \big) \nonumber\\
& \qquad{} + {\int_{\hDa}^{\Da}}(s-w)\big[G\big( \K(\hDa)- x \big)  -G\big( \K(x)-x \big)\big] \mathrm{d}{x}
\end{aligned}
\end{equation}
If $\Da > \hDa$, then the above equation is non-positive (because both $\K$ and $G$ are increasing) and hence $\Ums \big( \K(\Da),\P(\Da),\Da \big) \geq
\Ums \big( \K({\hDa}),\P(\hDa),\Da \big)$. This inequality also holds for $\Da < \hDa$ by a similar argument. Therefore, the two condition imply IC.

Evaluate $\Ums  (\Da)$ at $\underline{\Da}$ and using (\ref{eq:opt1}), we can get IR immediately. \end{proof}

\subsection{Proof for Proposition \ref{lemma:OCI}}\label{lemma:OCI-proof}
\begin{proof} Let $i=1,2$ and define ${\hDa}_i^{*} = \arg\max_{\hDa}\Ums \big( \K(\hDa),\P(\hDa),\Da_i  \big)$  We therefore have $\Ums \big( \K({\hDa}_1^{*}),\P({\hDa}_1^{*}),\Da_2  \big) \leq \Ums \big( \K({\hDa}_2^{*}),\P({\hDa}_2^{*}),\Da_2  \big)$. Noted that:
\begin{equation}
\begin{aligned}
\frac{\partial{\Ums \big( \K({\hDa}),\P(\hDa),\Da  \big)}}{\partial{\Da}} & = (r-s) + (s-w)G\big(\K(\hDa)-\Da  \big) \nonumber\\
& \geq 0.
\end{aligned}
\end{equation}

Hence, for any $\Da_1<\Da_2$, We have $\Ums \big( \K({\hDa}_1^{*}),\P({\hDa}_1^{*}),\Da_1  \big) \leq \Ums \big( \K({\hDa}_1^{*}),\P({\hDa}_1^{*}),\Da_2  \big)$. By using the IC constraint, we have ${\hDa}_i^{*} = \Da_i$, and therefore $\Ums \big( \K(\Da_1),\P(\Da_1),\Da_1  \big) \leq \Ums \big( \K(\Da_2),\P(\Da_2),\Da_2  \big)$. Hence IR only needs to be satisfied at $\Da = \underline{\Da}$ and the other participation constraints for $\Da > \underline{\Da}$ are redundant. Hence, we get (\ref{eq:opt1}).  \end{proof}

\subsection{Proof for Theorem \ref{thrm:optimal-contract}}\label{thrm:optimal-contract-proof}
This Theorem can be simply proved by using Proposition \ref{lemma:sufficent} and Proposition \ref{lemma:OCI}. Here we jus show that the optimal point of $\phidb (\K(\Da),\Da )$ is unique by proving $\phidb (\K(\Da),\Da ) $ is unimodal.

\begin{proof}
We let $z(\Da) = \K(\Da)-\Da$, then we rewrite $\phidb (\K(\Da),\Da ) $ as $\phidb (z(\Da))$ and get
\begin{equation}
\begin{aligned}
%\frac{\partial {\phidb (z )}}{\partial{z} } = & s \cdot [1-G ( z )  ]  - c \\& \textstyle  - \frac{1-F(\Da)}{f(\Da)} ( s-w  ) g( z  ).
\frac{\partial {\phidb (z )}}{\partial{z} } = & s \cdot [1-G ( z )  ]  - c   - \frac{1-F(\Da)}{f(\Da)} ( s-w  ) g( z  ).
\end{aligned}
\end{equation}

To prove that $\phidb (z) $ is unimodal, it suffices to show that ${\partial {\phidb (z )}}/{\partial{z} }$ changes the sign once. Note that $\Db$ follows chi-square distribution with support $[0, +\infty)$. Then we have $\lim_{z \rightarrow 0}{\partial {\phidb (z )}}/{\partial{z} } = s - c >0 $,and $\lim_{z \rightarrow +\infty}{\partial {\phidb (z )}}/{\partial{z} } = - c <0 $. Then we consider the second order derivative of $\phidb (z ) $ with respect to $z$ and we have
\begin{equation}
\begin{aligned}
\frac{\partial^2 {\phidb (z )}}{\partial{z^2} } &=  - s \cdot g ( z  )  - \frac{1-F(\Da)}{f(\Da)} ( s-w  ) g^{'}( z  ) \\
&= \frac{z^{\frac{n}{2} - 2} e^{-\frac{z}{2}} }{\Gamma(\frac{n}{2}) 2^{\frac{n}{2}}} \left[  \frac{A(\xi) - 2}{2} z - A(\xi)\left(\frac{n}{2} - 1\right) \right]
\end{aligned}
\end{equation}
where $A(\xi) = { \left( (1-F(\Da))(s-w)\right) }/{(s \cdot f(\Da))}$ and $n$ is the freedom of chi-square distribution. Note that $\lim_{z \rightarrow +\infty}{\partial^2 {\phidb (z )}}/{\partial{z^2} } <0 $, and
case 1: $\frac{A - 2}{2} < 0$, then ${\partial^2 {\phidb (z )}}/{\partial{z^2} } < 0$, and ${\partial {\phidb (z )}}/{\partial{z} }$ changes the sign once;
case 2: $\frac{A - 2}{2} > 0$, then the value of ${\partial^2 {\phidb (z )}}/{\partial{z^2} }$ is first negative, then becomes positive. However, as ${\partial^2 {\phidb (z )}}/{\partial{z^2} }$ is the linear function of $z$, then ${\partial {\phidb (z )}}/{\partial{z} }$ only changes the sign once.

If $\Db$ follows from normal distribution with mean $\mu$ and standard variance $\sigma$, then $\phidb (z) $ is also unimodal as long as $G(x) \rightarrow 0$ when $x<0$. In such case, $\lim_{z \rightarrow 0}{\partial {\phidb (z )}}/{\partial{z} } = s - c >0 $, and $\lim_{z \rightarrow +\infty}{\partial {\phidb (z )}}/{\partial{z} } = - c <0 $. And
\begin{equation*}
\begin{aligned}
\frac{\partial^2 {\phidb (z )}}{\partial{z^2} } &=  - s \cdot g ( z  )  - \frac{1-F(\Da)}{f(\Da)} ( s-w  ) g^{'}( z  ) \\
&= \frac{1 - F(\Da) }{f(\Da)} \frac{s-w}{s} \frac{1}{ \sigma \sqrt{2 \pi} } e^{- \frac{ (z - \mu )^2 }{ 2 \sigma^2  }} \left[ \frac{z}{\sigma^2} - \frac{\mu}{\sigma^2} - 1 \right]
\end{aligned}
\end{equation*}
The above equation shows that the value of ${\partial^2 {\phidb (z )}}/{\partial{z^2} }$ first is negative, then become positive. However, as ${\partial^2 {\phidb (z )}}/{\partial{z^2} }$ is the linear function of $z$, then ${\partial {\phidb (z )}}/{\partial{z} }$ only changes the sign once.
\end{proof}

\subsection{Proof for Proposition \ref{lemma:nece4_ad} } \label{lemma:nece4_ad-proof}
\begin{proof} By using IC constraint and the envelope theorem, we have:
\begin{equation}\label{asy_contract_master_profit_ad_proof}
\begin{aligned}
\frac{\dt{\Ums} (\Da )}{\dt{\Da}} & = \left.\frac{\partial{\Ums (\K({\hDa}),\P(\hDa),\Da_i )}}{\partial{\Da}}\right|_{\hDa = \Da} \\
& =  (r-s  ) + s \cdot G \big( \K( {\Da})-\Da \big).
\end{aligned}
\end{equation}
By integrating both sides, we get the Proposition \ref{lemma:nece4_ad}. \end{proof}

\subsection{Proof for Theorem \ref{thrm:optimal-contract-AD}}\label{thrm:optimal-contract-AD-proof}
This Theorem can be simply proved by using Proposition \ref{lemma:sufficent} and Proposition \ref{lemma:OCI}. Here we jus show that the optimal point of $\phims (\K(\Da),\Da )$ is unique by proving $\phims (\K(\Da),\Da ) $ is unimodal.

\begin{proof}
We let $z(\Da) = \K(\Da)-\Da$, then we rewrite $\phims (\K(\Da),\Da ) $ as $\phims (z(\Da))$ and get
\begin{equation}
\begin{aligned}
%\frac{\partial {\phims (z )}}{\partial{z} } = & s \cdot [1-G ( z )  ]  - c \\& \textstyle  - \frac{1-F(\Da)}{f(\Da)} \cdot  s \cdot g( z  ).
\frac{\partial {\phims (z )}}{\partial{z} } = & s \cdot [1-G ( z )  ]  - c  - \frac{1-F(\Da)}{f(\Da)} \cdot  s \cdot g( z  ).
\end{aligned}
\end{equation}

To prove that $\phims (z) $ is unimodal, it suffices to show that ${\partial {\phims (z )}}/{\partial{z} }$ changes the sign once. Note that $\Db$ follows chi-square distribution with support $[0, +\infty)$. Then we have $\lim_{z \rightarrow 0}{\partial {\phims (z )}}/{\partial{z} } = s - c >0 $,and $\lim_{z \rightarrow +\infty}{\partial {\phims (z )}}/{\partial{z} } = - c <0 $. Then we consider the second order derivative of $\phims (z ) $ with respect to $z$ and we have
\begin{equation}
\begin{aligned}
\frac{\partial^2 {\phims (z )}}{\partial{z^2} } &=  - s \cdot g ( z  )  - \frac{1-F(\Da)}{f(\Da)} \cdot s \cdot g^{'}( z  ) \\
&= \frac{z^{\frac{n}{2} - 2} e^{-\frac{z}{2}} }{\Gamma(\frac{n}{2}) 2^{\frac{n}{2}}} \left[  \frac{A(\xi) - 2}{2} z - A(\xi)\left(\frac{n}{2} - 1\right) \right]
\end{aligned}
\end{equation}
where $A(\xi) = { \left( (1-F(\Da)) \cdot s\right) }/{(s \cdot f(\Da))}$ and $n$ is the freedom of chi-square distribution. Note that $\lim_{z \rightarrow +\infty}{\partial^2 {\phims (z )}}/{\partial{z^2} } <0 $, and
case 1: $\frac{A - 2}{2} < 0$, then ${\partial^2 {\phims (z )}}/{\partial{z^2} } < 0$, and ${\partial {\phims (z )}}/{\partial{z} }$ changes the sign once;
case 2: $\frac{A - 2}{2} > 0$, then the value of ${\partial^2 {\phims (z )}}/{\partial{z^2} }$ is first negative, then becomes positive. However, as ${\partial^2 {\phims (z )}}/{\partial{z^2} }$ is the linear function of $z$, then ${\partial {\phims (z )}}/{\partial{z} }$ only changes the sign once.

If $\Db$ follows from normal distribution with mean $\mu$ and standard variance $\sigma$, then $\phims (z) $ is also unimodal as long as $G(x) \rightarrow 0$ when $x<0$. In such case, $\lim_{z \rightarrow 0}{\partial {\phims (z )}}/{\partial{z} } = s - c >0 $, and $\lim_{z \rightarrow +\infty}{\partial {\phidb (z )}}/{\partial{z} } = - c <0 $. And
\begin{equation*}
\begin{aligned}
\frac{\partial^2 {\phims (z )}}{\partial{z^2} } &=  - s \cdot g ( z  )  - \frac{1-F(\Da)}{f(\Da)} \cdot s \cdot g^{'}( z  ) \\
&= \frac{1 - F(\Da) }{f(\Da)} \frac{s-w}{s} \frac{1}{ \sigma \sqrt{2 \pi} } e^{- \frac{ (z - \mu )^2 }{ 2 \sigma^2  }} \left[ \frac{z}{\sigma^2} - \frac{\mu}{\sigma^2} - 1 \right]
\end{aligned}
\end{equation*}
The above equation shows that the value of ${\partial^2 {\phims (z )}}/{\partial{z^2} }$ first is negative, then become positive. However, as ${\partial^2 {\phims (z )}}/{\partial{z^2} }$ is the linear function of $z$, then ${\partial {\phims (z )}}/{\partial{z} }$ only changes the sign once.
\end{proof}

\subsection{Proof for Lemma \ref{lemma:contract_compare} } \label{lemma:contract_k_compare-proof}
\begin{proof}
By comparing (\ref{asy_optimal_K_in_contract}) and (\ref{asy_optimal_K_in_contract_AD}), we can easily get the conclusion.
\end{proof}

\end{document}